\documentclass[11pt,letterpaper]{article}
\usepackage[T1]{fontenc}
\usepackage[utf8]{inputenc}
\usepackage{mathpazo}
\usepackage[margin=1in]{geometry}
\usepackage{amsmath,amssymb,amsthm,mathtools,bm}
\usepackage{microtype,booktabs,array,tabularx,enumitem,float,needspace}
\usepackage{xcolor}
\definecolor{deepred}{rgb}{0.58,0,0}
\definecolor{darkgreen}{rgb}{0,0.31,0.11}
\usepackage{tikz}
\usetikzlibrary{arrows.meta,positioning,fit,decorations.pathreplacing}
\usepackage[pagebackref,colorlinks=true,linkcolor=deepred,citecolor=darkgreen,urlcolor=blue]{hyperref}
\usepackage{aliascnt,etoolbox}
\usepackage[capitalise,nameinlink,noabbrev]{cleveref}
\setlist{itemsep=1pt,topsep=3pt}
\allowdisplaybreaks[1]
\newtheorem{theorem}{Theorem}[section]
\newaliascnt{lemma}{theorem}\newtheorem{lemma}[lemma]{Lemma}\aliascntresetthe{lemma}
\newaliascnt{proposition}{theorem}\newtheorem{proposition}[proposition]{Proposition}\aliascntresetthe{proposition}
\newaliascnt{corollary}{theorem}\newtheorem{corollary}[corollary]{Corollary}\aliascntresetthe{corollary}
\theoremstyle{definition}
\newaliascnt{definition}{theorem}\newtheorem{definition}[definition]{Definition}\aliascntresetthe{definition}
\newaliascnt{remark}{theorem}\newtheorem{remark}[remark]{Remark}\aliascntresetthe{remark}
\crefname{lemma}{Lemma}{Lemmas}
\crefname{proposition}{Proposition}{Propositions}
\crefname{corollary}{Corollary}{Corollaries}
\crefname{definition}{Definition}{Definitions}
\crefname{remark}{Remark}{Remarks}
\newcommand{\F}{\mathbb F}
\newcommand{\Z}{\mathbb Z}
\newcommand{\R}{\mathbb R}

\newcommand{\E}{\mathbb E}
\newcommand{\Prob}{\mathbb P}
\newcommand{\wt}{\operatorname{wt}}
\newcommand{\supp}{\operatorname{supp}}

\newcommand{\diag}{\operatorname{diag}}

\newcommand{\dist}{\operatorname{dist}}

\newcommand{\problemname}[1]{\text{\normalfont #1}}
\newcommand{\CircuitSAT}{\problemname{CircuitSAT}}
\newcommand{\UCS}{\problemname{UniqueCircuitSAT}}
\newcommand{\NP}{\mathsf{NP}}
\newcommand{\UP}{\mathsf{UP}}
\newcommand{\coNP}{\mathsf{coNP}}
\newcommand{\RP}{\mathsf{RP}}

\newcommand{\poly}{\operatorname{poly}}
\newcommand{\ones}{\mathbf 1}

\newcommand{\norm}[1]{\lVert #1\rVert}
\newcommand{\Dp}[1]{\mathsf D^{(p)}_{#1}}
\newcommand{\Up}[1]{\mathsf U^{(p)}_{#1}}
\newcommand{\GapSVP}{\mathsf{GapSVP}}

\newcommand{\BDDc}[1]{\mathsf{BDD}^{(q)}_{#1}}

\title{Hardness of Unique Shortest Vectors\\ and Bounded Distance Decoding}

\author{Sergei Kudria\\
  {\small CUHKSZ}\\
  {\small\texttt{sergeikudria@link.cuhk.edu.cn}}
  \and
  Jason Luo\\
  {\small MIT}\\
  {\small\texttt{luojason@mit.edu}}
}
\date{}

\hypersetup{pdftitle={Hardness of Unique Shortest Vectors and Bounded Distance Decoding},pdfauthor={Sergei Kudria, Jason Luo},bookmarksdepth=2}

\makeatletter
\renewcommand{\@maketitle}{\newpage\null\vskip 1em
  \begin{center}
    {\LARGE\@title\par}
    \vskip 1.5em
    {\normalsize\lineskip .5em
      \begin{tabular}[t]{c}\@author\end{tabular}\par}
  \end{center}\vskip 1em}
\makeatother

\begin{document}
\hypersetup{pageanchor=false}
\pagenumbering{gobble}
\maketitle
\begin{abstract}
We prove randomized NP-hardness of the unique shortest vector problem (uSVP) at every constant uniqueness factor greater than one, and of bounded distance decoding (BDD) at every fixed positive fraction less than one half of the minimum distance, in both their threshold decision and search versions. Both results hold in every fixed $\ell_p$ norm with $1\le p<\infty$. Previous Euclidean NP-hardness results for the search versions reached uniqueness factors tending to one and decoding radii greater than the minimum distance divided by $\sqrt2$.

For linear codes over every fixed finite field and each fixed ${0<\rho\le1/2}$, the decision version of BDD with radius strictly less than $\rho$ times the minimum distance is equivalent under deterministic polynomial-time reductions to satisfiability for circuits promised to have at most one satisfying input. It is therefore NP-hard under randomized polynomial-time reductions. Earlier hardness for decoding general codes applied beyond half the minimum distance.

For every fixed $p>2$ and for $p=\infty$, there is a constant $\epsilon>0$ such that, under randomized polynomial-time reductions, uSVP in the $\ell_p$ norm is NP-hard at uniqueness factor $n^\epsilon$, and BDD in the $\ell_p$ norm is NP-hard at radius $\lambda_1/n^\epsilon$, where $n$ is the lattice rank and $\lambda_1$ is its minimum distance.
\end{abstract}
\clearpage
\tableofcontents
\setcounter{tocdepth}{2}
\clearpage
\pagenumbering{arabic}
\setcounter{page}{1}
\hypersetup{pageanchor=true}

\section{Introduction}\label{sec:intro}

Finding short lattice vectors and recovering lattice points from nearby targets are central problems in lattice algorithms and cryptography~\cite{MG02}. A promise that all lattice vectors outside the span of a shortest vector are much longer could make finding that vector easier. Likewise, a promise that at most one lattice point lies within a given radius of the target could make decoding easier. We study the computational difficulty that remains under such geometric promises.

Given linearly independent basis vectors, a lattice is the set of their integer combinations. Its rank is the number of vectors in a basis. The shortest vector problem (SVP) asks for a shortest nonzero vector in this set and is NP-hard even to approximate \cite{Ajt98,Mic01,Kho05}. The unique shortest vector problem (uSVP) asks for a shortest nonzero vector $v$ under the promise that every lattice vector that is not a real multiple of $v$ is longer than $v$ by a given factor, called the uniqueness factor. For the search version of uSVP in the Euclidean norm, NP-hardness under randomized polynomial-time reductions was known for factors tending to one with the dimension \cite{AD16,SD16}. Hardness of uSVP at constant uniqueness factors was known under nonstandard or fine-grained complexity assumptions \cite{BGPS23,JX24}, while NP-hardness in the Euclidean norm at any fixed constant factor greater than one remained open.

Bounded distance decoding (BDD) asks whether a target lies within a given radius of a lattice point and, in its search form, asks for that point. The length of a shortest nonzero vector equals the minimum distance between distinct lattice points, so a radius less than half this distance guarantees at most one solution. The computational question is how hard it is to find that solution when it exists. Previous NP-hardness results for finding a lattice point within a given Euclidean distance of a target applied at radii greater than $1/\sqrt2$ times the minimum distance~\cite{LLM06}. NP-hardness of Euclidean BDD at any fixed positive fraction less than one half of the minimum distance remained open. Dadush, Regev, and Stephens-Davidowitz asked whether the search version of Euclidean BDD remains NP-hard at arbitrarily small constant fractions of the minimum distance~\cite[Section~1.2]{DRS14}.

Linear codes provide an analogue of bounded distance decoding. A code represents messages by codewords, and its minimum distance is the smallest number of coordinates in which two codewords differ. In Hamming distance, a ball of radius less than half the minimum distance contains at most one codeword. For some families, such as Reed--Solomon codes, algebraic decoders find this codeword if one exists \cite{Gao03}. For general binary linear codes specified by a matrix, Vardy conjectured that decoding is NP-hard even when the number of errors is less than half the minimum distance~\cite{Var97}. Dumer, Micciancio, and Sudan proved hardness at radii greater than half the minimum distance~\cite{DMS03}.

We prove randomized NP-hardness of Euclidean uSVP at every fixed constant uniqueness factor greater than one and of Euclidean lattice BDD at every fixed positive fraction less than one half of the minimum distance. Both results extend to every fixed $\ell_p$ norm with $1\le p<\infty$ and hold for decision and search (Corollary~\ref{cor:main-search} and Table~\ref{tab:results}).

For every fixed finite field and every fixed ${0<\rho\le1/2}$, decision BDD for linear codes at radius strictly less than $\rho$ times the minimum distance is equivalent under deterministic polynomial-time reductions to $\UCS$, satisfiability for circuits promised to accept at most one input. In particular, this decoding problem is NP-hard under randomized polynomial-time reductions. For binary codes, this confirms Vardy's conjecture under randomized reductions.

{\clubpenalty=10000 For every fixed $p>2$ and for $p=\infty$, we also prove randomized NP-hardness of uSVP with a uniqueness factor that is a positive power of the lattice rank, and of BDD at the minimum distance divided by a positive power of that rank. Hecht and Safra proved conditional hardness at constant factors for every fixed $p>2$ and at superconstant factors for sufficiently large $p$, using deterministic reductions that run in subexponential or quasipolynomial time~\cite{HS26}. Their assumptions concern the time needed by deterministic algorithms to solve 3SAT with at most one satisfying assignment (Section~\ref{sec:related}). Aggarwal and Dubey proved deterministic NP-hardness of finding a shortest vector in the infinity norm when it is unique up to sign, without guaranteeing a constant uniqueness factor~\cite{AD16}.\par}

\subsection{Problems}

\begin{definition}[Problems]\label{def:problems}
Fix a finite field $\F_q$ and $1\le p\le\infty$. For finite $p$, we measure vector lengths in the norm $\norm{x}_p=(\sum_i|x_i|^p)^{1/p}$, and we set $\norm{x}_\infty=\max_i|x_i|$. For a lattice $\Lambda$, let $\lambda_1$ be the length of a shortest nonzero vector and $\lambda_2$ the smallest radius of a ball centered at the origin containing two linearly independent lattice vectors. For a linear code $C$, let $d(C)$ be the minimum Hamming weight, or number of nonzero coordinates, of a nonzero codeword.
\begin{enumerate}[label=(\roman*)]
\item For $\Gamma>1$, the decision version $\Up\Gamma$ of uSVP receives a basis of a lattice $\Lambda$ and a threshold $r>0$ and distinguishes
\[
 \lambda_1\le r<\Gamma r<\lambda_2
 \quad\text{(YES)}
 \qquad\text{from}\qquad
 \lambda_1>\Gamma r
 \quad\text{(NO)}.
\]
\item For ${0<\rho<1/2}$, BDD for lattices, denoted $\Dp\rho$, receives a basis of a lattice $\Lambda$, a target $a$, and a threshold $r>0$ satisfying $r<\rho\lambda_1$. It asks whether $\dist_p(a,\Lambda)\le r$, where $\dist_p(a,\Lambda)=\min_{v\in\Lambda}\norm{a-v}_p$.
\item For ${0<\rho\le1/2}$, BDD for linear codes, denoted $\BDDc\rho$, uses a matrix $H$ over $\F_q$ to specify the code $C=\ker H$. Such a matrix is called a parity-check matrix, and the vector $Hz$ is called the syndrome of $z$. Given $H$, a syndrome $\sigma$, and a nonnegative integer $w$ with $w<\rho d(C)$, the problem asks whether there is an error vector $e$ of Hamming weight at most $w$ satisfying $He=\sigma$.
\end{enumerate}
\end{definition}

Adding an error $e$ of weight at most $w$ to a codeword $c\in C$ produces $c+e$, whose syndrome is $He$. Subtracting this unique error vector recovers $c$.

The search versions ask for a shortest nonzero lattice vector when $\lambda_2>\Gamma\lambda_1$, and for a closest lattice point when $\dist_p(a,\Lambda)<\rho\lambda_1$. The search version of code BDD takes $(H,\sigma,w)$ with $w<\rho d(\ker H)$ and asks for an error $e$ of weight at most $w$ satisfying $He=\sigma$, under the promise that such an error exists. An oracle for any of these search problems must return a solution on inputs satisfying these conditions and may return an arbitrary string otherwise (Section~\ref{sec:search-convention}). The lattice parameters $\Gamma$ and $\rho$ may also depend on the rank. The threshold in decision uSVP makes its YES promise stronger than the search promise: $\lambda_1\le r$ and $\lambda_2>\Gamma r$ imply $\lambda_2>\Gamma\lambda_1$.

In a YES instance of $\Up\Gamma$, the shortest nonzero vector is unique up to sign. In a YES instance of $\Dp\rho$, exactly one lattice point lies within distance $r$ of the target. For code BDD, two errors with the same syndrome and weight at most $w$ differ by a codeword of weight at most $2w<d(C)$, so they must be equal.

A deterministic polynomial-time many-one reduction between promise problems is promise-preserving if it maps every promised YES input to a promised YES input and every promised NO input to a promised NO input.

\begin{definition}[Randomized reductions]\label{def:randomized-reduction}
A randomized polynomial-time many-one reduction from a problem $\Pi$ to a promise problem $\Pi'$ has success probability $\pi$ if every NO instance of $\Pi$ maps to a promised NO instance of $\Pi'$ for every random seed, and every YES instance of $\Pi$ maps to a promised YES instance of $\Pi'$ with probability at least $\pi$. The remaining random choices on a YES input may produce any output.
\end{definition}

Valiant and Vazirani reduce \CircuitSAT{} to $\UCS$ by adding random constraints on the circuit's input bits~\cite{VV86}.  If the original circuit is satisfiable, the resulting circuit accepts exactly one input with inverse-polynomial probability, as in Definition~\ref{def:randomized-reduction}. A polynomial-time algorithm for $\UCS$ would thus imply $\NP\subseteq\RP$. Dumer, Micciancio, and Sudan use the same kind of randomized reduction in their hardness result for decoding beyond half the minimum distance~\cite[Theorem~16]{DMS03}.

\subsection{Main results}

\begin{theorem}[Lattices, informal]\label{thm:main-lattice}\label{thm:main-polynomial}
\begin{enumerate}[label=(\roman*)]
\item For every fixed $1\le p<\infty$, $\Gamma>1$, and ${0<\rho<1/2}$, there are randomized polynomial-time many-one reductions in the sense of Definition~\ref{def:randomized-reduction} from \CircuitSAT{} to the decision problem $\Up\Gamma$ and to $\Dp\rho$ in $\ell_p$, with inverse-polynomial success probability on satisfiable inputs.
\item For every fixed rational $p>2$ and every fixed $0<\epsilon<\min\{(p-2)/(4p),1/8\}$, there are deterministic polynomial-time promise-preserving reductions from $\UCS$ to $\Up{n^\epsilon}$ and $\Dp{n^{-\epsilon}}$ in $\ell_p$, where $n$ is the rank of the output lattice. The same holds in $\ell_\infty$ for every fixed $0<\epsilon<1/8$.
\end{enumerate}
\end{theorem}

Consequently, all these problems are NP-hard under randomized polynomial-time reductions. Increasing $\Gamma$ or decreasing $\rho$ strengthens the promise, leaving an algorithm fewer instances to solve. Hardness nevertheless holds for arbitrarily large fixed $\Gamma$ and arbitrarily small fixed $\rho>0$. See Theorem~\ref{thm:formal-lattice} for part (i), Theorem~\ref{pa2:main} for part (ii), and Remark~\ref{pa2:rem-real-p} for the extension of part (ii) to every fixed real $p>2$.

\begin{theorem}[Code BDD, informal]\label{thm:main-equivalence}
For every fixed finite field $\F_q$ and constant $\rho\in(0,1/2]$,
$\BDDc\rho$ and $\UCS$ are equivalent under deterministic promise-preserving polynomial-time reductions. Consequently, $\BDDc\rho$ is NP-hard under randomized polynomial-time reductions.
\end{theorem}

See Theorem~\ref{thm:formal-code} for the formal statement and proof. Our reductions from $\UCS$ to code BDD and to the lattice problems in Theorem~\ref{thm:main-lattice}(ii) are deterministic. To extend these reductions to \CircuitSAT{}, we first add the random constraints of Valiant and Vazirani~\cite{VV86}. The lattice reductions in Theorem~\ref{thm:main-lattice}(i) instead start from \CircuitSAT{} and use random parity-check equations (Section~\ref{sec:overview-tensors}).

\begin{corollary}[Search versions, informal]\label{cor:main-search}
For the respective parameters in Theorems~\ref{thm:main-lattice} and~\ref{thm:main-equivalence}, the search versions of uSVP, lattice BDD, and code BDD are NP-hard under randomized polynomial-time oracle reductions that make at most one oracle call, have no false positives, and succeed with inverse-polynomial probability on satisfiable circuits.
\end{corollary}

To solve \CircuitSAT{} using a search oracle, we recover a candidate input to the circuit from what the oracle returns and test it on the original circuit. When the random choices succeed, the instance satisfies the search promise and every solution lets us recover an accepted input. The test prevents false positives even if the oracle returns an arbitrary string (Corollary~\ref{cor:formal-search-usvp}).

For reductions from SAT to code BDD or to the decision versions of Euclidean uSVP and BDD, removing randomness while preserving the promises would imply $\NP=\UP$ (Theorem~\ref{thm:formal-code} for code BDD, Proposition~\ref{prop:lattice-up} for the lattice problems). Here $\UP$ consists of languages with polynomial-time verifiers having at most one accepting witness per input. Each of these problems has such a verifier on its promised inputs. For code BDD, $\NP=\UP$ also suffices for a deterministic polynomial-time reduction from SAT that preserves the promise.

\begin{table}[H]
\centering
\renewcommand{\arraystretch}{1.18}
\begin{tabularx}{\textwidth}{@{}>{\raggedright\arraybackslash}p{0.28\textwidth}>{\raggedright\arraybackslash}p{0.31\textwidth}>{\raggedright\arraybackslash}X@{}}
\toprule
Problem & Earlier hardness & This work \\
\midrule
Euclidean uSVP, factor $\Gamma$ & $1+1/n^{\varepsilon}$ for every fixed $\varepsilon>0$ \cite[Corollary~3.5]{SD16} & Every constant $\Gamma>1$ \\
\addlinespace
Euclidean BDD, radius $\rho\lambda_1$ & $\rho>1/\sqrt2$ \cite{LLM06} & Every ${0<\rho<1/2}$ \\
\addlinespace
Code BDD, radius $<\rho d$ & $\rho>1/2$ \cite{DMS03} & Every ${0<\rho\le1/2}$ \\
\bottomrule
\end{tabularx}
\caption{Selected NP-hardness results under randomized polynomial-time reductions. The first two rows concern Euclidean lattices of rank $n$. The earlier lattice results in the table concern the search versions. Our results hold for both decision and search (Corollary~\ref{cor:main-search}).}
\label{tab:results}
\end{table}

Euclidean SVP is NP-hard to approximate within any fixed constant factor under deterministic polynomial-time reductions~\cite{Wan26b}. Lyubashevsky and Micciancio reduce GapSVP, the decision version of approximate SVP, to the search version of uSVP~\cite{LM09}. The resulting uniqueness factor is the approximation factor in GapSVP divided by a quantity of order $\sqrt{n/\log n}$. Because this divisor grows with $n$, their reduction cannot transfer hardness at a fixed constant factor to uSVP at a factor greater than one.

Bennett asked whether the search version of uSVP in the Euclidean norm is NP-hard at a constant factor greater than one under randomized reductions \cite[Open Problem~4.6]{Ben23}. Hecht and Safra conjectured hardness for the decision version at a fixed factor $1+\varepsilon$ under randomized subexponential-time reductions \cite[Conjecture~7.2, arXiv version]{HS26}. Corollary~\ref{cor:main-search} answers Bennett's question and the Euclidean BDD question of Dadush, Regev, and Stephens-Davidowitz. Theorem~\ref{thm:main-lattice}(i) proves the Hecht--Safra conjecture using randomized polynomial-time reductions.

\noindent\textbf{Techniques and relation to prior work.} We encode inputs that the circuit accepts as short vectors that satisfy a linear system. We add further linear constraints and choose a linear function $\mu$ equal to one on encodings. We seek a lower bound on weight or norm far above what an encoding would have. If no encoding satisfies the constraints, we need this bound for every nonzero solution. If exactly one encoding $v$ satisfies them, we bound the weight or norm of nonzero solutions $x$ with $\mu(x)=0$. This excludes nonzero multiples of $v$, since $\mu(cv)=c$. We use this bound to prove uniqueness for uSVP and to lower-bound the minimum distance for BDD.
\begin{itemize}

\needspace{3\baselineskip}
\item {\clubpenalty=10000 For hardness at constant factors, we arrange symmetric tensors with $r$ indices as matrices. If no encoding satisfies the constraints, every nonzero tensor over the field satisfying them gives a matrix of rank at least $r$. The matrix of an encoding has rank one. When exactly one encoding satisfies the constraints, we bound the weight or norm of a tensor that is nonzero over the field and has $\mu=0$ by fixing an index at a label the encoding does not use. We show that some such label gives a nonzero tensor. Fixing the index imposes an equation that excludes the rank-one matrices arising from the encoding. We build on the matrix constructions and bounds on codeword weights of Austrin--Khot and Bhattiprolu--Guruswami--Lee--Ren~\cite{AK14,BGLR25}. For codes, the map in~\cite[Lemma~3.2]{BGLR25} makes nonzero vectors nearly equal in weight, letting the lower bound on weight grow faster than the encoding's weight. For norms of integer tensors, we adapt Wan's counting argument~\cite{Wan26b}. For lattices, we copy the assignments at each gate so that several vectors can represent one input. We choose the number of copies and a code's parity-check matrix at random. If the circuit is satisfiable, then with inverse-polynomial probability the code contains just one vector representing an input that the circuit accepts. The code's minimum distance is close to that vector's weight.\par}

\needspace{3\baselineskip}
\item For hardness at polynomial factors when $p>2$ or $p=\infty$, we use OpenAI's method based on interpolation and moments for the closest vector problem (CVP)~\cite[Chapter~7]{Ope26}. We also use Hair and Sahai's reconstruction from weighted moments and the Hadamard map in their ordinary SVP construction~\cite{HairSahai26}, adapting their arguments to enforce agreement between gates. For a sufficiently short solution, reconstruction expresses its moment polynomials as weighted sums of powers of algebraic functions. Several functions could describe the same input even when the circuit accepts only one, and their weights could cancel. With two or more functions, the moment polynomials form a matrix with a nonzero determinant. The interpolation constraints force the determinant to have more zeros, counted with multiplicity, than its degree allows. At most one term remains, with weight $\mu\bmod P$ over the prime field $\F_P$. When $\mu=0$, all moment polynomials vanish, forcing nonzero solutions to have large norm (Appendix~\ref{sec:poly-above-two}).
\end{itemize}
Section~\ref{sec:overview} explains these two arguments and the resulting reductions.

\section{Technical overview}\label{sec:overview}

\begingroup
\clubpenalty=10000
\widowpenalty=10000

We reduce circuit satisfiability to uSVP and BDD for lattices, and to BDD for linear codes. Inputs that the circuit accepts have short encodings satisfying linear equations, but the equations may also admit unwanted solutions. We add further constraints to control these other solutions.

\noindent\textbf{Soundness.} If no encoding satisfies the constraints, every nonzero solution must have much greater weight or norm than an encoding would have. In particular, if the circuit accepts no input, there is no short nonzero solution.

\noindent\textbf{Uniqueness.} Even if only one encoding satisfies the constraints, other short solutions may remain. For uSVP, vectors linearly independent of the encoding must be much longer. For BDD, the encoding must provide a solution within the decoding radius, while the code or lattice has minimum distance much larger than that radius. We first relate bounds on weight and norm to these properties, then prove the bounds using tensors for constant factors and interpolation for polynomial factors.

\subsection{From a circuit to uSVP and BDD}\label{sec:overview-encoding}

We first represent an input that the circuit accepts by a vector, so that agreement between gates can be expressed by linear equations. For each gate of a circuit $\phi$, create a \emph{block} with one \emph{label} for each assignment to its input and output wires that is consistent with the gate's operation. For an AND gate, these assignments are $(0,0,0)$, $(0,1,0)$, $(1,0,0)$, and $(1,1,1)$, with the output last. At the output gate, keep only labels setting the output to one. For each input bit that no gate uses, add a block with labels for zero and one.

A \emph{selection} chooses exactly one label in each block. It is \emph{compatible} if the chosen assignments agree on shared wires. The assignments then describe one circuit input, and the restriction at the output gate ensures that the circuit accepts it. Thus compatible selections correspond bijectively to inputs that the circuit accepts. We represent a selection by its indicator $y$, with one coordinate per label, equal to one at each chosen label and zero elsewhere. If there are $g$ blocks, $y$ has Hamming weight $g$.

The entries of a selection's indicator sum to one in each block. In a block, summing the entries for labels that assign one to a wire gives the chosen label's value on that wire. We form an integer matrix $A_\phi$ from equations equating these sums for each shared wire and equations equating the total sums in the blocks. The indicator of every compatible selection satisfies $A_\phi y=0$. Conversely, a nonzero solution with at most one nonzero entry per block has the same nonzero sum in every block. Dividing the solution by this common sum gives the indicator of a compatible selection (Lemma~\ref{lem:onehot-source}). General solutions may have several nonzero entries per block and need not encode an input. Further constraints must rule out unwanted short solutions.

{\tolerance=1000 \looseness=-1 We encode each selection as a vector of weight or norm at most $R$. We impose linear constraints on vectors in this space. For encodings, these constraints enforce agreement on shared wires. We construct a linear function $\mu$ that equals one on encodings and is integer-valued for lattices. We seek $\Delta$ much larger than $R$ with two properties. \textbf{Soundness:} if no encoding satisfies the constraints, every nonzero solution has weight or norm at least $\Delta$. \textbf{Uniqueness:} if exactly one encoding satisfies the constraints, every nonzero solution with $\mu=0$ has weight or norm at least $\Delta$. For an encoding $v$, $\mu(cv)=c$, so $\mu=0$ excludes its nonzero multiples. Distinct solutions with the same value of $\mu$ differ by a nonzero solution with $\mu=0$, so their distance is at least $\Delta$. The ratio $\Delta/R$ controls the uSVP uniqueness factor, while its reciprocal controls the BDD radius as a fraction of minimum distance.\par}

For code BDD, we start with an instance of $\UCS$ and construct a code with large minimum distance. Form its parity-check matrix $H$ from the rows defining the constraints and $\mu$. We ask whether an error of weight at most $R$ has syndrome $(0,1)$. An encoding satisfying the constraints has this syndrome and weight at most $R$. Whether the circuit accepts zero or one input, nonzero codewords have weight at least $\Delta$ because they satisfy the constraints and $\mu=0$. If no encoding satisfies the constraints, soundness rules out any error $e$ of weight at most $R$ with this syndrome, since $\mu(e)=1$ implies $e\ne0$. When $R<\rho\Delta$, both cases meet the decoding promise (Theorem~\ref{thm:formal-code}).

For the converse in Theorem~\ref{thm:main-equivalence}, take a code BDD instance with threshold $w$ and construct a circuit checking whether a candidate error has the prescribed syndrome and weight at most $w$. The decoding promise ensures at most one such error: two distinct errors would differ by a nonzero codeword of weight at most $2w$, below the minimum distance. The circuit is therefore an instance of $\UCS$ (Theorem~\ref{thm:formal-code}).

{\clubpenalty=10000 \widowpenalty=10000 For lattice uSVP, an encoding in the lattice must be much shorter than every lattice vector linearly independent of it. Suppose the lattice contains only one encoding $v$. For $x$ not proportional to $v$, the vector $x-\mu(x)v$ is nonzero with $\mu=0$, so its norm is at least $\Delta$. The norm of $x$ itself could still be small if $\|\mu(x)v\|_p$ is large. To account for this term, we append the coordinate $h\mu(x)$, with $R\le h=O(R)$. Since $h\ge R\ge\|v\|_p$, this coordinate has magnitude at least $\|\mu(x)v\|_p$. By the triangle inequality, either $x$ or $\mu(x)v$ has norm at least $\Delta/2$, so $(x,h\mu(x))$ does too. The encoding $(v,h)$ has norm $O(R)$, giving a uniqueness factor $\Omega(\Delta/R)$. If no encoding lies in the lattice, soundness ensures that every nonzero lattice vector has norm at least $\Delta$ (Lemma~\ref{lem:normalized-output}).\par}

For lattice BDD, we need large minimum distance and a target close to an encoding. Suppose at most one encoding satisfies the constraints. Append the coordinate $h\mu(x)$ to each solution $x$, with $h>\Delta>R$, and use target $(0,h)$. Nonzero vectors with $\mu(x)=0$ have norm at least $\Delta$. For a nonzero integer $\mu(x)$, the added coordinate has magnitude $|h\mu(x)|\ge h$. Thus the minimum distance is at least $\Delta$. An encoding $v$ satisfying the constraints has $\mu(v)=1$, so $(v,h)$ lies within distance $R$ of the target. Conversely, any lattice point $(x,h\mu(x))$ within distance $R$ of the target has $\mu(x)=1$, since $\mu(x)$ is an integer and $h>R$. Then $x\ne0$ and $\|x\|_p\le R$, which soundness rules out when no encoding satisfies the constraints. When $R<\rho\Delta$, both cases meet the decoding promise (Lemma~\ref{lem:normalized-output}).

\subsection{Constant factors using tensors}\label{sec:overview-tensors}

We use tensors to make $\Delta/R$ exceed any fixed constant. Each new index increases the weight of an encoding, so we need other solutions to grow faster. We first consider the case in which no encoding satisfies the constraints, working over a fixed finite field. For lattices, the norm depends on how large the entries are, as well as on how many are nonzero.

{\tolerance=1000 \looseness=-1 \noindent\textbf{From matrices to tensors.} We represent pairs of chosen labels by $yy^T$, whose entry at $(i,j)$ is one exactly when both labels are chosen. Its diagonal is $y$, and the columns from any block sum to $y$ because the block has exactly one chosen label. To ensure that nonzero columns have large weight, we choose a linear code $C$ over a finite field $\F_q$, with distance $\delta$ and coordinates indexed by labels. We restrict to compatible selections whose indicators lie in $C$. Following the matrix approach of~\cite{AK14,BGLR25}, we impose symmetry on a matrix $M$, require its columns to lie in $C$, require the columns from each block to sum to its diagonal, and require $A_\phi\operatorname{diag}M=0$. For codes we also set entries between distinct labels in a block to zero. For lattices this extra constraint is unnecessary over a prime field $\F_P$ with $P>N$, where $N$ is the number of labels. In either case, the constraints force a rank-one solution to choose one label per block. The condition $A_\phi\operatorname{diag}M=0$ makes these choices agree. Thus the rank-one solutions are the nonzero multiples of $yy^T$ where $y\in C$ is the indicator of a compatible selection (Proposition~\ref{prop:partition-main}). Write $V$ for the resulting space of matrices. If no compatible selection has its indicator in $C$, every nonzero matrix in $V$ has two independent columns in $C$.\par}

The union of the supports of two independent columns contains at least $(1+1/q)\delta$ rows \cite[Lemma~2.7]{BGLR25}. By symmetry, each nonzero row is a codeword of weight at least $\delta$. Thus the matrix has weight at least $(1+1/q)\delta^2$, while $yy^T$ has weight $g^2$.

To repeat this improvement, we extend the matrix constraints to symmetric tensors with $r$ indices. In the encoding $y^{\otimes r}$, an entry is one if all its indices are chosen labels, and zero otherwise. Its weight is $g^r$. We require every matrix formed by fixing all but two indices to lie in $V$. Fixing all but one index then gives a codeword in $C$. Applying the minimum distance bound successively to the $r$ indices shows that every nonzero solution has weight at least $\delta^r$. Since $\delta\le g$ when $y\in C$, distance alone cannot prove that other solutions have greater weight than the encoding.

{\clubpenalty=10000 We arrange a nonzero order-$r$ tensor as a matrix with rows indexed by the first tensor index and columns indexed by tuples of the remaining indices. This matrix is called a \emph{flattening}. When $V$ has no nonzero rank-one matrix, this flattening has rank at least $r$. We use $r$ independent columns to bound the norms of integer tensors. Fixing an index at a label $i$ gives an order-$(r-1)$ tensor, or \emph{slice}. Slices at distinct labels use disjoint sets of tensor coordinates.\par}

We prove the rank bound by finding a nonzero slice with smaller rank. Let $U$ be the span of the columns of the original flattening. By symmetry, the columns $v$ of the flattening of the slice at $i$ lie in $U$. They satisfy $v_i=\sum_{j\in B_1}v_j$ because the columns in the first block $B_1$ sum to the diagonal. If every nonzero slice had the original rank, its columns would span $U$ and the corresponding equation would hold throughout $U$. For a zero slice at $i$, every $v\in U$ has $v_i=0$. Thus the sum over $B_1$ would determine every coordinate of $v$. All columns would then be proportional, so any nonzero matrix formed by fixing all but the first two indices would have rank one, contrary to the assumption on $V$. Thus some nonzero slice has smaller rank. Its rank is at least $r-1$ by induction, so the original rank is at least $r$ (Lemma~\ref{lem:compatible-conditioning}).

For Hamming weight, two independent columns suffice. When no compatible selection has its indicator in $C$, the union of their supports contains at least $(1+1/q)\delta$ rows. Fixing the first index at the label of each row gives nonzero slices on disjoint sets of coordinates. Each slice satisfies the same constraints, so we can repeat the argument within it and add the resulting weights. This yields weight at least $(1+1/q)^{r-1}\delta^r$, compared with $g^r$ for an encoding (Lemma~\ref{lem:no-growth}). If $(1+1/q)\delta>g$, the lower bound grows faster than the encoding's weight, making $\Delta/R$ exceed any fixed constant for sufficiently large fixed $r$.

\noindent\textbf{Uniqueness.} If exactly one compatible selection has its indicator $y$ in $C$, then $y^{\otimes r}$ is itself short and $yy^T$ lies in $V$. Since $V$ contains a nonzero rank-one matrix, we cannot apply the preceding argument directly. Define $\mu(T)=\sum_{i\in B_1}T_{i,\ldots,i}$, summing the entries whose indices all equal a label in the first block. Then $\mu(y^{\otimes r})=1$. To bound the weight or norm of a tensor $T$ that is nonzero over the field and has $\mu(T)=0$, we find a nonzero slice at a label outside the selection. Fixing an index at that label adds a constraint that rules out the rank-one matrices in $V$.

If $r=2$ and $T$ uses only labels chosen by $y$, each column $j$ has at most one nonzero entry per block. By symmetry, its entries in every block sum to $T_{jj}$. Each column is therefore a multiple of $y$, and symmetry gives $T=\mu(T)yy^T=0$, a contradiction. The same reasoning at higher orders shows that a tensor using only the selected labels is a multiple of $y^{\otimes r}$. Since $\mu(T)=0$, this multiple must be zero. Thus some nonzero slice is at an unchosen label $i$. Each column $v$ of the slice satisfies $v_i=\sum_{j\in B_1}v_j$. For $y$, the left side is zero and the right side is one. This excludes the multiples of $yy^T$, the only rank-one matrices in $V$, so we can apply the rank argument to the slice (Lemma~\ref{lem:compatible-conditioning}(iii)).

{\tolerance=1000 \looseness=-1 \noindent\textbf{Codes.} For a circuit accepting at most one input, take $C$ to be the full space so that it contains the indicator of every compatible selection. But $C$ has minimum distance one, whereas each indicator has Hamming weight $g$. We choose an injective linear map $E$ that sends every nonzero vector to a vector of Hamming weight between $L_{\rm bal}$ and $U_{\rm bal}$, with $U_{\rm bal}/L_{\rm bal}$ as close to one as needed~\cite[Lemma~3.2]{BGLR25}. Applying $E$ to each tensor index preserves the rank of each flattening. Every nonzero vector in $E(C)$ has weight at least $L_{\rm bal}$. Each additional index multiplies the lower bound on the tensor's weight by $(1+1/q)L_{\rm bal}$, while multiplying the encoding's weight by at most $U_{\rm bal}$. When the circuit accepts exactly one input, the lower bound for nonzero tensors with $\mu=0$ is smaller by a factor of \mbox{$1+1/q$} because we apply the earlier bound to a slice. Choosing $(1+1/q)L_{\rm bal}>U_{\rm bal}$ still makes $\Delta/R$ exceed any fixed constant for large fixed $r$ (Theorem~\ref{thm:compatible-hamming}).\par}

The uniqueness argument finds a nonzero slice $T_i$ at a label $i$ not chosen by $y$. To evaluate $\mu$ on the encoded tensor, recover the original tensor by applying a left inverse of $E$ in every index. Applying that inverse only in the first index and then fixing that index at $i$ gives $E^{\otimes(r-1)}T_i$. Each nonzero entry of this slice requires a distinct nonzero column of the encoded tensor's flattening. Each of these columns has weight at least $L_{\rm bal}$, so the encoded tensor has weight at least $L_{\rm bal}$ times that of the slice (Theorem~\ref{thm:compatible-hamming}).

\needspace{3\baselineskip}

\noindent\textbf{Lattices.} For lattices, we need the same separation in norm rather than Hamming weight. Our lattice consists of integer tensors satisfying the matrix constraints modulo a prime $P$. A tensor with one entry equal to $P$ and all other entries zero satisfies these constraints, so $P$ must be larger than the required norm threshold. A nonzero tensor divisible by $P$ already has norm at least $P$. For lattices $q=P$, so for a tensor that is nonzero modulo $P$ the argument for Hamming weight gains only $1+1/P$ per index, too close to one for large $P$. The map $E$ used for codes makes nonzero images nearly equal in Hamming weight, but integer norms also depend on entry values. We adapt Wan's counting argument for equal entries to these tensors~\cite[Lemma~5.1]{Wan26b}.

For fixed finite $p$, the norm bound is $\min\{P,\Omega_p(r\delta^{r/p})\}$ for every nonzero tensor if no compatible selection has its indicator in $C$, and for every nonzero tensor with $\mu=0$ if exactly one does. Dividing the second term by the encoding's norm $g^{r/p}$ yields $\Omega_p(r(\delta/g)^{r/p})$. The factor $r$ grows, while $(\delta/g)^{r/p}$ decreases when $\delta<g$. Choose fixed $r$ large, then $\delta/g$ close enough to one, and finally $P$ large relative to the encoding's norm and required factor. This makes $\Delta/R$ exceed any fixed constant while the dimension $N^r$ remains polynomial (Theorem~\ref{thm:linear-tensor}).

To bound the norm, we compare how often columns agree and disagree within the rows where they have nonzero entries. With no compatible selection whose indicator lies in $C$, the flattening of a tensor nonzero modulo $P$ has rank at least $r$ modulo $P$. If the tensor has small norm, we can discard columns of large norm without reducing that rank, by the distance argument in Lemma~\ref{lem:no-growth}. Among $r$ remaining independent columns modulo $P$, most entries are small integers, forcing many pairs of equal entries in the same row. Yet any two columns differ by a nonzero codeword modulo $P$ and disagree in at least $\delta$ rows. If all $r$ columns together had nonzero entries in only slightly more than $\delta$ rows, each pair could agree in only a few of those rows. Counting the agreements shows that the columns together have nonzero entries in more than $\delta$ rows. The lower bound exceeds $\delta$ by a factor independent of $P$ (Lemma~\ref{lem:many-short}). These rows index disjoint slices. Repeating the count gives the norm bound in Theorem~\ref{thm:linear-tensor}.

For uniqueness, using a single slice would lose a factor $\delta^{1/p}$ because a slice has one fewer index. We recover this factor by finding many nonzero slices. Suppose only one compatible selection has its indicator $y$ in $C$, and $T\ne0$ satisfies $\mu(T)=0$. Reduce each entry of $T$ modulo $P$ and flatten the resulting tensor into a matrix. If this matrix is nonzero, it has a column $v$ with a nonzero entry at a label not chosen in the selection. Subtracting a multiple of $y$ makes the entries sum to zero in each block and changes only coordinates with $y_i=1$. Since $v,y\in C$, the result is a nonzero codeword of weight at least $\delta$. Each nonzero entry at a chosen label needs another nonzero entry in its block to make the sum zero. Thus $v$ has at least $\delta/2$ nonzero entries where $y_i=0$. These labels give disjoint slices whose constraints exclude nonzero multiples of $yy^T$. Adding the $p$th powers of the norms of these slices recovers the factor $\delta^{1/p}$, up to a constant (Theorem~\ref{thm:linear-tensor}).

\needspace{3\baselineskip}

{\widowpenalty=10000 \noindent\textbf{Isolating a selection for lattices.} We need $C$ to contain exactly one indicator of a compatible selection and have minimum distance close to $g$. This gives a short encoding and the lower bound on norms when $\mu=0$. Even if the circuit accepts only one input, the indicator of its compatible selection may lie outside $C$. We make several copies of each label, all representing the same assignment. Before copying, each accepted input corresponds to one compatible selection. Choosing among the copies in each block gives many selections for that input.\par}

{\clubpenalty=10000 Use a Vandermonde parity-check matrix with enough rows to give $C$ distance close to $g$. Multiplying its columns by random nonzero field elements preserves distance and changes which indicators lie in $C$. For a fixed selection, membership of its indicator in $C$ requires the multipliers to satisfy one independent equation per row, which occurs with small probability. Copies increase the number of selections per accepted input and hence the expected number of indicators in $C$. A suitable number of copies makes the expectation a small constant. Since we do not know how many inputs the circuit accepts, we list polynomially many possibilities for the number of copies.\par}

To bound the chance of more than one compatible selection with its indicator in $C$, we count pairs of selections with both indicators in $C$. Selections sharing labels also share multipliers. If two selections differ in at least as many blocks as the parity-check matrix has rows, the equations requiring both indicators to lie in $C$ are linearly independent. For pairs that differ in fewer blocks, the Vandermonde columns on the labels chosen only by the second selection are independent. Once the first selection's multipliers are fixed, at most one assignment to the other multipliers works. A large $P$ makes the chance of drawing this assignment small enough to offset the number of choices for the other labels. For a suitable number of copies, these estimates imply a positive constant probability of exactly one compatible selection whose indicator lies in $C$. Choosing a list entry uniformly succeeds with inverse-polynomial probability (Appendix~\ref{sec:multiplier}). The norm estimates give the $\Delta/R$ ratio required in Section~\ref{sec:overview-encoding}, proving Theorem~\ref{thm:main-lattice}(i).

\subsection{\texorpdfstring{Polynomial factors for $p>2$}{Polynomial factors for p > 2}}\label{sec:poly-overview}

For lattices, increasing tensor order to $r$ gives a factor proportional to $r$ while the dimension grows as $N^r$, so it cannot give a polynomial factor in polynomial dimension. For $p>2$, including $p=\infty$, interpolation spreads an encoding across many groups of coordinates indexed by labels and points. A linear map makes many nonzero entries in one group contribute more to the norm than the same number of ones spread across separate groups, as in an encoding. Thus a short lattice vector comes from an array with many points at which every group has few nonzero entries. This lets us use moments to reconstruct inputs that the circuit accepts. We use OpenAI's interpolation and moment method~\cite[Chapter~7]{Ope26} and Hair and Sahai's reconstruction from weighted moments and Hadamard map~\cite{HairSahai26}, adapting reconstruction to enforce agreement between gates and adding an argument for uniqueness.

{\tolerance=2000 \looseness=-1 \noindent\textbf{Encoding by interpolation.} We use one polynomial to link the chosen gate labels across many points. Work over a prime field $\F_P$, give each block a distinct field element as its anchor, and assign distinct values to its labels. A selection with indicator $y$ determines a polynomial $f_y(X)$ of degree at most $g-1$ taking the chosen label's value at each anchor. We define the encoding using the values of $f_y$ at many other points of $\F_P$. The degree bound limits how independently these values can vary. For each label and each point $\xi$ where we evaluate $f_y$, we create a vector with one entry for every $w\in\F_P$. If the selection chooses the label, we put a one at $w=f_y(\xi)$ and zeros elsewhere; otherwise the vector is zero. These vectors form the encoding array. For any array with the same indices (label, point and $w$), we call the vector obtained by fixing a label and a point and letting $w$ vary a \emph{fiber}.\par}

{\widowpenalty=10000 We follow Hair and Sahai~\cite{HairSahai26} and multiply each integer fiber by a Hadamard matrix with orthogonal $\pm1$ rows. The image of a fiber with $k$ nonzero entries has norm at least $\sqrt{k}$ times a row's norm. For $k$ ones in separate fibers, the image has norm $k^{1/p}$ times a row's norm. The gap grows with $k$ for $p>2$. For the first estimate, a fiber with $k$ nonzero integer entries has Euclidean norm at least $\sqrt{k}$. Orthogonality and the comparison of $\ell_p$ and $\ell_2$ norms give the lower bound relative to a row. For the second, each one maps to a row in a separate fiber. At $p=2$, the factors agree (Section~\ref{pa2:sec-euclidean-obstruction}).\par}

{\widowpenalty=10000 The reductions in Section~\ref{sec:overview-encoding} also need an integer-valued linear function $\mu$ equal to one on encodings. We choose the Hadamard matrix with an all-ones column, so one output coordinate sums the original fiber entries. Summing that coordinate over labels in the first block at one fixed point defines $\mu$: this sum is one on an encoding.\par}

{\widowpenalty=10000 We need linear constraints that link the polynomial values across points and enforce gate agreement. Requiring the positions of the ones to be values of one low-degree polynomial is not a linear constraint on the array. The $j$th \emph{moment} $\sum_w v_w w^j$ is linear in the fiber entries and, for a chosen label in an encoding, equals $f_y$ at that point raised to the $j$th power. For each label and order $j$ used by the construction, we require its moments across points to be evaluations of a polynomial of degree at most $j(g-1)$, called a \emph{moment polynomial}. For an initial range of orders, we also require the vector $u$ of moments across labels to satisfy $A_\phi u=0$ at each point, enforcing agreement between gates. In the encoding of a compatible selection, $A_\phi y=0$ and $u$ is a scalar multiple of $y$. Thus $A_\phi u=0$.\par}

{\clubpenalty=10000 \tolerance=2000 \looseness=-1 The moments must also reflect each label's value at its block's anchor. For a chosen label of value $\zeta$ at anchor $a$, the encoding satisfies $f_y(a)=\zeta$, so $(f_y(X)-\zeta)^j$ is divisible by $(X-a)^j$. For each label, the sum $\sum_w v_w(w-\zeta)^j$ is a linear combination of the first $j+1$ moments of the fiber $(v_w)_w$. As the point varies, these sums are values of a polynomial. Requiring divisibility by $(X-a)^j$ imposes linear constraints on the array. Take the integer arrays whose reductions modulo $P$ satisfy all the constraints above, and apply the Hadamard map to every fiber. Their images form the lattice.\par}

{\clubpenalty=10000 In the analysis, reconstruction shows that the moment polynomials of a sufficiently short vector vanish when the circuit accepts no input, or when it accepts only one input and $\mu=0$. Any fiber nonzero modulo $P$ must then have many nonzero entries, so its image under the Hadamard map has large norm.\par}

{\tolerance=2000 \looseness=-1 \noindent\textbf{Soundness.} Fix a threshold $\kappa$. The lower bound on the norm after applying the Hadamard map implies that, for a sufficiently short vector, at more than half the points every label's fiber has at most $\kappa$ entries that are nonzero modulo $P$. At each such point, a label's moments are weighted sums of powers of at most $\kappa$ field values, which may vary with the point. Bounded degrees and sufficiently many points and moment orders allow reconstruction from weighted moments~\cite{HairSahai26}: for each label, its moment polynomials have the form $\sum_s\omega_s\alpha_s^j$ as identities in $X$ with at most $\kappa$ distinct \emph{nodes} $\alpha_s$. The same nodes and nonzero weights $\omega_s$ represent all these moment polynomials (Lemma~\ref{pa2:lem-reconstruction}). Nodes play the role of $f_y$ in an encoding, but they and the weights can be algebraic functions of $X$ rather than polynomials.\par}

To recover a selection from a node, we first show that two labels in the same block cannot share a node. For each label, the monic polynomial in $Z$ with its nodes as roots has coefficients in $\F_P[X]$. At the block's anchor $X=a$, the divisibility constraints make this polynomial a power of $Z-\zeta$, where $\zeta$ is the label's value. If two labels in the block shared a node, their polynomials would share a nonconstant monic factor. At $X=a$, this factor would keep its degree but would have to divide two coprime powers because the labels have different values (Lemma~\ref{pa2:lem-anchor-separation}).

{\widowpenalty=10000 We must also show that labels in different blocks that share a node agree on shared wires. Collect the weights of a node $\alpha$ across labels into a vector $v_\alpha$, with zero for labels where that node is absent. Each polynomial on the left of $A_\phi u=0$ vanishes at more points than its degree, so the equations hold identically in $X$. At each moment order, the equations sum contributions from all the nodes. With enough moment orders, inverting the Vandermonde matrix on the nodes separates these contributions, so $A_\phi v_\alpha=0$ for every node. Since $v_\alpha$ has at most one nonzero entry per block, it is a multiple of the indicator of a compatible selection (Lemma~\ref{lem:onehot-source}). Thus each node describes an input that the circuit accepts. If the circuit accepts no input, there are no nodes, so the moment polynomials used by the construction vanish (Theorem~\ref{pa2:thm-rigidity}).\par}

{\clubpenalty=10000 \noindent\textbf{Uniqueness.} Even when the circuit accepts only one input, we must still rule out several distinct nodes in the reconstruction. Their weights can cancel in the zeroth moment even when higher moments are nonzero, so $\mu=0$ alone does not rule them out. With two or more nodes, we can form a matrix of moment polynomials with nonzero determinant. The constraints at the anchors force the determinant to have more zeros, counted with multiplicity, than its degree allows.\par}

{\widowpenalty=10000 Each node's weights are proportional to the indicator of the unique compatible selection, so the chosen labels share moment polynomials $M_j$. For $\nu\ge2$ distinct nodes, the Vandermonde matrix on the nodes is invertible and the weights are nonzero, so $D=\det((M_{i+j})_{0\le i,j<\nu})$ is nonzero. Its degree is at most $(g-1)\nu(\nu-1)$. At an anchor $a$ with chosen label value $\zeta$, changing from powers of $w$ to powers of $w-\zeta$ adds multiples of earlier rows and columns to later ones and preserves $D$. The new entry $(i,j)$ is divisible by $(X-a)^{i+j}$, so every determinant term is divisible by $(X-a)^{\nu(\nu-1)}$. The $g$ anchors thus contribute more zeros, counted with multiplicity, than $\deg D$. At most one node remains. Its weight would be the zeroth moment $M_0=\mu\bmod P$. Since node weights are nonzero, $\mu=0$ forces the moment polynomials to vanish (Theorem~\ref{pa2:thm-rigidity}).\par}

{\clubpenalty=10000 \widowpenalty=10000 \noindent\textbf{The ratio $\Delta/R$.} We compare each lower bound on the norm with $R$, the upper bound on the encoding's norm, omitting fixed powers of the circuit size. First, suppose at least half the points have a fiber with more than $\kappa$ entries that are nonzero modulo $P$. Each such fiber maps to a vector of norm at least $\sqrt\kappa$ times that of one row of the Hadamard matrix. At each point an encoding maps to one row per chosen label. The number of points contributes the same factor to both norms, giving a ratio of $\sqrt\kappa$.\par}

{\tolerance=1000 \looseness=-1 Second, suppose fewer than half the points have a fiber with more than $\kappa$ entries that are nonzero modulo $P$. The moment polynomials vanish if the circuit accepts no input, or if it accepts exactly one input and $\mu=0$. Evaluating the polynomials at each point shows that the first $\Theta(\kappa^2)$ moments in each fiber vanish modulo $P$. Since the Vandermonde matrix on distinct field elements is invertible, any fiber nonzero modulo $P$ has $\Omega(\kappa^2)$ nonzero entries. For large $P$, a nonzero integer fiber with these moments zero modulo $P$ maps to a vector of norm at least $\kappa$ times a row's norm. One fiber suffices for this lower bound, while the encoding uses $\Theta(\kappa^2)$ points for interpolation and reconstruction. Dividing the norm bound by the encoding's norm leaves the factor $\kappa/\kappa^{2/p}=\kappa^{1-2/p}$. This ratio grows only for $p>2$, with $1/p=0$ at infinity. After division by a fixed power of the circuit size, $\min\{\sqrt{\kappa},\kappa^{1-2/p}\}$ is a lower bound on $\Delta/R$. Taking $\kappa$ to be a large enough fixed power of the circuit size makes $\Delta/R$ polynomially large while keeping the dimension polynomial. The reductions in Section~\ref{sec:overview-encoding} then prove Theorem~\ref{thm:main-lattice}(ii) (Theorem~\ref{pa2:main} and Remark~\ref{pa2:rem-real-p}).\par}

\par\endgroup

\section{Related work}\label{sec:related}

\paragraph{Codes.}
{\widowpenalty=10000 Dumer, Micciancio, and Sudan prove randomized hardness of deciding whether a vector lies within Hamming distance $t$ of some codeword in $C$, under the promise $t<\rho d(C)$, for every $\rho>1/2$ \cite{DMS03}, while Bhattiprolu, Guruswami, Lee, and Ren prove deterministic hardness for $\rho>1$ \cite[Remark~D.4]{BGLR25}. Wan and Zhang prove deterministic NP-hardness of deciding whether a vector lies within a specified Hamming distance of some codeword in a Reed--Solomon code over growing finite extension fields, at a radius above half the minimum distance and without a uniqueness promise~\cite{WZ26}. Our code BDD reduction covers every fixed ${0<\rho\le 1/2}$ with $t<\rho d(C)$, where at most one codeword can lie within distance $t$. Other work studies nearest codeword, minimum distance, and their parameterized variants \cite{ABSS97,DFVW99,BBE21,BCGR23,GR26}.\par}

\needspace{2\baselineskip}
Our use of rank-one matrices to represent assignments follows Austrin and Khot~\cite{AK14}. For a matrix whose rows and columns lie in a code $C$, two independent columns imply Hamming weight greater than $d(C)^2$~\cite[Lemma~2.4]{BGLR25}. Bhattiprolu, Guruswami, Lee, and Ren combine this bound, which they attribute to Austrin and Khot, with an injective map that sends all nonzero vectors to nearly the same Hamming weight~\cite{BGLR25}. After establishing hardness at a constant factor, they use tensor products to increase the factor by multiplying minimum distances. The binary case of the matrix bound appears in Gopalan, Guruswami, and Raghavendra~\cite{GGR11}. Section~\ref{sec:symmetric-tensors} imposes the matrix constraints of Proposition~\ref{prop:partition-main} on every matrix formed by fixing all but two indices of a symmetric tensor. When the constraints exclude nonzero rank-one matrices, a nonzero tensor of order $r$ has a flattening of rank at least $r$. When only one encoding satisfies the constraints, slices also establish the estimates for nonzero tensors with $\mu=0$ (Lemma~\ref{lem:compatible-conditioning}).

\paragraph{Lattices.}
Ajtai, Micciancio, and Khot established randomized hardness for SVP \cite{Ajt98,Mic01,Kho05}. For the search version of uSVP in the Euclidean norm, Kumar and Sivakumar proved hardness without a constant uniqueness factor \cite{KS01}. Aggarwal and Dubey proved randomized hardness for a factor $1+1/\poly(n)$, where $n$ is the lattice rank \cite{AD16}, and Stephens-Davidowitz proved hardness for a factor $1+1/n^{\varepsilon}$ for every fixed $\varepsilon>0$ \cite[Corollary~3.5]{SD16}. Bennett, Ganju, Peetathawatchai, and Stephens-Davidowitz reduce ZSVP, the shortest vector problem on rotations of $\Z^n$, to uSVP at every constant factor \cite{BGPS23}. Their reduction establishes hardness under an assumption on ZSVP. Jin and Xue prove a fine-grained lower bound for a constant factor under GapSETH, a strong exponential-time hypothesis for distinguishing satisfiable formulas from those far from satisfiable \cite{JX24}.

For fixed $p>2$, Hecht and Safra prove hardness for uSVP at a factor $\sqrt2-o(1)$ by a deterministic subexponential-time reduction from Unambiguous-3SAT, the problem of distinguishing 3CNF formulas with one satisfying assignment from unsatisfiable ones \cite{HS26}. Their result rules out polynomial-time algorithms under the assumption that Unambiguous-3SAT on $n$ variables has no deterministic algorithm running in time $2^{O(n^{2/3}\log n)}$. They also prove conditional hardness at superconstant factors for sufficiently large $p$. Aggarwal and Dubey prove deterministic hardness for $\ell_\infty$ uSVP without guaranteeing a fixed uniqueness factor greater than one \cite{AD16}.

For lattice decoding at radius $\rho\lambda_1$, Liu, Lyubashevsky, and Micciancio, using the Regev--Rosen norm embedding for $p>2$, prove hardness when $\rho>\min\{1/\sqrt2,2^{-1/p}\}$ \cite{LLM06,RR06}. For $p>4.2773$, Bennett and Peikert prove hardness for smaller values of $\rho$, with a threshold that tends to $1/2$ as $p$ grows \cite{BP20}. The decoding radii in these results exceed $\lambda_1/2$ for finite $p$. Bennett and Peikert also prove randomized NP-hardness of the search version of BDD in $\ell_\infty$ at radius $\lambda_1/2$, where the target's distance to the lattice may equal the radius and the target may have several closest lattice points~\cite[Theorem~1.1]{BP20}.

{\clubpenalty=10000 \widowpenalty=10000 For ordinary SVP, Hair and Sahai prove conditional deterministic subexponential-time hardness in every fixed $\ell_p$ norm with $1\le p<\infty$, and Wan proves deterministic Euclidean results \cite{HaS26,Wan26a,Wan26b}. For every fixed $p>2$ and $\epsilon>0$, Hair and Sahai also prove deterministic NP-hardness of approximating SVP in $\ell_p$ within a factor of $2^{\log^{1-\epsilon} n}$~\cite{HairSahaiSTOC26}. OpenAI introduced an encoding for the closest vector problem (CVP) based on polynomial interpolation and moments \cite[Chapter~7]{Ope26}, which Hair and Sahai adapted using weighted moments to prove deterministic hardness of approximating ordinary SVP within a polynomial factor in $\ell_p$ with $p>2$~\cite{HairSahai26}. Appendix~\ref{sec:poly-above-two} combines their reconstruction from weighted moments with a determinant argument to construct full-rank uSVP and BDD instances.\par}

Appendix~\ref{sec:amplification} explains how we adapt Wan's counting argument~\cite[Lemma~5.1]{Wan26b} to bound norms of integer tensors. Tensor amplification for SVP also appears in Haviv--Regev and Bennett--Cheraghchi--Guruswami--Ribeiro \cite{HR12,BCGR23}, whose analyses bound the length of a shortest vector in particular lattice families.

\needspace{2\baselineskip}
The reduction of Lyubashevsky and Micciancio from GapSVP to the search version of uSVP sets the uniqueness factor to the approximation factor for GapSVP divided by a quantity of order $\sqrt{n/\log n}$ \cite{LM09}. For the Euclidean norm, Aggarwal and Dubey's deterministic reduction from SVP to uSVP guarantees a uniqueness factor $1+2^{-\poly(\ell)}$ relative to $\lambda_1$, where $\ell$ is the input length \cite{AD16}.

{\widowpenalty=10000
Lyubashevsky and Micciancio also relate the search versions of BDD and uSVP \cite{LM09}. Regev and Rosen transfer Euclidean results to other norms by randomized embeddings \cite{RR06}. We construct direct reductions with one-sided error to decision uSVP and decision BDD in every finite $\ell_p$ norm.\par}

\paragraph{Isolation and parameterized hardness.}
Valiant and Vazirani isolate witnesses by hashing, and Mulmuley, Vazirani, and Vazirani isolate them by random weights \cite{VV86,MVV87}. Dell, Kabanets, van Melkebeek, and Watanabe show that $\coNP\subseteq\mathsf{NP}/\poly$ would follow from a randomized polynomial-time procedure that preserves unsatisfiability for every random choice and, on every satisfiable circuit of description length $s$, succeeds with probability greater than $2/3+1/\poly(s)$~\cite{DKMW13}. Success means that the output circuit accepts exactly one input and a fixed deterministic polynomial-time algorithm, given both circuits and that input, recovers an input accepted by the original circuit. Appendix~\ref{sec:limits} applies this result to witness recovery in our reductions. Parameterized hardness for coding and lattice problems is studied in \cite{BBE21,BCGR23,HO26,GR26}.

\section{Preliminaries}\label{sec:prelim}

\subsection{Notation and conventions}

Lattice bases, targets, and thresholds are rational and encoded in binary. We set $\lambda_2=\infty$ for lattices of rank one and $d(\{0\})=\infty$. For fixed real parameters that the reductions need to approximate, we assume that an approximation with $t$ correct bits can be computed in time polynomial in $t$.

The randomized reductions in Definition~\ref{def:randomized-reduction} follow the convention used in \cite{MG02,Mic12}.
Circuits use AND gates with two inputs and NOT gates with one input. Circuit size counts input bits and gates. Circuits with no gates are handled directly.

For a vector or array $x$, $\wt(x)$ counts nonzero coordinates and $\supp(x)$ is their set. For a symmetric matrix $M$, the weight counts $M_{ij}$ and $M_{ji}$ separately when $i\ne j$. A tensor flattening arranges an array as a matrix, with rows indexed by tuples of some of its indices and columns indexed by tuples of the remaining indices. It preserves weight and $\ell_p$ norm. We use the successive minima $\lambda_1,\lambda_2$ from Definition~\ref{def:problems}. We identify a Boolean vector over $\F_P$ with the integer vector having the same zero and one entries.

\subsection{Isolating inputs accepted by a circuit}

We apply the following randomized reduction from \CircuitSAT{} to $\UCS$ before our reduction to code BDD and the lattice reductions in Theorem~\ref{thm:main-lattice}(ii).

\begin{lemma}[Valiant--Vazirani isolation, {\normalfont\cite{VV86}}]\label{lem:vv-isolation}
There is a randomized polynomial-time transformation that maps a Boolean circuit $F$ of size $s$ to a Boolean circuit $F'$ on the same input variables. For every random choice, every input accepted by $F'$ is also accepted by $F$, and hence unsatisfiability is preserved. If $F$ is satisfiable, then $F'$ accepts exactly one input with probability at least $1/\operatorname{poly}(s)$.
\end{lemma}

The next two propositions relate the success probability of isolation procedures to complexity assumptions. We use them in the discussion in Appendix~\ref{sec:limits}.

\begin{proposition}[Isolation with recovery, {\normalfont\cite[Theorem~3.8(v)]{DKMW13}}]\label{prop:isolation-recovery-limit}
Suppose a randomized polynomial-time procedure maps every Boolean circuit $F$ to a Boolean circuit $F'$ and preserves unsatisfiability for every random choice. Suppose also that a fixed deterministic polynomial-time algorithm takes $F$, $F'$, and an input accepted by $F'$, and that, for every satisfiable $F$ of description length $s$, with probability greater than $2/3+1/\poly(s)$ the circuit $F'$ accepts exactly one input and the algorithm returns an input accepted by $F$ from that unique input. Then $\coNP\subseteq\mathsf{NP}/\poly$.
\end{proposition}

\begin{proposition}[Isolation on the same variables, consequence of {\normalfont\cite[Theorem~1.1]{BD26}}]\label{prop:isolation-pruning-limit}
Suppose a randomized polynomial-time procedure maps every circuit $F$ to a circuit $F'$ on the same input variables, and for every random choice every input accepted by $F'$ is also accepted by $F$. If there is a fixed constant $\epsilon>0$ such that, for every satisfiable $F$, the output $F'$ accepts exactly one input with probability at least $\epsilon$, then $\NP\subseteq\mathsf{P}/\poly$.
\end{proposition}

\subsection{Balanced linear encodings}

For code BDD, we use the following map to send nonzero vectors to vectors of nearly equal Hamming weight. As the tensor order increases, this lets the minimum distance of the code we construct grow faster than the weight of a vector representing an accepted input. A left inverse recovers the original vector from its image under the map.

\begin{lemma}[Balanced encoding, {\normalfont\cite[Lemma~3.2]{BGLR25}}]\label{lem:balanced-encoding}
Fix $\F_q$ and a constant $\varepsilon\in(0,1/2)$. Given $n_{\rm var}$, one can construct in deterministic polynomial time an injective linear map
\[
 \begin{gathered}
 E:\F_q^{n_{\rm var}}\to\F_q^N,\\
N\le\left(\frac{qn_{\rm var}}{\varepsilon}\right)^2,
 \end{gathered}
\]
such that every nonzero $x$ satisfies
\[
 \begin{gathered}
 L_{\rm bal}\le\wt(Ex)\le U_{\rm bal},\\
L_{\rm bal}=(1-\varepsilon)\left(1-\frac1q\right)N,\\
U_{\rm bal}=\left(1-\frac1q\right)N.
 \end{gathered}
\]
\end{lemma}
\begin{proof}
Lemma~3.2 of \cite{BGLR25} constructs $E$ in deterministic polynomial time with length $N=Q^2$ for a power $Q$ of $q$. Thus $U_{\rm bal}=(1-1/q)N$ is an integer. Since $E$ is injective, Gaussian elimination computes a left inverse $E_{\rm left}$ with $E_{\rm left}E=I$ in deterministic polynomial time.
\end{proof}

\subsection{Reconstruction from weighted moments}

To analyze short vectors in the lattice reductions in Theorem~\ref{thm:main-lattice}(ii), we need to combine moment information from many points, where the field elements and weights may differ from point to point. For field elements $w$ with weights $v_w$, their $j$th weighted moment is $\sum_w v_w w^j$. Suppose that, at sufficiently many points of $\F_P$, the evaluations of a sequence of polynomials are the weighted moments of a few field elements. The next lemma represents all these polynomials by moments of the same nodes and weights, in an algebraic extension of $\F_P(X)$. Here $\F_P(X)$ is the field of ratios of polynomials in $X$ over $\F_P$ with nonzero denominators, and each node and weight is a root of a nonzero polynomial with coefficients in $\F_P(X)$.

In the formulas for moments, we use $w^0=1$, including when $w=0$.

\begin{lemma}[Moment representations, {\normalfont\cite[Lemma~5.1]{HairSahai26}}]
\label{pa2:lem-reconstruction}
Let $P$ be prime, let $d,\kappa$ be positive integers, and let $T$ be a nonnegative integer. For each $\xi\in\Xi'\subseteq\F_P$, suppose there are at most $\kappa$ distinct elements $w\in\F_P$ with nonzero weights $v_{\xi,w}\in\F_P$, and their moments satisfy $M_j(\xi)=\sum_w v_{\xi,w}w^j$ for $0\le j\le T$. Suppose $M_j\in\F_P[X]$, $\deg M_j\le dj$ for $0\le j\le T$, and
\[
 \begin{gathered}
 T\ge2\kappa-1,\\
|\Xi'|>d\bigl(T+2\kappa(\kappa-1)\bigr).
 \end{gathered}
\]
Then, in an algebraic closure of $\F_P(X)$, these moment polynomials have a representation
\[
 M_j=\sum_{s=1}^\nu\omega_s\alpha_s^j,\qquad0\le j\le T,
\]
where $\nu\le\kappa$, the nodes $\alpha_s$ are distinct, and the weights $\omega_s$ are nonzero. For $\nu=0$, use the empty representation and zero moments.
\end{lemma}

\section{Encoding inputs accepted by a circuit}\label{sec:witness-encoding}

The reductions for constant factors use linear constraints on matrices whose rank-one solutions encode inputs that the circuit accepts. We construct these constraints, extend them to tensors, and then show how bounds on weights and norms yield the uSVP and BDD instances.

\subsection{From circuit inputs to rank-one matrices}\label{sec:common-source}

We describe a circuit evaluation by assignments to the wires at each gate that agree on shared wires. These compatible assignments will let us encode inputs accepted by the circuit as rank-one matrices. For each gate, create a block: a set with one label for each assignment to its input and output wires that is consistent with the gate. In the block for the designated output gate, require the designated output wire to equal one. For each input bit that no gate uses, include a block with one label for each of the values zero and one. The blocks $B_1,\ldots,B_g$ have at most four labels each. If a block is empty, there is no compatible selection, and the reduction outputs a promised NO instance. A compatible selection chooses exactly one label from each block, with agreement on every shared wire. Circuit evaluation defines a bijection between the inputs accepted by the circuit and these selections, including the blocks for input bits that no gate uses. Thus $g\le s$ and $L=\sum_b|B_b|\le4g$ for a size-$s$ circuit.

More generally, a block can list allowed assignments to specified Boolean variables, and compatibility means agreement on shared variables. This also allows constraints describing a clique in a graph, as used in the parameterized reduction. The instances used here have description length polynomial in $gL$. When required, choose a positive integer and make that many copies of each label. Each copy assigns the same values to the same variables as the original label. For the lattice reductions with constant factors, copying labels gives each accepted input multiple compatible selections. We impose random parity-check equations on their indicator vectors. For a satisfiable circuit, with inverse-polynomial probability over the random choices, exactly one compatible selection of copied labels has an indicator satisfying these equations. Let $\mathcal I$ be the resulting coordinate set, $N=|\mathcal I|$, let $B_b$ denote the coordinates in block $b$ after copying, and let $f=\ones_{B_1}$ be the indicator of the first block, so that $f^Tu$ is the sum of the entries of $u$ in that block.

We use linear equations to recognize compatible selections among vectors with at most one nonzero entry per block. For a vector $u$ indexed by labels, equation~\eqref{eq:source-mass} equates the sum of its entries in each block with the sum in the first block. For each shared Boolean variable $x_i$, equation~\eqref{eq:source-marginal} equates the sums over labels assigning it the value $1$ in blocks $b$ and $c$ containing that variable. These equations define an integer matrix $A_\phi$. If a nonzero solution has at most one nonzero entry per block, the sum in each block cannot be zero. Each block therefore has one nonzero entry equal to that sum. Dividing by this value gives the indicator of one selected label in each block, and~\eqref{eq:source-marginal} ensures agreement on shared variables. The equations are
\begin{align}
 \sum_{\ell\in B_b}u_\ell&=\sum_{\ell\in B_1}u_\ell,
       &&1\le b\le g,\label{eq:source-mass}\\
 \sum_{\substack{\ell\in B_b\\\ell(x_i)=1}}u_\ell
 &=\sum_{\substack{\ell\in B_c\\\ell(x_i)=1}}u_\ell.
       \label{eq:source-marginal}
\end{align}
For each variable it suffices to impose the second equation along a spanning tree on the blocks containing it. Write $\mathcal U_\phi(F)=\ker_F A_\phi$ over a field $F$. When the blocks come from a circuit and the labels are not copied, $A_\phi$ has $O(g)$ rows and each row has at most $r_A=8$ nonzero entries. We use this limit on the number of nonzero entries in each row in the construction in Appendix~\ref{sec:poly-above-two}.

\begin{lemma}[Vectors supported on at most one label per block]\label{lem:onehot-source}
A nonzero $u\in\mathcal U_\phi(F)$ supported on at most one coordinate
per block is $cy$ for a nonzero scalar $c$, where $y$ is the Boolean
indicator of a compatible selection. The indicator of every compatible selection
belongs to $\mathcal U_\phi(F)$.
\end{lemma}
\begin{proof}
The entries sum to $c=f^Tu$ in each block. If $c=0$, the restriction to at most one nonzero entry per block forces $u=0$. Otherwise
every block has exactly one nonzero entry, equal to $c$.
After division by $c$, each block has one entry equal to $1$ and all others zero. Each side of an equation for a shared variable is therefore the value assigned to that variable by the selected label, either $0$ or $1$. Since $0\ne1$ in the field, the two selected assignments agree on this variable.
Each selected label is an allowed assignment for its block. Conversely, the entries of the indicator of a compatible selection sum to $1$ in each block, and the sums for each shared variable agree, so it satisfies the equations.
\end{proof}

\begin{proposition}[Rank-one matrices and compatible selections]\label{prop:partition-main}
Let $C\subseteq F^N$ be any linear code. Let $V$ be the space of symmetric matrices $M$ with columns in $C$ satisfying
\begin{equation}\label{eq:compatible-parent}
 \begin{gathered}
 M\ones_{B_b}=\diag M\quad(1\le b\le g),\\
A_\phi\diag M=0,
 \end{gathered}
\end{equation}
and one of the following conditions:
\begin{enumerate}[label=(\roman*)]
\item each block $M[B_b,B_b]$ is diagonal.
\item $F=\F_P$ and $P>N$, without requiring $M_{ij}=0$ for distinct labels in the same block.
\end{enumerate}
In either case the nonzero rank-one matrices in $V$ are precisely
$a yy^T$, where $a\ne0$ and $y\in C$ is the Boolean indicator of a compatible selection.
\end{proposition}
\begin{proof}
Write a nonzero symmetric rank-one matrix as $a vv^T$ with $a\ne0$. A nonzero column is a nonzero scalar multiple of $v$. Since that column belongs to $C$, linearity implies $v\in C$.
Under (i), diagonality of $M[B_b,B_b]$ forces at most one nonzero coordinate of $v$ in each block. If $v_i\ne0$, the equations $M\ones_{B_b}=\diag M$ imply
$\ones_{B_b}^Tv=v_i$ for every $b$. Thus every block has one nonzero entry, and all these entries equal the same nonzero scalar $c$.
Under (ii), $M\ones_{B_b}=\diag M$ gives $v_i(\ones_{B_b}^Tv)=v_i^2$ for every $b$ and $i$. At any nonzero coordinate, we have $v_i=\ones_{B_b}^Tv$ for every block. Thus all nonzero coordinates equal $c=f^Tv\ne0$. Dividing the sum in each block by $c$ shows that the number of nonzero coordinates in the block is congruent to one modulo $P$.
This number is less than $P$, so each block has one nonzero coordinate.
In either case $v=cy$ for the Boolean indicator $y$ of a selection, and the equation $A_\phi\operatorname{diag}M=0$ implies $A_\phi y=0$. By Lemma~\ref{lem:onehot-source}, the selection is compatible. Conversely, $a yy^T$ belongs to $V$ whenever $a\ne0$ and $y\in C$ is the indicator of a compatible selection.
\end{proof}

\subsection{Symmetric tensors}\label{sec:symmetric-tensors}

We extend the matrix constraints to tensors to prove lower bounds on weight and norm that grow with the tensor order. An order-$r$ tensor is an array with $r$ indices. It is symmetric if permuting the indices leaves its entries unchanged. Fixing one index produces a slice of order $r-1$, fixing all but two produces a matrix slice, and fixing all but one produces a fiber. We require every matrix slice to lie in the space $V$ of matrices defined in Proposition~\ref{prop:partition-main}. When $V$ contains no nonzero rank-one matrix, the flattening of a nonzero symmetric order-$r$ tensor satisfying these constraints has rank at least $r$. Its rows are indexed by one tensor index and its columns by tuples of the remaining indices. Each nonzero column is a codeword in $C$ and has Hamming weight at least the minimum distance of $C$. After fixing an index $i$, the equation $M\ones_{B_1}=\diag M$ for each matrix slice forces coordinate $i$ of each remaining fiber to equal its sum over the first block. This extra linear equation restricts the span of the columns and enables induction on its dimension. The next lemma relates these restrictions to the rank-one matrices in $V$.

Let $F$ be the field and $C\subseteq F^N$ the code used to define $V$ in Proposition~\ref{prop:partition-main}. We also allow $V$ to be a subspace of that space defined by additional homogeneous linear equations. Fixing a tensor index in the induction adds equations of this kind. Set $S_1(V)=C$. For $r\ge2$, let $S_r(V)$ consist of the fully symmetric order-$r$ tensors with entries in $F$ whose matrix slices all belong to $V$. These are linear spaces, and every fiber belongs to $C$. For fixed $r$, given equations defining $C$ and $V$, Gaussian elimination computes bases for these spaces using polynomially many field operations in $N^r$ and the number of input equations.

For the indicator $y$ of a compatible selection, the diagonal entries of $y^{\otimes r}$ sum to one over the first block. Requiring this sum to be zero excludes every nonzero multiple of $y^{\otimes r}$.

\begin{lemma}[Fixing an index of a tensor]
\label{lem:compatible-conditioning}\label{lem:flattening-rank}
For $T\in S_r(V)$, let $U(T)$ be the span of the columns of the matrix whose rows are indexed by the first tensor index and whose columns are indexed by all remaining indices. Write $T_i=T(i,\cdot,\ldots,\cdot)$. For an order-one tensor, let $U(T)$ be its linear span. Write $e_i$ for the $i$th standard basis vector, and let
\[
 \begin{gathered}
 C_i=C\cap\ker(e_i-f)^T,\\
V_i=\{M\in V:\hbox{columns in $C_i$}\}.
 \end{gathered}
\]
\begin{enumerate}[label=(\roman*)]
\item For $r\ge2$, $T_i\in S_{r-1}(V_i)$, with $S_1(V_i)=C_i$, and
\[
 U(T_i)\subseteq U(T)\cap\ker(e_i-f)^T.
\]
\item If $V$ has no nonzero rank-one matrix, every nonzero
      $T\in S_r(V)$ has $\dim U(T)\ge r$.
\item Suppose the rank-one matrices in $V$ are precisely the nonzero scalar multiples of $yy^T$ for a
      Boolean indicator $y$ of a compatible selection. For $r\ge2$, if $T\ne0$ and
      $\sum_{j\in B_1}T_{j,\ldots,j}=0$, some $i\notin\supp y$
      has $T_i\ne0$. For each such index $i$, $V_i$ contains no rank-one matrix.
\end{enumerate}
\end{lemma}
\begin{proof}
A column $v$ of $T_i$ has $v_j=T_{i,j,I}$ for fixed remaining indices
$I$. By \eqref{eq:compatible-parent} for the matrix slice with remaining indices $I$,
$v_i=T_{i,i,I}=\sum_j f_jT_{i,j,I}=f^Tv$.
By symmetry, $v\in U(T)$. Every matrix slice of $T_i$ is a matrix slice of $T$, so it lies in $V$, and it has columns in $C_i$. Hence it lies in $V_i$. For $r=2$, $T_i$ is a vector, and this says $T_i\in C_i$. 

For (ii), induct on $r$. For $r=1$, the span of a nonzero tensor has dimension one. Let $U=U(T)$. We have $T_i\ne0$ exactly when $e_i^T$ is nonzero on $U$. Suppose no nonzero $T_i$ has $e_i^T|_U\ne f^T|_U$. For each index $i$ with $T_i=0$, the coordinate $u_i$ is zero for all $u\in U$. For every other index, $u_i=f^Tu$ by the supposition. Thus $f^Tu=0$ implies $u=0$, so $u\mapsto f^Tu$ is injective. Since $U$ is nonzero, it is one-dimensional. Let $v$ span $U$ and choose $k$ with $v_k\ne0$. Symmetry and the fact that $U$ has dimension one imply
\[
 T_{i_1,\ldots,i_r}=T_{k,\ldots,k}\prod_{j=1}^r(v_{i_j}/v_k).
\]
Thus $T$ is a nonzero scalar multiple of $v^{\otimes r}$. A nonzero matrix slice would have rank one in $V$, a contradiction. Hence some nonzero slice $T_i$ has $e_i^T|_U\ne f^T|_U$. Since $V_i\subseteq V$, it has no nonzero rank-one matrix. The induction hypothesis applied to $T_i\in S_{r-1}(V_i)$ gives $\dim U(T_i)\ge r-1$, while (i) gives $\dim U(T_i)\le\dim U-1$.

For (iii), suppose $T_i=0$ for every $i\notin\supp y$. By symmetry $T$ would be supported on $(\supp y)^r$. Each column of the flattening can be nonzero only at the label chosen by $y$ in each block. The matrix constraints require equal sums in all blocks, so each column is a multiple of $y$. By symmetry this holds with any index left free, so $T=c y^{\otimes r}$ with $c\ne0$, contradicting $\sum_{j\in B_1}T_{j,\ldots,j}=0$ in part~(iii). A rank-one matrix in $V_i\subseteq V$ would be a scalar multiple of $yy^T$, but $y_i=0\ne1=f^Ty$ excludes every nonzero scalar multiple of $yy^T$.
\end{proof}

The diagonal entries recover the indicator of a selection from its tensor. Their sum over the first block defines a linear function:
\begin{equation}\label{eq:source-readout}
 \begin{gathered}
 \mathcal R_r(T)=(T_{i,\ldots,i})_{i=1}^N,\\
\mu_r(T)=f^T\mathcal R_r(T).
 \end{gathered}
\end{equation}
Let $y$ be the Boolean indicator of a selection. For its tensor, $\mathcal R_r(y^{\otimes r})=y$ and
$\mu_r(y^{\otimes r})=1$. For an integer tensor, the diagonal entries and their sum over the first block are integers. When we apply Lemma~\ref{lem:compatible-conditioning} to an integer tensor reduced modulo a prime, $\mu_r$ is reduced modulo the same prime.

\subsection{Constructing uSVP and BDD instances}

Both lattice reductions construct an integer lattice $\Lambda\subseteq\Z^D$, given by a basis and possibly of rank less than $D$, together with an integer-valued linear function $\mu:\Lambda\to\Z$. When the reduction succeeds on a YES input, $\Lambda$ contains a vector $y$ with $\mu(y)=1$ and $\norm y_p\le R$. The following lemma constructs full-rank uSVP and BDD instances from two kinds of lower bound. When the construction succeeds on a YES input, we need a lower bound on the norms of nonzero vectors in $\Lambda\cap\ker\mu$. On a NO input, we need a lower bound on the norms of all nonzero vectors in $\Lambda$.

\begin{lemma}[Constructing uSVP and BDD instances]
\label{lem:normalized-output}
Fix $1\le p\le\infty$, $\Gamma>1$, and ${0<\rho<1/2}$, and set
$\tau=\max\{\Gamma,1/\rho\}$. Let $R\ge1$ and let $\Delta>16\tau R$ be rational. Suppose the lattice $\Lambda$ and integer-valued linear function $\mu$ satisfy one of
these two conditions:
\begin{enumerate}[label=(\roman*)]
\item every nonzero vector of $\Lambda$ has norm greater than $\Delta$.
\item there is $y\in\Lambda$ with $\mu(y)=1$ and $\norm y_p\le R$, and
      every nonzero vector of $\Lambda\cap\ker\mu$ has norm greater than
      $\Delta$.
\end{enumerate}
Given a rational $u$ with $R\le u\le5R/4$, let
$d=\lceil4u\rceil$. One can construct a full-rank integer basis and threshold for $\Up\Gamma$, and a full-rank integer basis, target, and threshold for $\Dp\rho$, in $\R^{D+1}$. Both use the same integer threshold $d$ and are promised NO instances in case (i) and YES instances in case (ii).
The construction takes as input a basis of $\Lambda$, the values of $\mu$ on its basis vectors, and the rationals $\Delta$ and $u$. It is deterministic and runs in polynomial time in their description lengths. In case (ii), it also recovers $y$ from any valid solution of the constructed search problem, up to sign for uSVP.
\end{lemma}
For uSVP, append $h\mu(x)$ to each $x\in\Lambda$, with $R\le h=O(R)$. In case (ii), $x-\mu(x)y$ is a nonzero vector in $\Lambda\cap\ker\mu$ whenever $x$ is not proportional to $y$, so its norm exceeds $\Delta$. Either $\|x\|_p$ or $\|\mu(x)y\|_p$ exceeds $\Delta/2$. The magnitude of the added coordinate is at least $\|\mu(x)y\|_p$, so $(x,h\mu(x))$ has norm greater than $\Delta/2$, while $(y,h)$ has norm of order $R$.

For BDD, use a separate $h>\Delta$ and target $(0,h)$. Nonzero vectors with $\mu=0$ have norm greater than $\Delta$, while integer-valued $\mu$ makes every other vector's added coordinate at least $h$ in magnitude. The embedded lattice therefore has minimum distance at least $\Delta$. In case (ii), $(y,h)$ is within distance $R$ of the target. In case (i), a point $(x,h\mu(x))$ within distance $\Delta$ of the target would have $\mu(x)=1$ and hence a nonzero $x$ of norm at most $\Delta$, contradicting case (i). We add vectors to make the lattice full rank, choosing them to preserve these guarantees.
Section~\ref{sec:output-details} contains the full construction and proof.

\section{The code theorem}\label{sec:codes-body}

\begin{theorem}\label{thm:formal-code}
For every fixed finite field $\F_q$ and every fixed polynomial-time computable $\rho\in(0,1/2]$, there are deterministic polynomial-time many-one reductions in both directions between $\UCS$ and $\BDDc\rho$. Both reductions preserve the promises. Consequently, a deterministic polynomial-time promise-preserving many-one reduction from SAT to $\BDDc\rho$ exists if and only if $\NP=\UP$.
\end{theorem}

To construct the decoding instance, we encode the tensors satisfying the constraints from Section~\ref{sec:symmetric-tensors}. The map sends the tensor of a compatible selection to a vector of Hamming weight at most $w$. A left inverse of the encoding map recovers the original tensor from an encoded tensor. A codeword is an encoded tensor whose original tensor has $\mu_r=0$. The prescribed syndrome requires an encoded tensor whose original tensor has $\mu_r=1$. Two errors with this syndrome therefore differ by a codeword. When the circuit accepts exactly one input, the lower bound on the weight of a nonzero codeword exceeds $w/\rho$, establishing the required minimum distance. When the circuit accepts no input, the bound for all nonzero encoded tensors also establishes this distance and excludes errors of weight at most $w$, since the prescribed syndrome requires $\mu_r=1$ on the original tensor, which excludes the zero error.

\subsection{Hamming weight of encoded tensors}\label{sec:code-amplification}

We bound the Hamming weight of encoded tensors in two cases. When the space $V$ defined below has no nonzero rank-one matrix, the bound applies to every nonzero encoded tensor. When the rank-one matrices in $V$ are precisely the nonzero scalar multiples of $yy^T$, where $y$ is the indicator of a compatible selection, the bound applies to nonzero encoded tensors in the kernel of the linear function induced by $\mu_r$. The encoded tensors on which this function is zero form the code whose minimum distance we bound for BDD.

Let $V$ be a space of matrices as in Section~\ref{sec:symmetric-tensors}, defined using a code $C\subseteq\F_q^{N_0}$.
Let $E:\F_q^{N_0}\to\F_q^N$ be an injective linear map with a known linear left inverse
$E_{\rm left}:\F_q^N\to\F_q^{N_0}$, and suppose $EC$ has distance at least $\delta>0$.
Let $r\ge1$ be an integer and let
\[
 \begin{gathered}
 W_r=E^{\otimes r}S_r(V),\\
\alpha=1+1/q,\\
\ell_r(Z)=\mu_r(E_{\rm left}^{\otimes r}Z).
 \end{gathered}
\]
For $Z\in W_r$, $\ell_r(Z)$ is $\mu_r$ evaluated on the original tensor.
\begin{theorem}[Hamming weight of encoded tensors]
\label{thm:compatible-hamming}
If $V$ has no nonzero rank-one matrix, every nonzero $Z\in W_r$ has
\[
 \wt(Z)\ge\alpha^{r-1}\delta^r.
\]
If instead the rank-one matrices in $V$ are precisely the nonzero scalar multiples of $yy^T$ for the Boolean indicator $y$ of a compatible selection, every nonzero $Z\in W_r\cap\ker\ell_r$, $r\ge2$, has
\[
 \wt(Z)\ge\alpha^{r-2}\delta^r.
\]
\end{theorem}
\begin{proof}
For independent codewords $u,v$, the union of their supports has size at least $\alpha\delta$ \cite[Lemma~2.7]{BGLR25}. Indeed, the $q$ words $u+tv$ for $t\in\F_q$, together with $v$, each have weight at least $\delta$, while each coordinate in the union of their supports is nonzero in exactly $q$ of these $q+1$ words. Counting nonzero entries gives a union of size at least $(q+1)\delta/q=\alpha\delta$. Any $r\ge2$ independent columns in $EC$ contain such a pair, so the union of their supports has size at least $\alpha\delta$. Lemma~\ref{lem:no-growth} with $A_r=\alpha^{r-1}$ propagates this factor across tensor orders, proving the first bound.

For the case $\ell_r(Z)=0$, write $Z=E^{\otimes r}T$. By the definition of $\ell_r$ and the left inverse of $E$, $\mu_r(T)=\ell_r(Z)=0$. Lemma~\ref{lem:compatible-conditioning}(iii) gives an index $i\notin\supp y$ with a nonzero slice $T_i\in S_{r-1}(V_i)$ of $T$, where $V_i$ has no nonzero rank-one matrix. Since $EC_i\subseteq EC$, the code $EC_i$ has distance at least $\delta$. For $r\ge3$, apply the first assertion of Theorem~\ref{thm:compatible-hamming} at order $r-1$. For $r=2$, use the minimum distance of $EC_i$. In both cases,
\[
 \wt(E^{\otimes(r-1)}T_i)\ge\alpha^{r-2}\delta^{r-1}.
\]
Applying the row $e_i^TE_{\rm left}$ to the first index of $Z$ recovers $E^{\otimes(r-1)}T_i$. Each nonzero coordinate of this tensor fixes the remaining $r-1$ indices and identifies a distinct nonzero column of $Z$, with the first index free. Each such column lies in $EC$ and contributes at least $\delta$ nonzero entries. The columns are disjoint, so $\wt(Z)\ge\delta\wt(E^{\otimes(r-1)}T_i)$, proving the claim.
\end{proof}

\subsection{The two reductions}\label{sec:code-reduction}

\begin{proof}[Proof of Theorem~\ref{thm:formal-code}]
We first reduce $\UCS$ to code BDD using the tensor bounds, then reduce code BDD to $\UCS$ by checking a candidate error. Circuits with no gates are handled by direct evaluation and fixed
promised outputs. Assume the circuit has a gate. Use the blocks and equations~\eqref{eq:source-mass}--\eqref{eq:source-marginal} from Section~\ref{sec:common-source}, with one
copy per label, and handle an empty block by a fixed promised NO
output. Take $C=\F_q^L$, and let $V$ be the space of matrices defined by case~(i) of Proposition~\ref{prop:partition-main}.

We choose a linear map $E$ to control the weight of the error from a compatible selection, then choose the tensor order large enough that the minimum distance exceeds the allowed error weight divided by $\rho$. Fix a rational $\rho_-$ with $\rho/2<\rho_-\le\rho$ by refining a rational interval $[a,b]$ known to contain $\rho$ until $a>0$ and $b<2a$, and taking $\rho_-=a$.
Set
\[
 \begin{gathered}
 \varepsilon=\frac1{2(q+1)},\\
\theta=1-\varepsilon,
 \end{gathered}
\]
and choose the least integer $t\ge2$ satisfying
\[
 \left(1+\frac1{2q}\right)^{t-2}\ge\frac2{\rho_-}.
\]
Then $t=O(q\log(2/\rho))$. Apply
Lemma~\ref{lem:balanced-encoding} with accuracy $\varepsilon$, so every
nonzero vector in the image of $E$ has weight in $[\theta U,U]$, where
$U=(1-1/q)N$ is an integer and $N\le(qL/\varepsilon)^2$.
Use $W=W_t$, $\ell=\ell_t$, and $w=U^t$.
The second bound in Theorem~\ref{thm:compatible-hamming}, with $\delta=\theta U$ and $r=t$, exceeds $w/\rho$. Indeed, $\alpha\theta=1+1/(2q)$ and $\theta^2>1/2$, so
\[
 (\theta U)^t\alpha^{t-2}
 =\theta^2(\alpha\theta)^{t-2}U^t
 >w/\rho_-\ge w/\rho.
\]

Compute a basis for $S_t(V)$ from its linear equations, then apply $E^{\otimes t}$ to the basis vectors. Their images form a basis for $W$. Append the row $\ell$ to a parity-check matrix for $W$ and use zeros
followed by one as the syndrome. The code is $C_0=W\cap\ker\ell$, and the vectors with the prescribed syndrome form $\mathcal A=\{e\in W:\ell(e)=1\}$. We take $d(\{0\})=\infty$.
When the original circuit accepts exactly one input, let $y$ be the indicator of its compatible selection. The error
$(Ey)^{\otimes t}$ has weight at most $w$ and satisfies $\ell((Ey)^{\otimes t})=1$.
By Theorem~\ref{thm:compatible-hamming}, $d(C_0)>w/\rho$. For an unsatisfiable circuit, the first bound in Theorem~\ref{thm:compatible-hamming} says every nonzero vector in $W$ has weight greater than $w/\rho$. Since $C_0\subseteq W$, its minimum distance is again greater than $w/\rho$. There is no error of weight at most $w$ with the prescribed syndrome, since $\ell(e)=1$ implies $e\ne0$. Thus $w<\rho d(C_0)$ holds in both cases. The output is YES when the circuit accepts exactly one input and NO when it is unsatisfiable.
For fixed $q,\rho$, the output length $N^t$ and the time needed for the linear algebra are polynomial in the input size. The threshold $w$ has polynomial bit length.

For the search version (Corollary~\ref{cor:formal-search-usvp}(iii)), we recover an input accepted by the circuit from an error of weight at most $w$ with the prescribed syndrome. When the circuit accepts exactly one input, $d(C_0)>w/\rho\ge2w$ ensures that at most one error of weight at most $w$ has the prescribed syndrome. The unique error of weight at most $w$ with this syndrome is therefore $e=(Ey)^{\otimes t}$. Taking the diagonal of $E_{\rm left}^{\otimes t}e$ recovers $y$. Reading the selected assignments, including those in blocks for input bits that no gate uses, recovers the accepted input.

For the reduction from code BDD to $\UCS$, take a promised code BDD instance $(H,\sigma,w)$ with $w<\rho d(\ker H)$. Build a Boolean circuit whose input is a binary representation of an error $e$: assign each field element a distinct fixed-length binary string and reject unused strings. The circuit checks whether $He$ equals the prescribed syndrome and $\wt(e)\le w$. Each error has one binary representation. Two distinct accepted errors would differ by a nonzero codeword of weight at most $2w<d(\ker H)$. The verifier is therefore unambiguous on every promised input.

If SAT has a deterministic polynomial-time promise-preserving reduction to $\BDDc\rho$, composition with the reverse reduction shows SAT $\le_m\UCS$ and hence $\NP=\UP$. Conversely, if $\NP=\UP$, an unambiguous verifier for SAT reduces to $\UCS$, and the forward reduction completes the composition.
\end{proof}

Composing Valiant and Vazirani's reduction from SAT to $\UCS$ with the forward reduction in Theorem~\ref{thm:formal-code} yields randomized polynomial-time hardness of code BDD with inverse-polynomial success probability and no false positives~\cite{VV86}. If the circuit after adding the random constraints accepts at least two inputs, the affine space $\mathcal A$ in the forward construction contains two distinct corresponding errors of weight at most $w$. Their difference is a nonzero codeword of weight at most $2w$, so $d(C_0)\le2w$ and the required inequality $w<\rho d(C_0)$ fails. Definition~\ref{def:randomized-reduction} allows outputs outside the promise on these random choices.
\section{The lattice theorem}\label{sec:lattice-main}

\begin{theorem}\label{thm:formal-lattice}
Fix polynomial-time computable constants $1\le p<\infty$, $\Gamma>1$, and ${0<\rho<1/2}$, and let $\tau=\max\{\Gamma,1/\rho\}$. There are randomized polynomial-time reductions from \CircuitSAT{} to $\Up\Gamma$ and $\Dp\rho$ with no false positives and success probability at least $0.20/(2s+1)$ on a satisfiable circuit of size $s$. Separate randomized polynomial-time reductions from $\UCS$ to these problems have no false positives and succeed with probability greater than $0.24$ on a YES input. The output dimension is $s^{O_p(\tau^2)}$, and the bases, targets, and rational thresholds have polynomial bit length.
\end{theorem}

For a satisfiable circuit, we choose random parity checks defining a code $C$ with large minimum distance and prove that, with inverse-polynomial probability, exactly one compatible selection of copied labels has its indicator in $C$. The tensor encoding this selection is a short lattice vector. We define an integer-valued linear function $\mu$ that equals one on every tensor encoding a compatible selection. We bound the norm of every nonzero lattice vector when no compatible selection has its indicator in $C$. When exactly one does, we bound the norm of every nonzero lattice vector with $\mu=0$. In both cases the lower bound is much larger than the norm of an encoding. Lemma~\ref{lem:normalized-output} turns these lower bounds on norms into the uSVP and BDD instances. Appendix~\ref{sec:multiplier} analyzes the probability that exactly one compatible selection has its indicator in $C$, and Appendix~\ref{sec:amplification} proves the lower bounds on norms of integer tensors.

\subsection{The lattice of integer tensors}

We choose the tensor order $m$ to make both lower bounds large compared with $R=g^{m/p}$, the norm of a tensor encoding a selection. The lower bounds include a term proportional to $m\delta^{m/p}$, where $\delta$ is the minimum distance of the code we construct. Dividing this term by $R=g^{m/p}$ introduces the factor $(\delta/g)^{m/p}$, so we choose $\delta$ close enough to $g$ to keep the gain from $m$. Set $G=2\tau$. As in Section~\ref{sec:lattice-parameters}, choose the tensor order $m$ as a sufficiently large constant times $G$, where the constant may depend on $p$. Set $\varepsilon=1/(10m)$. Choose a prime $P>16\tau R$ large enough for the probability estimates in the isolation argument. Every nonzero integer tensor vanishing modulo $P$ then has norm at least $P>16\tau R$. Then construct the code $C\subseteq\F_P^N$ using the random parity checks of Proposition~\ref{prop:lattice-isolation-main} with this $\varepsilon$. Choose $V$ as the space of matrices for $C$ in the prime-field case of Proposition~\ref{prop:partition-main}. Let $D=N^m$.
The lattice of integer tensors is
\begin{equation}\label{eq:mod-lattice}
 \begin{gathered}
 \Lambda(W,P)=\{z\in\Z^D:z\bmod P\in W\},\\
W=S_m(V).
 \end{gathered}
\end{equation}
Its integer-valued linear function is $\mu(T)=\mu_m(T)$, the sum of the diagonal entries over the first block in \eqref{eq:source-readout}. A basis for this lattice is
computable in polynomial time (Lemma~\ref{lem:small-basis}).

The two isolation procedures in Appendix~\ref{sec:multiplier} construct a code $C$ of distance
$\delta=\lfloor(1-\varepsilon)g\rfloor+1$ for every seed. For an arbitrary satisfiable circuit, the first procedure has inverse-polynomial probability of producing a code that contains the indicator of exactly one compatible selection of copied labels. Under the promise that the original circuit accepts exactly one input, the second procedure has constant probability of producing a code with the same property (Proposition~\ref{prop:lattice-isolation-main} in Appendix~\ref{sec:multiplier}). The first procedure also chooses the number of copies at random. In both procedures, random nonzero field elements scale the columns of the parity-check matrix and thereby determine the code. The remaining steps of the lattice construction are deterministic.

Theorem~\ref{thm:linear-tensor} bounds every nonzero $T\in\Lambda(W,P)$ according to how many compatible selections have their indicators in $C$:
\[
 \begin{aligned}
 \norm{T}_p&\ge\min\{P,c_p^{-1}m\delta^{m/p}\}
 &&\text{if none exist},\\
 \norm{T}_p&\ge\min\{P,d_p^{-1}m\delta^{m/p}\}
 &&\text{if exactly one exists and }\mu(T)=0,
 \end{aligned}
\]
for constants $c_p,d_p$ depending only on $p$. The choices of $m$ and $P$ in the proofs of Propositions~\ref{prop:lp-separation-main} and~\ref{prop:compatibility-lattice} imply $(\delta/g)^m>0.9$ and make both lower bounds exceed $16\tau R$. Choose a rational $\Delta$ between $16\tau R$ and these lower bounds. When no compatible selection has its indicator in $C$, every nonzero lattice vector has norm greater than $\Delta$, as required by Lemma~\ref{lem:normalized-output}(i). When $C$ contains exactly one indicator $y$ of a compatible selection, the tensor $y^{\otimes m}$ has norm $R$ and $\mu=1$, while every nonzero lattice vector with $\mu=0$ has norm greater than $\Delta$. These are the hypotheses of Lemma~\ref{lem:normalized-output}(ii). The lemma constructs both instances. Section~\ref{sec:lattice-parameters} contains the parameter choices and bounds on the bit lengths of the output bases, targets, and thresholds for the reduction from instances with blocks of allowed assignments and compatibility requirements.

To analyze the success probability, write $\ell_i$ for the original block sizes and $J=\lceil\log_2\prod_i\ell_i\rceil$, as in Proposition~\ref{prop:lattice-isolation-main}.

\begin{proof}[Proof of Theorem~\ref{thm:formal-lattice}]
Circuits with no gates are handled by direct evaluation and fixed promised outputs. Otherwise convert the circuit to a constraint satisfaction instance with blocks and labels as in Section~\ref{sec:common-source}, and apply Proposition~\ref{prop:compatibility-lattice} with $G=2\tau$. Use construction (i) for arbitrary \CircuitSAT{} inputs and construction (ii) for promised $\UCS$ inputs. If the original circuit has size $s$, the construction has $g\le s$ blocks, at most four labels per block, and $L\le4g$. Thus
\[
 J\le2g\le2s.
\]
The success probabilities follow from Proposition~\ref{prop:lattice-isolation-main}. The dimension is $s^{O_p(\tau^2)}$, and the running time and bit lengths of bases, targets, and thresholds are polynomial by Proposition~\ref{prop:compatibility-lattice}.
\end{proof}

\section{Discussion}\label{sec:open}

\paragraph{Larger factors.}
To make the lower bounds on norms exceed larger multiples $G$ of the encoding's norm, we increase the tensor order. This raises the dimension and cost of the lattice construction. For a circuit of size $s$, the reduction in Proposition~\ref{prop:compatibility-lattice} uses tensor order $O_p(G)$ and has dimension $s^{O_p(G^2)}$. A quasipolynomial-time reduction reaches the factor $(\log(n+2))^\beta$ for decision uSVP, where $n$ is the output rank and ${0<\beta<1/2}$ (Corollary~\ref{cor:formal-growing}). For every fixed rational $p>2$ and for $p=\infty$, a different construction reaches uniqueness factors that are fixed positive powers of the lattice rank (Theorem~\ref{pa2:main}). Section~\ref{pa2:sec-euclidean-obstruction} explains an obstruction to using that construction in the Euclidean norm. Proving randomized polynomial-time NP-hardness for any superconstant uniqueness factor in the Euclidean norm remains open.

\paragraph{Parameterized hardness.}
Applying the construction in Proposition~\ref{prop:compatibility-lattice} to $k$-Clique shows that GapSVP at every fixed constant approximation factor in every fixed finite $\ell_p$ norm is $\mathsf W[1]$-hard under randomized FPT reductions with one-sided error, with the integer threshold on the vector norm as the parameter. This includes approximation factors greater than $2$ in $\ell_1$, answering questions of Hirahara and Ogitsuka~\cite{HO26} and Bennett~\cite[Open Problem~4.7]{Ben23}. See Corollary~\ref{cor:formal-gapsvp} for the result and proof.

\paragraph{Deterministic reductions.}
By Theorem~\ref{thm:formal-code}, code BDD is NP-hard under deterministic promise-preserving reductions if and only if $\NP=\UP$. The decision versions of Euclidean uSVP and BDD also have verifiers with at most one accepting witness on promised inputs. Thus a deterministic polynomial-time promise-preserving reduction from SAT to either problem would imply $\NP=\UP$ (Proposition~\ref{prop:lattice-up}). For fixed rational $p>2$ and for $p=\infty$, the reductions of Theorem~\ref{pa2:main} are deterministic from $\UCS$. The lattice reduction for constant factors uses random parity checks, including for circuits accepting at most one input (Theorem~\ref{thm:formal-lattice}). Whether $\UCS$ reduces deterministically to either Euclidean lattice problem remains open.

\paragraph{Success probability.}
Our reductions have no false positives, so independent repetition amplifies their success probability for decision: run the reduction several times, call the decision algorithm on each output, and accept if any call accepts. Proposition~\ref{prop:isolation-recovery-limit} instead concerns producing one output circuit with exactly one accepted input, together with a fixed deterministic polynomial-time algorithm that recovers an input accepted by the original circuit from that unique input. Under the other hypotheses of the proposition, if this joint success probability exceeds $2/3+1/\poly(s)$ for every satisfiable circuit of description length $s$, then $\coNP\subseteq\mathsf{NP}/\poly$. See Appendix~\ref{sec:limits} for discussion.

\phantomsection
\addcontentsline{toc}{section}{References}
\begin{sloppypar}

\end{sloppypar}
\clearpage
\appendix
\addtocontents{toc}{\protect\setcounter{tocdepth}{1}}
\section{Isolation calculations}\label{sec:multiplier}

The random parity checks in this appendix define a code with large minimum distance for every random choice. If the instance has a compatible selection, then with inverse-polynomial probability exactly one compatible selection of copied labels has its indicator in that code, as needed for the reductions to uSVP and BDD.

\begin{proposition}[Isolation]\label{prop:lattice-isolation-main}
Fix a rational number $0<\varepsilon\le1/4$. Consider the blocks and requirements of Section~\ref{sec:common-source} for an instance with $g$ nonempty blocks, and call a compatible selection a solution. Suppose block $i$ contains $\ell_i\ge1$ labels, let $L=\sum_i\ell_i$, and let $Q_0$ be an integer satisfying
\[
 Q_0\ge(2^{24}gL)^{1/\varepsilon}.
\]
Set
\[
 \begin{gathered}
 J=\left\lceil\log_2\prod_{i=1}^g\ell_i\right\rceil,\\
\delta=\lfloor(1-\varepsilon)g\rfloor+1.
 \end{gathered}
\]
There is a universal constant $g_0$ such that, for $g\ge g_0$, one can deterministically find a prime $P\in[Q_0,2Q_0]$ and run either of the following randomized procedures in time polynomial in the numerical value of $Q_0$, $g$, and $L$. Each procedure chooses a number $b$ of copies per label and a code $C\subseteq\F_P^{bL}$. We count compatible selections of copied labels whose indicators lie in $C$. Each copy represents the same assignment as its original label, so compatibility is unchanged by the choice of copies.
\begin{enumerate}[label=(\roman*)]
\item If the instance has a solution, exactly one such selection exists under the first procedure with probability at least
\[
 \frac{0.20}{J+1}.
\]
\item If the instance has exactly one solution, exactly one such selection exists under the second procedure with probability greater than $0.24$.
\end{enumerate}
For every random seed, $C$ has minimum distance $\delta$ and length $N=bL<P\le2Q_0$. Let $V$ be the space of matrices constructed from $C$ as in case~(ii) of Proposition~\ref{prop:partition-main}. If the instance has no solution, $V$ has no nonzero rank-one matrix for any seed.
\end{proposition}

Let the instance have $g\ge g_0$
nonempty label blocks of sizes $\ell_1,\ldots,\ell_g$, and let
$L=\sum_i\ell_i$, where $g_0$ is the constant in
Lemma~\ref{lem:multiplier-margins}. Set
\[
 \begin{gathered}
 h=\lfloor(1-\varepsilon)g\rfloor,\\
\delta=h+1.
 \end{gathered}
\]
We make $b$ copies of each label, so block $i$ has $n_i=b\ell_i$ coordinates and $N=bL$. We use a Vandermonde matrix so that the code has minimum distance $\delta$ for every choice of nonzero scalars, one for each column. Choosing these scalars at random changes which indicators of compatible selections belong to the code. For $h<N\le P-1$, as ensured by the choices of $b$ below, choose distinct $\alpha_1,\ldots,\alpha_N\in\F_P^*$ and set $H_{j,i}=\alpha_i^j$ for $0\le j<h$. Choose the scalars $\beta_i\in\F_P^*$ independently and uniformly. They define
\begin{equation}\label{eq:multiplier-code}
 C_\beta=\ker\bigl(H\operatorname{diag}(\beta)\bigr).
\end{equation}
Any $h$ columns of the Vandermonde matrix are independent, and multiplying columns by nonzero scalars preserves independence. Since $N>h$, some $h+1$ columns are dependent. The code $C_\beta$ therefore has distance $h+1=\delta$ for every choice of $\beta$ (Lemma~\ref{lem:multiplier-mds}).

Let $\mathcal F$ consist of the sets of coordinates chosen by compatible selections of copied labels. Let $X$ count the sets in $\mathcal F$ whose indicator vectors lie in $C_\beta$. If the original instance has $\nu$ solutions, then $|\mathcal F|=\nu b^g$. Write
\[
 \begin{gathered}
 \vartheta=\nu b^gP^{-h},\\
A=(1-1/P)^{-g},\\
\eta=\prod_{i=1}^g\left(1+\frac{n_i-1}{P-1}\right)-1.
 \end{gathered}
\]
Since $\Pr[X=1]\ge \mathbb E[X]-\mathbb E[X(X-1)]$, we lower-bound $\mathbb E[X]$, the expected number of compatible selections of copied labels whose indicators lie in $C_\beta$, and upper-bound $\mathbb E[X(X-1)]$, the expected number of ordered pairs of distinct compatible selections whose indicators lie in $C_\beta$. When a compatible selection exists, a suitable number of copies makes the lower bound on $\mathbb E[X]$ exceed the upper bound on $\mathbb E[X(X-1)]$ by a positive constant.

\begin{lemma}[First and second moments]
\label{lem:multiplier-survival}\label{lem:multiplier-second-moment}
The random variable $X$ satisfies
\begin{align}
 A(1-g/P)\vartheta&\le\E[X]\le A\vartheta,
       \label{eq:multiplier-survival-bounds}\\
 \E[X(X-1)]&\le A^2\vartheta^2+A\vartheta\eta,
       \label{eq:multiplier-second-moment}\\
 \Prob[X=1]&\ge A\vartheta\bigl(1-g/P-A\vartheta-\eta\bigr).
       \label{eq:one-survivor-lower}
\end{align}
Moreover, $\eta\le\exp((N-g)/(P-1))-1$.
\end{lemma}
\begin{proof}
Let $S\in\mathcal F$, so $|S|=g$. Its indicator lies in $C_\beta$ when $H_S\beta_S=0$, where $H_S$ is the submatrix of $H$ with columns indexed by $S$ and $\beta_S$ is the corresponding vector of scalars. Every $\beta_i$ is nonzero. Any $h$ columns of $H$ are independent, so $H_S$ has rank $h$. Over $\F_P$, the equation $H_S\beta_S=0$ therefore has $P^{g-h}$ solutions before excluding zero coordinates. Since $g-1\ge h$, deleting any column of $H_S$ leaves rank $h$, and each coordinate is zero in $P^{g-h-1}$ solutions. By a union bound over the $g$ coordinates, between $P^{g-h}-gP^{g-h-1}$ and $P^{g-h}$ solutions have all coordinates nonzero. Dividing by the $(P-1)^g$ equally likely choices of $\beta_S$ gives~\eqref{eq:multiplier-survival-bounds}.

Take distinct $S,T\in\mathcal F$ and let
$r=|S\setminus T|=|T\setminus S|$, the number of blocks in which they select different coordinates.
If $r\ge h$, view the two systems $H_S\beta_S=0$ and $H_T\beta_T=0$ as one system in the variables indexed by $S\cup T$. The rows for $T$ vanish on the columns indexed by $S\setminus T$. Since any $h$ Vandermonde columns are independent, a linear dependence among the rows must have zero coefficients on the rows for $S$. The columns indexed by $T\setminus S$ similarly force the coefficients on the rows for $T$ to be zero. Thus the combined system has rank $2h$.
Thus the probability that both indicators belong to $C_\beta$ is at most
\[
 \frac{P^{g+r-2h}}{(P-1)^{g+r}}\le A^2P^{-2h}.
\]
If $r<h$, fix the scalars on $S$ to any nonzero values satisfying $H_S\beta_S=0$. The scalars on $T\setminus S$ remain independent and uniform in $\F_P^*$. The columns of $H$ indexed by $T\setminus S$ are linearly independent, so at most one assignment to the scalars on $T\setminus S$ satisfies the equations for $T$. With the scalars on $S$ fixed, the indicator of $T$ belongs to $C_\beta$ with probability at most $(P-1)^{-r}$.
For a fixed set $I$ of blocks on which the selections differ, there are at most $\prod_{i\in I}(n_i-1)$ choices of $T$ for each $S$. Summing the corresponding joint probabilities over ordered pairs with $r<h$ bounds their contribution to $\E[X(X-1)]$ by $A\vartheta\eta$. There are at most $|\mathcal F|^2$ pairs with $r\ge h$, and their contribution is at most $A^2\vartheta^2$. This proves \eqref{eq:multiplier-second-moment}.
Inequality~\eqref{eq:one-survivor-lower} follows from
$\mathbf1_{\{X=1\}}\ge X-X(X-1)$. Finally use $1+x\le e^x$
and $\sum_i(n_i-1)=N-g$.
\end{proof}

\subsection{Choosing the number of copies}
$\mathbb E[X]$ is the expected number of compatible selections of copied labels whose indicators lie in $C_\beta$. When a compatible selection exists, a suitable number of copies makes $\mathbb E[X]$ lie between positive constants. Since the number of original solutions is unknown, we choose an index uniformly at random from a list of possibilities and use the corresponding number of copies. Under a promise of at most one solution, we fix the number of copies.

Set $J=\lceil\log_2\prod_i\ell_i\rceil$. The number of compatible selections before copying is at most $\prod_i\ell_i$, so if it is positive, one of the powers $2^j$ for $j\in\{0,\ldots,J\}$ approximates it within a factor of two. The algorithm chooses $j$ uniformly. If there are $\nu>0$ solutions, the analysis uses $j=\lceil\log_2\nu\rceil$. For each $j$, we choose $b_j$ to make $b_j^g/P^h$ roughly the reciprocal of $2^j$, which makes $\vartheta$ a positive constant for a suitable guess. Specifically, let $b_j$ be the largest integer satisfying
\begin{equation}\label{eq:multiplier-Bj}
 3\cdot2^j b_j^g\le2P^h.
\end{equation}
For an instance promised to have at most one solution, use instead
the largest integer $b_*$ satisfying
\begin{equation}\label{eq:multiplier-Bstar}
 2b_*^g\le P^h.
\end{equation}
The next lemma controls the effect of rounding. Binary search computes these integers.

\begin{lemma}[Bounds for the number of copies]\label{lem:multiplier-margins}
There is a universal $g_0$ such that, if $g\ge g_0$ and
$P\ge(2^{24}gL)^{1/\varepsilon}$, every choice $b=b_j$ or $b=b_*$ satisfies
\[
 \begin{gathered}
 h<N=bL\le P-1,\\
1-g/P\ge0.999,\\
A\le1.001,\\
\eta\le10^{-6}.
 \end{gathered}
\]
Let $\widehat b$ be the positive real solution when the inequality defining $b$ in
\eqref{eq:multiplier-Bj} or \eqref{eq:multiplier-Bstar} is replaced by equality.
Then $b^g\ge0.999\,\widehat b^g$.
\end{lemma}
\begin{proof}
Each $b\le P^{h/g}\le P^{1-\varepsilon}$, so
\[
 \frac{N}{P-1}\le\frac{2L}{P^\varepsilon}
 \le\frac{1}{2^{23}g}.
\]
In particular, $N\le P-1$ and $1-g/P\ge0.999$. The bounds on $A$ and $\eta$ follow from $\log A\le g/(P-1)$ and $\eta\le\exp(N/(P-1))-1$.
Also $2^J\le2L^g$, so $\widehat b\ge P^{h/g}/(3L)$ in either case. Take $g_0$ large enough that
$h/g\ge1/2$. The lower bound on $P$ then implies
$\widehat b\ge P^{1/2}/(3L)>10^6g+1$. Consequently
\[
 (b/\widehat b)^g\ge(1-1/\widehat b)^g
 \ge1-g/\widehat b\ge0.999.
\]
Since $b=\lfloor\widehat b\rfloor$ and $\widehat b>10^6g+1$, we have $b>10^6g$ and hence $N>h$.
\end{proof}

\begin{proof}[Proof of Proposition~\ref{prop:lattice-isolation-main}]
Find $P\in[Q_0,2Q_0]$ by Lemma~\ref{lem:prime-interval}.
The first procedure guesses $j$ uniformly and uses $b_j$ copies.
The second uses $b_*$. First suppose the scalars are independent and uniform on $\F_P^*$.
For $\nu>0$, the choice $j=\lceil\log_2\nu\rceil$ ensures
$0.333\le\vartheta\le2/3$. Lemma~\ref{lem:multiplier-margins} makes the factor $1-g/P-A\vartheta-\eta$ in \eqref{eq:one-survivor-lower} positive on this interval. Since $A\ge1$, \eqref{eq:one-survivor-lower} gives
\begin{equation}\label{eq:factorial-moment-simple}
 \Prob[X=1]\ge
 \vartheta\bigl(0.999-1.001\vartheta-10^{-6}\bigr).
\end{equation}
This concave quadratic exceeds $0.22$ at both endpoints.
When the original instance has exactly one compatible selection, $b_*$ ensures $0.4995\le\vartheta\le0.5$,
and the same quadratic exceeds $0.249$ throughout that interval.

Both procedures use a fixed number of fair bits as follows.
For the first, let
$N_{\max}=\max_j b_jL$ and
$K=(J+1)(P-1)^{N_{\max}}$. A uniform residue modulo $K$ determines a uniformly random choice of $j$ and $N_{\max}$ independent uniform nonzero scalars. Use the first $N=b_jL$ scalars.
Let $t=\lceil\log_2(100K)\rceil$, choose $U$ uniformly from $\{0,\ldots,2^t-1\}$ using $t$ fair bits, and use $U\bmod K$.
Every residue has probability at least
\[
 \frac{\lfloor2^t/K\rfloor}{2^t}\ge\frac{0.99}{K},
\]
so every set of residues has probability at least $0.99$ times its probability under the uniform distribution modulo $K$.
For the second procedure use $K=(P-1)^{b_*L}$.
These integers have polynomial bit length, and each procedure uses polynomially many random bits.
The success probabilities with fair bits therefore exceed
\[
 \begin{gathered}
 \frac{0.99\cdot0.22}{J+1}>\frac{0.20}{J+1},\\
0.99\cdot0.249>0.24,
 \end{gathered}
\]
respectively.

For every choice of random bits, all scalars are nonzero.
Lemma~\ref{lem:multiplier-margins} implies
$h<N\le P-1\le2Q_0$, and Lemma~\ref{lem:multiplier-mds}
establishes distance $\delta$ for the code $C_\beta$ defined in \eqref{eq:multiplier-code}.
Proposition~\ref{prop:partition-main}(ii) shows that $V$ contains no nonzero rank-one matrix when the original instance has no compatible selection, for every random choice.
\end{proof}

\section{Tensor amplification for codes and lattices}\label{sec:amplification}

\subsection{Lower bounds on weight and norm from slices and columns}

We prove the lower bounds on weights and norms of tensors used to establish minimum distances and uniqueness factors in the code reduction and the lattice reductions for constant factors. The proofs add contributions from disjoint slices. The induction needs independent short columns at each order. If the code has minimum distance at least $\delta$, a nonzero order-$d$ array whose fibers lie in the code has at least $\delta^d$ nonzero entries. For the tensors in the induction, this count ensures that deleting sufficiently few columns preserves the span of the flattening's columns, leaving independent short columns.

\begin{lemma}[Support in a product code]\label{lem:product-support}
Let $F$ be a field and let $C\subseteq F^N$ be a linear code with minimum distance at least $\delta$. Let $A$ be a nonzero order-$d$ array, with $d\ge1$, such that every one-dimensional fiber of $A$, formed by fixing all but one index, belongs to $C$. Then
\[
 |\supp A|\ge\delta^d.
\]
\end{lemma}
\begin{proof}
For $d=1$, this is the minimum distance of $C$. For $d>1$, choose a nonzero fiber along the first index. It has at least $\delta$ nonzero positions. Fixing the first index at each such position gives a nonzero order-$(d-1)$ slice satisfying the same constraint on its fibers. By induction, each of these disjoint slices has at least $\delta^{d-1}$ nonzero entries, giving at least $\delta^d$ nonzero entries in all.
\end{proof}

Use a code $C$ and the space $V$ of matrices from Proposition~\ref{prop:partition-main}, and the spaces $S_r(V)$ from Section~\ref{sec:symmetric-tensors}. Let $\delta>0$ be a finite lower bound on the minimum distance of a code containing each tensor fiber. The following lemma shows that the union of the supports of sufficiently many short integer columns with distinct residues in $C$ contains more than $\delta$ rows. In the induction that follows, each such row fixes a disjoint slice with one fewer index. The inductive bounds on Hamming weights or on $p$th powers of norms add over these disjoint slices, yielding the bound at the next tensor order.

\begin{lemma}[Many short columns]\label{lem:many-short}
Let $C\subseteq\F_P^N$ have distance at least $\delta$, and let
$p\ge1$ and $K\ge1$. Let $v_1,\ldots,v_q\in\Z^N$ have pairwise distinct
residues modulo $P$ in $C$ and satisfy
\[
 \begin{gathered}
 \norm{v_j}_p^p\le K\delta
 \quad(1\le j\le q),\\
q\ge25K^{1/p}.
 \end{gathered}
\]
If $n_0$ is the size of the union of their supports, then
\[
 n_0\ge\left(1+\frac1{32K^{1/p}}\right)\delta.
\]
\end{lemma}
\begin{proof}
We adapt the argument of \cite[Lemma~5.1]{Wan26b} for agreements among integer entries. Let
\[
 \begin{gathered}
 A=\#\{z\in\Z:|z|^p<4K\},\\
m=\lceil8A/3\rceil.
 \end{gathered}
\]
Then $3\le A\le9K^{1/p}$ and $m\le25K^{1/p}\le q$.
Select $m$ columns, and let $u$ be the number of rows in the union of their supports.
For any $m$ integers $z_1,\ldots,z_m$,
\[
 \sum_{j=1}^m|z_j|^p
 +4K\#\{i<j:z_i=z_j\}\ge4K(m-A).
\]
At most $A$ entries are first occurrences of values with $|z|^p<4K$. Each entry with $|z|^p\ge4K$ contributes at least $4K$ to the first term. Each remaining entry repeats an earlier value. Pair it with the first occurrence of that value. These pairs are distinct and each contributes $4K$ to the second term.

Two selected columns have distinct residues in $C$, so their difference modulo $P$ is a nonzero codeword with at least $\delta$ nonzero entries. Thus they disagree on at least $\delta$ coordinates and agree on at most $u-\delta$ rows in the union of the supports of all selected columns. Sum the preceding inequality over these $u$ rows and use $\norm{v_j}_p^p\le K\delta$ for each selected column:
\[
 mK\delta+4K\binom m2(u-\delta)\ge4K(m-A)u.
\]
Consequently
\[
 u-\delta\ge
 \frac{3m-4A}{2m^2-6m+4A}\delta
 \ge\frac9{32A}\delta
 \ge\frac{\delta}{32K^{1/p}}.
\]
For the middle inequality, write $m=8A/3+e$, $0\le e<1$.
The difference between the two coefficients of $\delta$, multiplied by the
positive product $32A(2m^2-6m+4A)$, is $108A+54e-18e^2>0$. Since $n_0\ge u$, this proves the lemma.
\end{proof}

\begin{lemma}[Weight and norm from disjoint slices]\label{lem:no-growth}
Suppose $V$ has no nonzero rank-one matrix. Consider either
\begin{enumerate}[label=(\roman*)]
\item $X=E^{\otimes r}T$ over $\F_q$, with $T\in S_r(V)$, $E$
      injective, and $EC$ of distance at least $\delta>0$, where
      $\mathsf w(X)=\wt(X)$, or
\item for $1\le p<\infty$, integer $X$ with $X\bmod P\in S_r(V)$ and $C$ of distance at
      least $\delta>0$, where $\mathsf w(X)=\norm X_p^p$.
\end{enumerate}
Let $A_1=1$ and $A_r\ge1$. For each $r\ge2$, take any $r$ vectors $v_j$ with $\mathsf w(v_j)\le A_r\delta$. In (i), require the vectors to be independent in $EC$. In (ii), require integer vectors whose reductions modulo $P$ are independent in $C$. Suppose the union of the supports of the $v_j$ has size at least $\delta A_r/A_{r-1}$ for every choice satisfying these requirements. Then every nonzero tensor in (i) or (ii) satisfies 
\[
 \begin{gathered}
 \mathsf w(X)\ge A_r\delta^r\quad\hbox{in (i)},\\
\mathsf w(X)\ge\min\{P^p,A_r\delta^r\}\quad\hbox{in (ii)}.
 \end{gathered}
\]
\end{lemma}
\begin{proof}
We induct on $r$. For $r=1$, the minimum distance of $EC$ proves (i). For (ii), let $X$ be a nonzero integer vector whose residue lies in $C$. If its residue is nonzero, it has at least $\delta$ nonzero entries by the minimum distance of $C$. If its residue is zero, some entry is a nonzero multiple of $P$, so $\norm X_p\ge P$. In (ii), assume $\norm X_p<P$, so every nonzero entry has nonzero residue. We seek short columns that span the columns of $X$ over $\F_q$ in (i), or whose reductions modulo $P$ span the columns of $X\bmod P$ over $\F_P$ in (ii). Suppose $\mathsf w(X)<A_r\delta^r$ and discard columns $v$ with $\mathsf w(v)>A_r\delta$. Fewer than $\delta^{r-1}$ columns are discarded, since otherwise their contribution to $\mathsf w(X)$ would exceed $A_r\delta^r$.

These deletions preserve the span in each case. Otherwise, choose a linear function that vanishes on the remaining columns but not on a discarded column. Its values on all columns form a nonzero order-$(r-1)$ array supported only at discarded positions. Each fiber is a linear combination of fibers of $X$ in (i), or of $X\bmod P$ in (ii), so it lies in $EC$ or $C$, respectively. Lemma~\ref{lem:product-support} gives at least $\delta^{r-1}$ nonzero entries, but fewer than $\delta^{r-1}$ columns were deleted, a contradiction. When $V$ has no nonzero rank-one matrix, Lemma~\ref{lem:compatible-conditioning} gives rank at least $r$ for the flattening of $T$ in (i) and of $X\bmod P$ in (ii). The flattening of $X$ in (i) is $E$ times the flattening of $T$ times $(E^{\otimes(r-1)})^T$, and both multiplications preserve rank because $E$ is injective. Thus at least $r$ independent short columns remain in (i), and at least $r$ columns with independent reductions modulo $P$ remain in (ii).

Together, these columns have nonzero entries in at least $\delta A_r/A_{r-1}$ rows.
Fixing the first coordinate at one of these rows produces a nonzero slice of order $r-1$ satisfying the inductive hypotheses. In (i), the slice at row $i$ is $X_i=E^{\otimes(r-1)}(\sum_k E_{i,k}T_k)$, where $T_k$ is the slice of $T$ at index $k$. Each $T_k$ belongs to $S_{r-1}(V)$, so the sum does too. In (ii) each slice has norm below $P$ and has nonzero residue. Add the inductive lower bounds for these disjoint slices:
\[
 \mathsf w(X)\ge
 \frac{\delta A_r}{A_{r-1}}\,A_{r-1}\delta^{r-1}
 =A_r\delta^r,
\]
a contradiction.
\end{proof}

\begin{theorem}[Norms of integer tensors]\label{thm:linear-tensor}
Let $P$ be prime, and let $C\subseteq\F_P^N$ and $V$ be as in Proposition~\ref{prop:partition-main}. Assume $d(C)\ge\delta$ for a finite $\delta>0$. Fix $1\le p<\infty$ and an integer $a_p\ge p$, and let $c_p=256a_p$, $d_p=4c_p$. Let $B_1$ be the coordinates in the first block. For $r\ge1$, let
\[
 \begin{gathered}
 b_r=\max\{1,r/c_p\},\\
a_r=\max\{1,r/d_p\},\\
\mu_r(T)=\sum_{j\in B_1}T_{j,\ldots,j}.
 \end{gathered}
\]
\begin{enumerate}[label=(\roman*)]
\item If $V$ has no nonzero rank-one matrix, every nonzero integer $T$
      with $T\bmod P\in S_r(V)$ satisfies
      $\norm T_p\ge\min\{P,b_r\delta^{r/p}\}$.
\item If the rank-one matrices in $V$ are precisely the nonzero scalar multiples of $yy^T$, with $y$ a
      Boolean indicator of a compatible selection, every nonzero integer $T$ with
      $T\bmod P\in S_r(V)$ and $\mu_r(T)=0$ satisfies
      $\norm T_p\ge\min\{P,a_r\delta^{r/p}\}$.
\end{enumerate}
\end{theorem}
The condition $\mu_r(T)=0$ in part~(ii) excludes the nonzero multiples of $y^{\otimes r}$, which may be short, since $\mu_r(y^{\otimes r})=1$.
\begin{proof}
For part~(i), apply Lemma~\ref{lem:no-growth} with
$A_r=b_r^p=\max\{1,(r/c_p)^p\}$.
For $r\le c_p$, the support of a nonzero codeword contains at least $\delta=\delta A_r/A_{r-1}$ rows. For $r>c_p$, let $K=(r/c_p)^p$.
Consider any $r$ integer columns whose residues modulo $P$ are independent and lie in $C$, each with $p$th power of its norm at most $K\delta$. Lemma~\ref{lem:no-growth} requires the union of their supports to contain at least $\delta A_r/A_{r-1}$ rows. Since $r\ge25K^{1/p}$, Lemma~\ref{lem:many-short} gives at least $(1+c_p/(32r))\delta$ rows. It remains to compare these quantities. Since $r-1\ge c_p$,
\[
 \frac{A_r}{A_{r-1}}=\left(\frac r{r-1}\right)^p
 \le e^{p/(r-1)}\le1+\frac{4p}{r}
 \le1+\frac{c_p}{32r}.
\]
By Lemma~\ref{lem:no-growth}, $\norm T_p\ge\min\{P,b_r\delta^{r/p}\}$. Its proof also shows $\norm T_p\ge b_r\delta^{r/p}$ when $\norm T_p<P$.

For part~(ii), it again suffices to consider integer tensors $T$ with $0<\norm{T}_p<P$. Reduction modulo $P$ then preserves the support of $T$, so $\overline T=T\bmod P$ is nonzero.
Lemma~\ref{lem:product-support} bounds the support of $\overline T$, implying $\norm T_p^p\ge\delta^r$, covering $r\le d_p$.
For larger $r$, reducing $\mu_r(T)=0$ modulo $P$ gives $\mu_r(\overline T)=0$.
Lemma~\ref{lem:compatible-conditioning}(iii) identifies a nonzero slice at a label outside $\supp y$.
Let $v$ be a column of the flattening of $\overline T$ with a nonzero coordinate outside $S=\supp y$. By symmetry and the equations requiring the columns in each block to sum to the diagonal, the entries of $v$ sum to the same value $c$ in each block. Subtracting $cy$ makes the entries sum to zero in every block and leaves the coordinates outside $S$ unchanged. Since $v,y\in C$ and $C$ is linear, the resulting vector $v-cy\in C$ is a nonzero codeword. Since $S$ contains one coordinate per block, every nonzero entry of $v-cy$ indexed by $S$ must be accompanied by a nonzero entry outside $S$ in the same block. The blocks are disjoint, so
\[
 \delta\le\wt(v-cy)\le2\wt(v|_{\overline S}).
\]
Consequently $v$ has at least $\delta/2$ nonzero coordinates outside $S$. Each such coordinate identifies a nonzero slice of $\overline T$, so at least $\delta/2$ disjoint slices at labels outside $S$ are nonzero. By Lemma~\ref{lem:compatible-conditioning}(i) and (iii), each belongs to $S_{r-1}(V_i)$, where $V_i$ has no nonzero rank-one matrix and $C_i$ has distance at least $\delta$.
Apply the bound from part~(i) for tensors of norm below $P$ to the integer slices at these labels outside $\supp y$. Their residues are nonzero, and summing the $p$th powers of their norms over these disjoint slices shows that
\[
 \norm T_p^p\ge\frac12 b_{r-1}^p\delta^r
 \ge\left(\frac r{4c_p}\right)^p\delta^r,
\]
using $2^{1/p}\le2$ and $r-1\ge r/2$.

\end{proof}

\begin{proposition}[Lower bounds on norms of integer tensors]\label{prop:lp-separation-main}
Fix polynomial-time computable parameters $1\le p<\infty$ and $G>1$. There is an integer $m=O_p(G)$, with $\varepsilon=1/(10m)$, such that for every integer $g\ge1$ one can compute an integer
\[
 Q_{\rm sep}\le (g+1)^{O_p(G)}
\]
with the following property. Fix an instance with $g$ blocks and the requirements of Section~\ref{sec:common-source}, and assign copies to its labels. Let $B_i$ be the coordinates for copies from block $i$, and set
\[
 \begin{gathered}
 \delta=\lfloor(1-\varepsilon)g\rfloor+1,\\
r_0=g^{m/p}.
 \end{gathered}
\]
Let $N$ be the number of copied coordinates. For a prime
$P\ge Q_{\rm sep}$ with $P>N$, let
$C\subseteq\F_P^N$ have distance at least $\delta$,
and let $V$ be the space of matrices defined by case~(ii) of Proposition~\ref{prop:partition-main}, including $A_\phi\diag M=0$. Use the
tensor spaces $S_r(V)$ defined in Section~\ref{sec:symmetric-tensors}.
If exactly one compatible selection has its indicator in $C$, let $y$ be that indicator. The tensor $y^{\otimes m}$ has residue in $S_m(V)$, norm $r_0$, and satisfies
$\mu_m(y^{\otimes m})=1$, where $\mu_m(T)=\sum_{j\in B_1}T_{j,\ldots,j}$. In this case, every nonzero integer tensor with residue in $S_m(V)$ and $\mu_m(T)=0$ over the integers has norm greater than $Gr_0$. If no compatible selection has its indicator in $C$, every nonzero integer tensor with residue in $S_m(V)$ has norm greater than $Gr_0$.
\end{proposition}

\begin{proof}[Proof of Proposition~\ref{prop:lp-separation-main}]
We choose the tensor order $m$ proportional to $G$ so that the factor $m/d_p$ in Theorem~\ref{thm:linear-tensor} is larger than $G$. Choosing $\varepsilon$ inversely proportional to $m$ keeps $(\delta/g)^m$ bounded below by a constant. Thus $\delta^{m/p}$ remains comparable to $g^{m/p}$, the norm of a tensor encoding a selection.
Choose an integer $a_p\ge p$, let $c_p=256a_p$, $d_p=4c_p$, and choose
an integer $2d_pG<m\le2d_pG+2$. Set $\varepsilon=1/(10m)$.
Bernoulli's inequality implies
\[
 (\delta/g)^m>(1-1/(10m))^m\ge0.9.
\]
Choose an integer $Q_{\rm sep}>2Gg^{m/p}$ by rational approximation.
Then $Q_{\rm sep}\le(g+1)^{O_p(G)}$.
When exactly one compatible selection has its indicator in $C$, let $y$ be that indicator. Proposition~\ref{prop:partition-main}(ii) says that the only nonzero rank-one matrices in $V$ are multiples of $yy^T$. Every matrix slice of the symmetric Boolean tensor $y^{\otimes m}$ is zero or $yy^T$, so its residue lies in $S_m(V)$. It has norm $r_0=g^{m/p}$ and satisfies $\mu_m(y^{\otimes m})=1$. Theorem~\ref{thm:linear-tensor} bounds the norm of every nonzero integer tensor $T$ with $T\bmod P\in S_m(V)$ and $\mu_m(T)=0$ from below by
\[
 \min\{P,(m/d_p)\delta^{m/p}\}>Gr_0,
\]
because $(m/d_p)\delta^{m/p}>2G\,0.9^{1/p}r_0>Gr_0$ and $P>2Gr_0$.
If no compatible selection has its indicator in $C$, Proposition~\ref{prop:partition-main}(ii) implies that $V$ has no nonzero rank-one matrix. Theorem~\ref{thm:linear-tensor}(i) then shows that every nonzero integer tensor with residue in $S_m(V)$ has norm greater than $Gr_0$.
\end{proof}

\section{Deterministic reductions and success probability}\label{sec:limits}

\subsection{Deterministic reductions}

Hecht and Safra discuss a related obstruction to deterministic NP-hardness of uSVP in terms of Promise-UP, the promise-problem analogue of $\UP$ \cite[Section~7, arXiv version]{HS26}. For the decision problems of Euclidean uSVP and BDD with rational input data (Definition~\ref{def:problems}), Proposition~\ref{prop:lattice-up} shows that deterministic promise-preserving NP-hardness would imply $\NP=\UP$.

\begin{proposition}[Unambiguous verification]\label{prop:lattice-up}
For every fixed polynomial-time computable ${0<\rho<1/2}$ and every fixed polynomial-time computable $\Gamma>1$, the Euclidean problems $\mathsf D^{(2)}_\rho$ and $\mathsf U^{(2)}_\Gamma$ of Definition~\ref{def:problems} reduce in deterministic polynomial time to $\UCS$, and every promised input maps to a promised input. Therefore, a deterministic polynomial-time many-one reduction from SAT to either lattice problem whose output satisfies the promise on every input would imply $\NP=\UP$.
\end{proposition}
\begin{proof}
For $\mathsf D^{(2)}_\rho$, a verifier guesses integer coefficients $z$ and checks $\norm{Bz-a}_2^2\le r^2$ using rational arithmetic. Two different accepted points both lie within $r$ of the target, so their nonzero difference is a lattice vector of length at most $2r$. The promise gives $2r<\lambda_1$, a contradiction. Since the basis columns are independent, each point has one coefficient vector. Hence there is at most one witness on every promised input.

For $\mathsf U^{(2)}_\Gamma$, the verifier checks that $z\ne0$, $\norm{Bz}_2^2\le r^2$, $\gcd(z_1,\ldots,z_d)=1$, and that the first nonzero coefficient is positive. On a promised YES input, the promise $\lambda_2>\Gamma r$ implies that all vectors of length at most $r$ lie on one real line. Its intersection with the lattice has one primitive generator up to sign. The gcd condition excludes nontrivial multiples. Of the two primitive generators, only one has a positive first nonzero coefficient, so the verifier accepts exactly one witness. On a promised NO input no nonzero vector has length at most $r$, so it accepts none.

To build a circuit with a fixed number of input bits, we show that every coefficient vector that the verifier accepts has polynomial bit length and can be encoded uniquely. Let $\ell$ be the input length, and multiply the basis and target by a common denominator $D$ of polynomial bit length. An accepted lattice vector has norm at most $r$ in uSVP. In BDD an accepted lattice point has norm at most $\norm a_2+r$. Both quantities are at most $2^{\poly(\ell)}$, so the integer coordinates of $DBz$ have polynomial bit length. Choose an invertible square submatrix of $DB$. Its determinant is a nonzero integer, and every cofactor has polynomial bit length. Cramer's rule bounds each coefficient of $z$ by $2^{\poly(\ell)}$.

Encode each coefficient by a fixed polynomial number of bits with one sign bit and no negative zero. Each coefficient vector then has one binary representation. Comparing rational squared norms, computing the gcd, and checking the sign take polynomial time. The verification circuit therefore accepts exactly one binary representation of each coefficient vector that the verifier accepts.

Compose this circuit with a deterministic polynomial-time reduction from SAT whose output satisfies the lattice promise on every input. The resulting SAT verifier is unambiguous on every input, so SAT belongs to $\UP$ and $\NP=\UP$.
\end{proof}

The deterministic converse reductions in the Euclidean norm remain open:
\[
 \begin{gathered}
 \UCS\le_m\mathsf D^{(2)}_\rho,\\
\UCS\le_m\mathsf U^{(2)}_\Gamma.
 \end{gathered}
\]

The uSVP verifier uses the promise $\lambda_2>\Gamma r$. The condition $\lambda_2>\Gamma\lambda_1$ alone does not make the same verifier unambiguous for an arbitrary threshold $r$. For example, $\Z e_1\oplus M\Z e_2$ has $\lambda_2>\Gamma\lambda_1$ when $M>\Gamma$, but the threshold $r=2M$ admits several primitive vectors.

\begin{remark}
The lattice construction in Proposition~\ref{prop:compatibility-lattice} can be strengthened to satisfy $\lambda_2>\Gamma\lambda_1$ on NO outputs as well. For an integer $A>\Gamma$, run the construction with uSVP uniqueness factor $\Gamma A$ and BDD parameter $\rho=1/4$. Write $\widehat r$ for the resulting uSVP threshold and add the summand $A\widehat r\,\Z$. On NO outputs, this summand spans the line containing all shortest nonzero vectors, since every nonzero vector of the original lattice is longer than $\Gamma A\widehat r$. When the construction succeeds on a YES input, the original vector representing the selection has length at most $\widehat r$. Any vector that is not a real multiple of it either has a nonzero coordinate in the added summand and thus has length at least $A\widehat r>\Gamma\widehat r$, or lies in the original lattice and has norm greater than $\Gamma\widehat r$ by the bound for vectors not proportional to the encoding. Thus $\lambda_2>\Gamma\lambda_1$ in both cases, with unchanged success probability.
\end{remark}

\subsection{Isolation probability}\label{sec:isolation-recovery}

When isolation succeeds, a solution of the output instance lets us recover an input accepted by the original circuit. This recovery lets us apply known limitations on the probability that isolation succeeds. For code BDD, the unique error
is $(Ey)^{\otimes t}$. Apply $E_{\rm left}^{\otimes t}$ and take the diagonal in~\eqref{eq:source-readout}. The row $\ell$ added to the parity-check matrix fixes the scalar to one, so this recovers the indicator $y$ of the compatible selection.

For either lattice problem in the construction of Theorem~\ref{thm:formal-lattice}, removing the last coordinate of a solution leaves $y^{\otimes m}$, up to sign for uSVP, by Lemma~\ref{lem:normalized-output}. The diagonal entry at $(i,\ldots,i)$ is nonzero exactly when $y_i=1$. For the construction in Theorem~\ref{pa2:main}, Section~\ref{pa2:sec-exponent-decoding} shows how to recover the selection from a solution. The selected labels, including those in the blocks for input bits that no gate uses, determine the circuit input. In the cases handled by direct evaluation or enumeration, that computation already finds an input accepted by the circuit.

To apply Proposition~\ref{prop:isolation-recovery-limit}, associate an output circuit to each original circuit $F$ and random seed. When the lattice construction copies labels and forms a code by multiplying the columns of a parity-check matrix by random nonzero field elements, the output circuit has one input bit per copied label, representing an indicator $y$. It checks that one label is selected in each block, that $A_\phi y=0$, and that
\[
 H\operatorname{diag}(\beta)y=0.
\]
The $N=bL$ copied coordinates are ordered by block, original label, and copy. The circuit $F$ determines $L$, and the input length of the output circuit gives $N$. We can therefore compute $b=N/L$ and read the original labels from the copied coordinates. For every seed, from any input accepted by the output circuit we can recover an input accepted by the original circuit $F$. When we decide satisfiability by direct evaluation or enumeration, output a circuit accepting no input if $F$ is unsatisfiable, and otherwise a circuit accepting only the first accepted input found by the evaluation or enumeration. Thus the output circuit is unsatisfiable whenever $F$ is unsatisfiable. Whenever exactly one selection of copied labels, one per block, has an indicator satisfying both equations above, the output circuit accepts exactly one input. From that input we can recover an input accepted by $F$.

For the code reduction, use the circuit that checks the original circuit $F$ together with the random constraints of Valiant and Vazirani. Its input is an assignment to $F$. Every accepted input satisfies $F$, and a successful random choice leaves exactly one accepted input. The recovery map is the identity.

For the circuits and deterministic recovery procedures constructed above, Proposition~\ref{prop:isolation-recovery-limit} implies $\coNP\subseteq\mathsf{NP}/\poly$ if the output circuit accepts exactly one input with probability greater than $2/3+1/\poly(s)$ for every satisfiable circuit of description length $s$. The reductions from $\UCS$ use the promise that the original circuit accepts at most one input to achieve constant success probability.

In the reduction from \CircuitSAT{} to code BDD, we first add the random constraints of Valiant and Vazirani to the original circuit's input variables. The lattice reduction in Theorem~\ref{pa2:main} uses the same initial procedure. For every random choice, each input accepted by the resulting circuit is also accepted by the original circuit. By Proposition~\ref{prop:isolation-pruning-limit}, $\NP\subseteq\mathsf{P}/\poly$ if, for every satisfiable circuit, the resulting circuit accepts exactly one input with any fixed positive probability. In the lattice reduction of Theorem~\ref{thm:formal-lattice}, the circuit for copied selections has different variables and uses a separate recovery map, so Proposition~\ref{prop:isolation-recovery-limit} applies there instead.

\section{\texorpdfstring{Polynomial factors for $p>2$}{Polynomial factors for p > 2}}
\label{sec:poly-above-two}

To make the uniqueness factor polynomial while keeping the dimension polynomial, we use polynomial interpolation and weighted moments instead of tensor powers. We build on the CVP construction of OpenAI using interpolation and moments \cite[Chapter~7]{Ope26} and its adaptation to ordinary SVP using weighted moments by Hair and Sahai \cite{HairSahai26}.

A selection determines a polynomial whose value at each block's anchor is the value assigned to its chosen label. The anchors are distinct elements of $\F_P$, where $P$ is prime. We encode the polynomial's values at the other field elements in an array indexed by a label, a point, and a possible value. Linear constraints ensure that the moments we use, as the point varies, are values of polynomials of bounded degree, called moment polynomials. These constraints also enforce the values at the anchors and agreement on shared wires. We apply a Hadamard matrix to each fiber, meaning the coordinates with a label and a point fixed. We construct an integer-valued linear function $\mu$ that equals one on the vector produced by applying the Hadamard map to the encoding of any selection. We prove a lower bound on the norm that is much larger than the norm of an encoding. When the circuit accepts no input, the bound applies to every nonzero lattice vector. When it accepts exactly one input, the bound applies to every nonzero lattice vector with $\mu=0$.

Many nonzero integer entries in a fiber force a large norm after applying the Hadamard map. Thus a sufficiently short lattice vector must come from an array in which every label's fiber has few entries that are nonzero modulo $P$ at each of more than half the points. Reconstruction from weighted moments at these points represents each label's moment polynomials as weighted sums of powers of a small number of algebraic quantities called nodes. The coefficients, called weights, and the nodes are the same for all the moment polynomials of a given label. For each node, collect its weights across labels into a vector. The constraints at the block anchors ensure that this vector has at most one nonzero entry per block. The compatibility equations then make it a scalar multiple of the indicator of a compatible selection. If no input is accepted, there can be no nodes, so the moment polynomials vanish. If exactly one input is accepted, a determinant argument shows that at most one node can occur. Its nonzero weight would then contradict $\mu=0$, so the moments vanish in this case too. If the moments vanish, a nonzero fiber with integer entries has many entries that are nonzero modulo $P$ or an entry of magnitude at least $P$. Both cases yield a lower bound on the norm after applying the Hadamard map. These bounds let Lemma~\ref{lem:normalized-output} construct full-rank uSVP and BDD instances. Section~\ref{pa2:sec-euclidean-obstruction} constructs a nonzero vector much shorter in Euclidean norm than the encoding of a compatible selection in this construction.

For $p=\infty$, use the same threshold problems as in
Definition~\ref{def:problems}, with the maximum norm in place of $\ell_p$.

\begin{theorem}\label{pa2:main}
Fix a rational $p>2$ and a constant
\[
 0<\epsilon<\min\left\{\frac{p-2}{4p},\frac18\right\}.
\]
There are deterministic polynomial-time promise-preserving reductions from
$\UCS$ to $\Up{n^\epsilon}$ and $\Dp{n^{-\epsilon}}$ in $\ell_p$.
The same conclusion holds for $p=\infty$ and every fixed
$0<\epsilon<1/8$. Here $n$ is both lattice rank and ambient dimension.
The bases and targets are integral, and the thresholds are rational.

An unsatisfiable circuit always maps to a promised NO instance. A circuit
that accepts exactly one input maps to a promised YES instance.
Applying the randomized reduction of Valiant and Vazirani before the deterministic construction produces randomized polynomial-time reductions from \CircuitSAT{} with inverse-polynomial success on satisfiable inputs and promised NO outputs on every seed for unsatisfiable inputs. 
\end{theorem}

For example, Theorem~\ref{pa2:main} applies with $p=4$ and $\epsilon=1/16$. Corollary~\ref{cor:formal-search-usvp}(ii) states the corresponding search results. Remark~\ref{pa2:rem-real-p} extends Theorem~\ref{pa2:main} to every fixed real $p>2$.

We use the result of Hair and Sahai on reconstruction from weighted moments~\cite[Lemma~5.1]{HairSahai26} as Lemma~\ref{pa2:lem-reconstruction}. Their Lemmas~6.1--6.2 show that the monic polynomial whose roots are the nodes has coefficients that are polynomials in $X$ and satisfies conditions at the anchors. Lemmas~\ref{pa2:lem-denominators} and~\ref{pa2:lem-anchor-separation} prove the versions needed for our labels: this polynomial has coefficients in $\F_P[X]$ and becomes a power of $Z-\zeta_\ell$ at the anchor of the block containing label $\ell$. The conclusion in Lemma~\ref{pa2:lem-anchor-separation} that representations for distinct labels in one block have no node in common is analogous to Hair and Sahai's conclusion for incompatible Boolean labels~\cite[Corollary~6.3]{HairSahai26}. Theorem~\ref{pa2:thm-rigidity} proves the additional conclusion that, when the circuit accepts exactly one input, the moments agree up to a scalar with those of its encoding. The cited Hair--Sahai lemma and corollary numbers are those in arXiv:2608.14529v3.

\subsection{Interpolation and constraints on moments}
\label{pa2:sec-code}

We define a linear code $\mathcal C$ over $\F_P$ containing the encodings of compatible selections and later form a lattice from integer arrays whose reductions lie in $\mathcal C$ (Section~\ref{pa2:sec-norm}). To bound the norms of lattice vectors, we need constraints on every codeword, including those that do not encode compatible selections. We encode selections by evaluating polynomials and impose linear constraints on moments. These constraints allow us to reconstruct the moments of codewords with few nonzero entries in every label's fiber at many common points, and relate those moments to compatible selections.

Use the blocks of allowed assignments and compatibility equations from Section~\ref{sec:common-source} without copying labels.
Handle gate-free circuits and empty blocks directly. If there are fewer than two blocks, add blocks with one allowed label until their number $g$ is at least two. Write
$\mathcal I=\bigsqcup_b B_b$, and use the matrix $A_\phi$ defined for these blocks, with at most $r_A\le8$ nonzero entries per row. Adding these blocks preserves satisfiability and uniqueness. Fix an integer $a\ge4$ sufficiently large depending only on the fixed $p$ and $\epsilon$. Choose a constant $S_0=S_0(a,p,\epsilon)\ge102$ large enough to ensure the numerical lower bounds in this appendix, including \eqref{pa2:eq-strict-gap} and the hypothesis of Lemma~\ref{pa2:lem-collision}.

Choose $S=\max\{S_0,\text{length of the circuit description}+g+|\mathcal I|+101\}$.
Let
\begin{equation}\label{pa2:eq-parameters}
 \begin{gathered}
 d=g-1,\\
\kappa=S^a,\\
T=4S^{2a+1}.
 \end{gathered}
\end{equation}
Choose a prime $P\in[S^{2a+3},2S^{2a+3})$, distinct anchors
$a_1,\ldots,a_g\in\F_P$, and, within each block, distinct values
$\zeta_\ell\in\F_P$ for its labels. Set $\Xi=\F_P\setminus\{a_1,\ldots,a_g\}$.
For fixed $a$, deterministic prime search takes polynomial time.
The coordinates are $x_{\ell,\xi,w}$, with
$\ell\in\mathcal I$, $\xi\in\Xi$, $w\in\F_P$.
Each pair $(\ell,\xi)$ indexes a fiber of $P$ coordinates, one for each $w\in\F_P$. In the encoding of a selection defined below, $w$ is a possible value of its polynomial at $\xi$.
Write
\begin{equation}\label{pa2:eq-sizes}
 \begin{gathered}
 G_{\rm fib}=|\mathcal I||\Xi|,\\
M=G_{\rm fib}P,\\
R_{\rm sel}=g|\Xi|.
 \end{gathered}
\end{equation}
The following bounds will let us express the ratio between the norm lower bounds and the encoding's norm as a power of the output dimension:
\begin{equation}\label{pa2:eq-size-bounds}
 \begin{gathered}
 \frac{S^{4a+6}}2<M<40S^{4a+7},\\
R_{\rm sel}<SP.
 \end{gathered}
\end{equation}

For $A\subseteq\F_P$, let $\operatorname{RS}(A,e)\subseteq\F_P^A$ be the subspace of evaluation vectors $(f(\xi))_{\xi\in A}$ of polynomials $f\in\F_P[X]$ of degree at most $e$. Define $\mathcal C\subseteq\F_P^M$ by two families of equations. The first reflects the value a selected label prescribes at its block's anchor: if a polynomial $f$ of degree at most $d$ satisfies $f(a_b)=\zeta_\ell$, then $(f(X)-\zeta_\ell)/(X-a_b)$ has degree at most $d-1$. For each $\ell\in B_b$, impose
\begin{equation}\label{pa2:eq-shifted}
 \left(\sum_wx_{\ell,\xi,w}
       \left(\frac{w-\zeta_\ell}{\xi-a_b}\right)^j\right)_{\xi\in\Xi}
 \in\operatorname{RS}(\Xi,(d-1)j),\qquad0\le j\le T.
\end{equation}
The second family applies the compatibility equations to vectors of moments across labels at each point. Once we show that different labels in a block cannot share a node, these equations will imply that, for each node, the vector of its weights across labels is a scalar multiple of the indicator of a compatible selection. For a fiber $(\ell,\xi)$, its $j$th moment is $m_j(\xi)_\ell=\sum_wx_{\ell,\xi,w}w^j$. Impose
\begin{equation}\label{pa2:eq-moment-consistency}
 A_\phi m_j(\xi)=0,
 \qquad\xi\in\Xi,\quad0\le j\le J_0:=r_A\kappa-1.
\end{equation}
All constraints are homogeneous and linear in $x$. For each Reed--Solomon evaluation space, a parity-check matrix gives linear equations for membership and can be computed by linear algebra. Gaussian elimination then computes a basis and a parity-check matrix for $\mathcal C$ in time polynomial in $M$ and $\log P$, which is polynomial in the input size for fixed $a$.

For a selection with indicator $y$, let $f_y$ be the unique polynomial of
degree at most $g-1$ with $f_y(a_b)=\zeta_{\ell_b}$ for its selected label
$\ell_b\in B_b$. Define
$x^y_{\ell,\xi,w}=y_\ell\mathbf1[w=f_y(\xi)]$.
It has $R_{\rm sel}$ entries equal to one and satisfies
\eqref{pa2:eq-shifted}. If $A_\phi y=0$, then $x^y$ also satisfies
\eqref{pa2:eq-moment-consistency} and belongs to $\mathcal C$.

\begin{remark}\label{pa2:rem-tight-degree}
The degree $d=g-1$ is essential for uniqueness. If we allowed degree $g$, the polynomials
\[
 f_y(X)+c\prod_{b=1}^g(X-a_b),\qquad c\in\F_P,
\]
would produce distinct encodings of the same selection, all with the same weight. The degree bound $g-1$ fixes the interpolating polynomial from its $g$ values at the anchors.
\end{remark}

Let $x\in\mathcal C$. Since $dT<|\Xi|$, evaluations on $\Xi$ uniquely determine polynomials of degree at most $dT$. For each $\ell\in B_b$ and $0\le j\le T$, let $\eta_{\ell,j}(X)$ be the unique polynomial of degree at most $(d-1)j$ whose values are the sums in \eqref{pa2:eq-shifted}. The binomial theorem gives the following polynomials, called moment polynomials, whose values at $\xi\in\Xi$ are the moments $m_j(\xi)_\ell$:
\begin{equation}\label{pa2:eq-ordinary}
 \begin{gathered}
 M_{\ell,j}(X)=\sum_{h=0}^j\binom jh\zeta_\ell^{j-h}
 (X-a_b)^h\eta_{\ell,h}(X),\\
\deg M_{\ell,j}\le dj.
 \end{gathered}
\end{equation}
Uniqueness of polynomial interpolation also implies
\begin{equation}\label{pa2:eq-shift-identity}
 (X-a_b)^j\eta_{\ell,j}(X)
 =\sum_{h=0}^j\binom jh(-\zeta_\ell)^{j-h}M_{\ell,h}(X).
\end{equation}
\subsection{Moment reconstruction and polynomial coefficients}
\label{pa2:sec-algebra}

To relate the nodes from Lemma~\ref{pa2:lem-reconstruction} to compatible selections, we first show that different labels in a block cannot share a node. For a label $\ell\in B_b$, consider the monic polynomial whose roots are the nodes representing its moment polynomials $M_{\ell,j}$. We show that its coefficients lie in $\F_P[X]$ and that it becomes a power of $Z-\zeta_\ell$ at $X=a_b$. If two labels in a block shared a node, their polynomials would have a nonconstant monic common factor. At the anchor, this factor would retain its degree and divide coprime powers of distinct linear polynomials, a contradiction. The following determinant identity removes denominators and, with $e=d-1$, recovers the conclusion of Hair and Sahai~\cite[Lemma~6.1]{HairSahai26} under the same condition on the number of moments.
\begin{lemma}[Polynomial coefficients]\label{pa2:lem-denominators}
Let $F$ be a field. Suppose $\nu\ge1$ distinct nodes $\alpha_s$ and nonzero weights $\omega_s$
lie in an algebraic closure of $F(X)$ and define the moments
$m_j=\sum_s\omega_s\alpha_s^j\in F[X]$ through order $T\ge2\nu-1$.
For $0\le j\le T-2\nu+2$, let $K_j=(m_{a+b+j})_{0\le a,b<\nu}$. Let $D=\det K_0$ and
$G(Z)=\prod_s(Z-\alpha_s)$. Then $D\ne0$ and $G\in F(X)[Z]$.
If $b(X)$ is the monic least common denominator of the coefficients
of $G$ and $r=T-2\nu+2$, then $b^r\mid D$.
In particular, if $e\ge0$, $\deg m_j\le ej$, and
$T\ge e\nu(\nu-1)+2\nu-1$, then $G\in F[X,Z]$.
\end{lemma}
\begin{proof}
For the Vandermonde matrix $V=(\alpha_s^a)_{a,s}$,
$K_j=V\diag(\omega_s\alpha_s^j)V^{\mathsf T}$, so $D\ne0$ and
$DG(Z)=\det(ZK_0-K_1)$. All $K_j$ for $0\le j\le r$ use moments
through $T$. By the binomial theorem,
\begin{equation}\label{pa2:eq-denominator-determinant}
 \det\left(\sum_{j=0}^r(-1)^j\binom rj Z^{r-j}K_j\right)=DG(Z)^r.
\end{equation}
By minimality of $b$, the polynomial $bG$ is primitive over $F[X]$: its coefficients as a polynomial in $Z$ have no common nonconstant factor in $F[X]$.
By Gauss's lemma, powers of primitive polynomials remain primitive. The coefficients of $(bG)^r$ therefore have greatest common divisor one, so the coefficients of $D(bG)^r$ have greatest common divisor $D$ up to a nonzero scalar in $F$. Multiplying~\eqref{pa2:eq-denominator-determinant} by $b^r$ shows that every coefficient of $D(bG)^r$ is divisible by $b^r$. Thus $b^r$ divides the greatest common divisor of these coefficients, giving $b^r\mid D$.
If $\deg m_j\le ej$, then $\deg D\le e\nu(\nu-1)$.
The bound on $T$ implies $r>\deg D$, forcing $b=1$.
\end{proof}

\begin{lemma}[Polynomial coefficients and evaluation at anchors]\label{pa2:lem-anchor-separation}
For a label $\ell\in B_b$, consider the representation of $M_{\ell,j}$ from Lemma~\ref{pa2:lem-reconstruction}, and suppose $M_{\ell,j}$ and $\eta_{\ell,j}$ arise from the same fibers of one array $x$ at the points in $\Xi'$, as in \eqref{pa2:eq-shift-identity}. Suppose also that
\[
 \begin{gathered}
 dT<|\Xi'|,\\
T\ge d\kappa^2.
 \end{gathered}
\]
Let $G_\ell=\prod_s(Z-\alpha_s)$ for the nodes in that representation. Then $G_\ell\in\F_P[X,Z]$ and $G_\ell(a_b,Z)=(Z-\zeta_\ell)^{\deg_ZG_\ell}$. If two labels in the same block both satisfy these hypotheses, their polynomials have disjoint root sets.
\end{lemma}
\begin{proof}
If there are no nodes, then $G_\ell=1$. For $\nu\ge1$, \eqref{pa2:eq-shift-identity} implies the polynomial identities
\[
 \begin{gathered}
 \eta_{\ell,j}=\sum_{s=1}^\nu\omega_s
 \left(\frac{\alpha_s-\zeta_\ell}{X-a_b}\right)^j,\\
\deg\eta_{\ell,j}\le(d-1)j.
 \end{gathered}
\]
The values $(\alpha_s-\zeta_\ell)/(X-a_b)$ remain distinct. Let
\[
 \widetilde G_\ell(Z)=\prod_{s=1}^\nu\left(Z-\frac{\alpha_s-\zeta_\ell}{X-a_b}\right).
\]
Since $(d-1)\nu(\nu-1)+2\nu-1\le d\nu^2\le d\kappa^2$, including when $d=1$, Lemma~\ref{pa2:lem-denominators} implies that $\widetilde G_\ell\in\F_P[X,Z]$. Hence
\[
 G_\ell(Z)=(X-a_b)^\nu
 \widetilde G_\ell\left(\frac{Z-\zeta_\ell}{X-a_b}\right)
 \equiv(Z-\zeta_\ell)^\nu\pmod{X-a_b}.
\]
If two labels shared a root, their monic nonconstant greatest common divisor over $\F_P(X)$ would lie in $\F_P[X,Z]$ by Gauss's lemma, since both polynomials are monic in $Z$ and have coefficients in $\F_P[X]$.
Evaluating at $a_b$ preserves its positive degree because it is monic, and yields a divisor of two coprime powers of distinct linear polynomials, a contradiction.
\end{proof}

\subsection{Reconstruction from fibers with few nonzero entries}\label{pa2:sec-rigidity}

To bound the norms of lattice vectors, we first determine which moment polynomials are possible for codewords with few nonzero entries per fiber. When every label's fiber has few nonzero entries at many common points, Lemma~\ref{pa2:lem-reconstruction} represents its moment polynomials by nodes and weights. For each node $\alpha$ in the representations, collect its weights across labels into a vector $v_\alpha$, using zero for a label whose representation does not contain $\alpha$. Lemma~\ref{pa2:lem-anchor-separation} shows that at most one coordinate of $v_\alpha$ in each block is nonzero. The equations on the moments and the invertibility of a Vandermonde matrix imply $A_\phi v_\alpha=0$ separately for every node. These equations force $v_\alpha$ to be a scalar multiple of the indicator of a compatible selection. Thus all moments vanish if the circuit is unsatisfiable. If the circuit accepts exactly one input, each node has the same weight at all selected labels and is absent from the other labels. A determinant argument at the anchors rules out two or more nodes. The moment polynomials through order $T$ then equal those of the encoding times a scalar.

For $x\in\mathcal C$, let $\Xi'\subseteq\Xi$ contain more than half the points, and suppose every label's fiber of $x$ at every point in $\Xi'$ has at most $\kappa$ nonzero entries. The parameters satisfy
\[
 \begin{gathered}
 d\bigl(T+2\kappa(\kappa-1)\bigr)<P/16<|\Xi|/2,\\
dT<|\Xi'|,\\
T\ge d\kappa^2,\\
T+1\ge r_A\kappa.
 \end{gathered}
\]
Indeed, using $d\le S$, $\kappa=S^a$, $T=4S^{2a+1}$, $P\ge S^{2a+3}$, and $S\ge102$, we have $d(T+2\kappa(\kappa-1))/P\le4/S+2/S^2<1/16$. Thus Lemmas~\ref{pa2:lem-reconstruction} and~\ref{pa2:lem-anchor-separation} apply, including when $d=1$.
The sum of the entries in the first block of $x^y$ at any point is one. On a multiple $cx^y$, this sum equals $c$ in $\F_P$. For arrays over $\F_P$, fix $\xi_0\in\Xi$ and define a linear function with values in $\F_P$ by
$\bar\mu(x)=\sum_{\ell\in B_1,w}x_{\ell,\xi_0,w}$.

\begin{theorem}[Moments when fibers have few nonzero entries]\label{pa2:thm-rigidity}
Let $x\in\mathcal C$, and suppose there is a set $\Xi'\subseteq\Xi$ containing more than half the points such that, for every $\xi\in\Xi'$ and every label $\ell$, the fiber $(x_{\ell,\xi,w})_{w\in\F_P}$ has at most $\kappa$ nonzero entries.
If the circuit is unsatisfiable, $M_{\ell,j}=0$ for every label $\ell$ and $0\le j\le T$. If the original circuit accepts exactly one input and $y$ is the indicator of its compatible selection, then the moment polynomials of $x$ equal those of $cx^y$ for $c=\bar\mu(x)\in\F_P$. In this case, for every $\xi\in\Xi'$, label $\ell$, and $w\in\F_P$, we have $x_{\ell,\xi,w}=c x^y_{\ell,\xi,w}$. The moments through $T$ of $x-cx^y$ vanish on every fiber at every point in $\Xi$.
\end{theorem}
\begin{proof}
Apply Lemma~\ref{pa2:lem-reconstruction} to each label's moment polynomials on $\Xi'$, choosing all representations in one algebraic closure of $\F_P(X)$. Let $\omega_{\ell,\alpha}$ be the weight for label $\ell$ at node $\alpha$, or zero if absent.
Fix a row $a$ of $A_\phi$. The labels with $a_\ell\ne0$ have at most $r_A\kappa$ distinct nodes between them, since $a$ has at most $r_A$ nonzero entries and each label has at most $\kappa$ nodes. For $0\le j\le J_0$, the polynomial $\sum_\ell a_\ell M_{\ell,j}$ has degree at most $dJ_0\le dT<|\Xi'|\le|\Xi|$. It vanishes at every point of $\Xi$ by \eqref{pa2:eq-moment-consistency}, so it is identically zero. In the power sum $\sum_\ell a_\ell M_{\ell,j}$, node $\alpha$ has weight $\sum_\ell a_\ell\omega_{\ell,\alpha}$. The first $J_0+1=r_A\kappa$ power sums are zero. Since the Vandermonde matrix on the distinct nodes is invertible, each of these weights is zero.

Let $K$ be an extension field of $\F_P(X)$ containing all the nodes and weights in these representations. Then $v_\alpha=(\omega_{\ell,\alpha})_\ell$ belongs to $\mathcal U_\phi(K)$.
Lemma~\ref{pa2:lem-anchor-separation} implies that at most one entry per block is nonzero.
Lemma~\ref{lem:onehot-source} therefore yields
$v_\alpha=c_\alpha y_\alpha$, where $c_\alpha\ne0$ and $y_\alpha$ is the indicator of a compatible selection.
If the circuit is unsatisfiable, there is no compatible selection, so no nodes occur.

When the original circuit accepts exactly one input, all $y_\alpha$ equal the indicator $y$ of its unique compatible selection. Since $v_\alpha=c_\alpha y$ with $c_\alpha\ne0$, every selected label has the same nodes $\alpha_1,\ldots,\alpha_\nu$ with the same nonzero weights $\omega_1,\ldots,\omega_\nu$. The other labels have no nodes. For $0\le j\le T$, the common moment polynomial of the selected labels is $M_j=\sum_{s=1}^\nu\omega_s\alpha_s^j$, while unselected labels have zero moments. Let $c=M_0=\bar\mu(x)$. We show that $\nu\le1$ by comparing the degree of a nonzero determinant formed from these moments with the zeros forced at the anchors. If $\nu\ge1$, the matrix $(M_{i+j})_{0\le i,j<\nu}$ (here $2\nu-2\le T$) factors as $V\operatorname{diag}(\omega_s)V^{\mathsf T}$, where $V$ is the Vandermonde matrix of the distinct nodes and the weights $\omega_s$ are nonzero. Hence
\[
 \begin{gathered}
 D=\det(M_{i+j})_{0\le i,j<\nu}\ne0,\\
\deg D\le(g-1)\nu(\nu-1).
 \end{gathered}
\]
For the selected label at anchor $a_b$, let
$Q_b=(\eta_{\ell_b,i+j})_{0\le i,j<\nu}$.
Let $L_b$ be the lower triangular matrix with entries
$(L_b)_{i,k}=\binom ik(-\zeta_{\ell_b})^{i-k}$ for $k\le i$.
Its diagonal entries are all one. By \eqref{pa2:eq-shift-identity},
\[
 L_b(M_{i+j})L_b^{\mathsf T}
 =\diag((X-a_b)^i)\,Q_b\,\diag((X-a_b)^i).
\]
The $(i,j)$ entry of $L_b(M_{i+j})L_b^{\mathsf T}$ is
\[
 \sum_s\omega_s(\alpha_s-\zeta_{\ell_b})^{i+j}
 =(X-a_b)^{i+j}\eta_{\ell_b,i+j}.
\]
Taking determinants shows that $(X-a_b)^{\nu(\nu-1)}\mid D$ for every $b$. The factors $X-a_b$ are pairwise coprime, so their product with these multiplicities divides $D$. If $\nu>1$, the product has degree $g\nu(\nu-1)$, exceeding the bound $(g-1)\nu(\nu-1)$ on $\deg D$, a contradiction. Thus $\nu\le1$. If $\nu=1$, the unique node has constant weight
$c=M_0=\bar\mu(x)\ne0$ and equals $M_1/c$, a polynomial of degree at most $g-1$. By Lemma~\ref{pa2:lem-anchor-separation}, it takes value $\zeta_{\ell_b}$
at every $a_b$. The defining polynomial $f_y$ has the same values at these $g$ points and degree at most $g-1$, so the two polynomials are equal. If $c=0$, there are no nodes.

In either case, for $0\le j\le T$, every selected label has moment $M_j=cf_y^j$, while every unselected label has zero moments. These are the moments of $cx^y$ at every label and point. For each label and point $\xi\in\Xi'$,
the fiber of $x-cx^y$ has at most $\kappa+1$ nonzero
entries. Its first $\kappa+1$ moments vanish, and the Vandermonde matrix
on its nonzero positions is invertible, so the fiber is zero.
\end{proof}

A nonzero fiber whose moments through $T$ vanish has at least $T+2$ nonzero coordinates. Indeed, the first $T+1$ moments determine the weights on any set of at most $T+1$ distinct coordinates because the Vandermonde matrix on those coordinates is invertible. This applies to a fiber of $x$ when the circuit is unsatisfiable and to a fiber of $x-cx^y$ when the circuit accepts exactly one input. The next subsection uses this lower bound on the number of nonzero coordinates to bound the norms of vectors in the constructed lattice.

\subsection{Constructing the lattice and bounding norms}\label{pa2:sec-norm}

To apply Theorem~\ref{pa2:thm-rigidity}, we show that a sufficiently short lattice vector comes from an integer array with the following property. At more than half the points, every label's fiber has at most $\kappa$ entries that are nonzero modulo $P$. Multiplying each fiber of the array by a Hadamard matrix bounds the norm of its image from below in terms of the number of nonzero integer entries in the fiber. Images of fibers at distinct points use disjoint coordinates, so the $p$th powers of their norms add for finite $p$. For $p=\infty$, one fiber with many nonzero entries already forces a large norm.

Take $\Lambda=\{z\in\Z^M:z\bmod P\in\mathcal C\}$.
Choose $W_H$ as the least power of two at least $P$, take $P$ rows $H_0$
of a Sylvester Hadamard matrix of order $W_H$, and apply $H_0$ to every
fiber indexed by $(\ell,\xi)$, as in Hair and Sahai~\cite[Section~8.2]{HairSahai26}. This defines an injective integer map $E:\Z^M\to\Z^{G_{\rm fib}W_H}$.
For $2\le p\le\infty$, with $1/\infty=0$, we have
\begin{equation}\label{pa2:eq-hadamard-bounds}
 \norm{vH_0}_p\ge W_H^{1/p}\norm v_2.
\end{equation}
For $p>2$ this inequality appears in \cite[Lemma~8.2]{HairSahai26}. It follows, including at $p=2$, from $\norm{vH_0}_2=\sqrt{W_H}\norm v_2$ and
$\norm w_p\ge W_H^{1/p-1/2}\norm w_2$. For a selection with indicator $y$, the vector $y_H=E(x^y)$ has $\ell_p$ norm
\begin{equation}\label{pa2:eq-radius}
 r_0=(R_{\rm sel}W_H)^{1/p},
\end{equation}
with $r_0=1$ when $p=\infty$. The first Hadamard column is constant one, so
\[
 \mu(E(z))=\sum_{\ell\in B_1,w}z_{\ell,\xi_0,w}
\]
defines a linear function $\mu$ with values in $\mathbb Z$ using the coordinates corresponding to the all-ones column. The sum over the first block is one for every selection, so $\mu(y_H)=1$. We also have $\mu(E(z))\bmod P=\bar\mu(z\bmod P)$.

If $z\in\Lambda$ and $\norm{E(z)}_p<A_1r_0$, with $A_1$ defined in \eqref{pa2:eq-gap}, then, at more than half the points, every label's fiber of $z$ has at most $\kappa$ entries nonzero modulo $P$. If a nonzero fiber $v$ has integer entries and its first $T+1$ moments vanish modulo $P$, then $\norm{vH_0}_p\ge B_0$, where $B_0$ is defined below. We compare $A_1$ with $B_0/r_0$ by letting
\begin{equation}\label{pa2:eq-gap}
 \begin{gathered}
 A_1=(2g)^{-1/p}\sqrt{\kappa+1},\\
B_0=W_H^{1/p}\min\{P,\sqrt{T+2}\},\\
\mathcal G=\min\{A_1,B_0/r_0\}.
 \end{gathered}
\end{equation}

\begin{theorem}\label{pa2:thm-integer-separation}
If the circuit is unsatisfiable, every nonzero vector of $E(\Lambda)$ has norm at least $\mathcal G r_0$. If the circuit accepts exactly one input, the same bound holds for every nonzero $v\in E(\Lambda)$ with $\mu(v)=0$.
\end{theorem}
\begin{proof}
Suppose $0\ne E(z)$ has norm less than $\mathcal G r_0$, and let
$u=z\bmod P$. Call a fiber of $z$ bad if its residue modulo $P$ has more than $\kappa$ nonzero entries. Call a point bad if at least one fiber at that point is bad. There are at most as many bad points as bad fibers. For finite $p$, each bad fiber contributes at least
$W_H(\kappa+1)^{p/2}$ to $\norm{E(z)}_p^p$, by
\eqref{pa2:eq-hadamard-bounds}. The definition of $A_1$ implies
\[
 (A_1r_0)^p=\frac{|\Xi|}{2}W_H(\kappa+1)^{p/2}.
\]
Since $\mathcal G\le A_1$, there are fewer than $|\Xi|/2$ bad fibers, hence fewer than that many bad points. For $p=\infty$, the bound
$\norm{E(z)}_\infty<A_1=\sqrt{\kappa+1}$ leaves no bad fiber.
Thus more than half the points have at most $\kappa$ nonzero entries in every fiber of $u$.

If the circuit is unsatisfiable, Theorem~\ref{pa2:thm-rigidity} implies $\sum_w u_{\ell,\xi,w}w^j=0$ for every label $\ell$, point $\xi$, and $0\le j\le T$. If it accepts exactly one input and $\mu(E(z))=0$, the scalar in that theorem is $c=\bar\mu(u)=\mu(E(z))\bmod P=0$, so the same sums vanish.
Since $E(z)$ is nonzero, some fiber of $z$ is nonzero. Choose one. If it is zero modulo $P$, all its coordinates are multiples of $P$, so one has magnitude at least $P$. Otherwise, at least $T+2$ entries are nonzero modulo $P$, since the first
$T+1$ moments determine the entries on any set of at most $T+1$ nonzero positions. By \eqref{pa2:eq-hadamard-bounds},
\begin{equation}\label{pa2:eq-zero-moment-floor}
 \norm{E(z)}_p\ge W_H^{1/p}\min\{P,\sqrt{T+2}\}
 =B_0\ge\mathcal G r_0,
\end{equation}
a contradiction.
\end{proof}

\subsection{Polynomial factors for uSVP and BDD}\label{pa2:sec-exponent-decoding}

We compare $\mathcal G$ with the rank of the lattice we construct to verify the exponent $\epsilon$ in Theorem~\ref{pa2:main}. We then apply Lemma~\ref{lem:normalized-output} to construct the uSVP and BDD instances.

Since $g\le S$ and $\kappa=S^a$,
\[
 A_1\ge2^{-1/p}S^{a/2-1/p}
 \ge2^{-1/p}S^{(a-2)/2}.
\]
Also $P>\sqrt{T+2}$ and $R_{\rm sel}<SP$, giving
\[
 \frac{B_0}{r_0}\ge
 2^{1-1/p}S^{a+1/2-(2a+4)/p}.
\]
Using $M<40S^{4a+7}$, we bound $A_1$ and $B_0/r_0$ below, up to constant factors, by powers of $M$ with respective exponents
\begin{equation}\label{pa2:eq-exponents}
 \begin{gathered}
 \frac{a-2}{2(4a+7)},\\
\frac{a+1/2-(2a+4)/p}{4a+7}.
 \end{gathered}
\end{equation}
As $a\to\infty$, these exponents tend to $1/8$ and $(p-2)/(4p)$, respectively.
By the choice of $a$, both exceed $\epsilon$ in Theorem~\ref{pa2:main}.
Each output lattice has rank and ambient dimension $n=G_{\rm fib}W_H+1\le2M$.
The choice of $S_0$ in Section~\ref{pa2:sec-code} ensures
\begin{equation}\label{pa2:eq-strict-gap}
 \begin{gathered}
 n^\epsilon>2,\\
\mathcal G>64n^\epsilon.
 \end{gathered}
\end{equation}

To apply Lemma~\ref{lem:normalized-output} with rational inputs, compute a rational $\Delta$ strictly below $\mathcal G r_0$ and a rational upper bound $u$ on $r_0$, the norm of the encoding of a compatible selection. Choose $\Delta$ between $\mathcal G r_0/2$ and $3\mathcal G r_0/4$ and $u$ with $r_0\le u\le5r_0/4$. By \eqref{pa2:eq-strict-gap}, these choices satisfy $\Delta>16n^\epsilon u$ and $n^{-\epsilon}<1/2$, as required by the lemma. These computations take polynomial time for fixed rational $p$, and when $p=\infty$. Lemma~\ref{lem:normalized-output}, applied to $E(\Lambda)$, whose rank may be smaller than its ambient dimension, constructs full-rank instances of uSVP at $\Gamma=n^\epsilon$ and BDD at $\rho=n^{-\epsilon}$, both with integer bases and threshold $\lceil4u\rceil$. The basis of $\Lambda$, the matrices defining the Hadamard map and $\mu$, and the basis after completion all have polynomial bit length.

\paragraph{Isolation and recovery.}
Add random constraints to the original circuit as in Valiant and Vazirani~\cite{VV86}, then apply the deterministic reduction from $\UCS$. Unsatisfiability is preserved on every seed, and when the original circuit is satisfiable, the resulting circuit accepts exactly one input with inverse-polynomial probability.  For Corollary~\ref{cor:formal-search-usvp}(ii), when the random constraints leave exactly one input accepted by the circuit, removing the last coordinate from a solution to either search problem leaves $E(x^y)$, up to sign for uSVP, by Lemma~\ref{lem:normalized-output}. At $\xi_0$, each label $\ell$ has one coordinate corresponding to the all-ones column of $H_0$. It is $y_\ell$ for BDD and $\pm y_\ell$, with one common sign, for uSVP. The nonzero coordinates among these identify the selected labels, and Lemma~\ref{lem:onehot-source} recovers the circuit input.

\begin{remark}[Every fixed real $p>2$]\label{pa2:rem-real-p}
The construction also gives hardness for uSVP with uniqueness factor $n^{\epsilon'}$ and for BDD with radius $n^{-\epsilon'}\lambda_1$ for every fixed real $p>2$ and a suitable constant $\epsilon'>0$. Choose a rational $q$ with $2<q<p$ close enough to $p$ that $\delta=1/q-1/p$ is smaller than some rational exponent $\epsilon_q$ allowed by Theorem~\ref{pa2:main} for $q$. Choose a rational $\epsilon'$ with $0<\epsilon'<\epsilon_q-\delta$. These are fixed constants, and every $x\in\mathbb R^n$ satisfies $\norm{x}_p\le\norm{x}_q\le n^\delta\norm{x}_p$.

Apply the construction for $q$ with exponent $\epsilon_q$. Write $r_q$ for the parameter $r_0=(R_{\rm sel}W_H)^{1/q}$ in this construction. This is the $q$-norm of every encoding of a compatible selection when one exists. The construction in Theorem~\ref{pa2:main} provides a threshold $d_q\le6r_q$ and a number $\Delta_q>16n^{\epsilon_q}r_q$ with the following norm and distance bounds. In the uSVP YES case, the vector produced from the encoding of the compatible selection, with its appended coordinate, has $q$-norm below $d_q$, while every lattice vector that is not a real multiple of it has $q$-norm above $\Delta_q/2$. In the NO case, every nonzero lattice vector has $q$-norm above $\Delta_q$. For BDD, the shortest nonzero lattice vector has $q$-norm above $\Delta_q$ in both cases. The YES target is within $r_q$ of the lattice point corresponding to the encoding, and in the NO case its $q$-norm distance from the lattice exceeds $\Delta_q$. Dividing these lower bounds by $n^\delta$ establishes the conditions in the $\ell_p$ norm for uSVP with uniqueness factor $n^{\epsilon'}$ and for BDD with decoding radius $\lambda_1/n^{\epsilon'}$, for sufficiently large $n$. For uSVP, $n^\delta d_q<\Delta_q/2$ ensures that every shortest vector is proportional to the vector constructed from the encoding, so the recovery procedure still applies. For BDD, $n^\delta d_q+r_q<\Delta_q$ excludes any other lattice point within $\ell_p$ distance $d_q$ of the target. Since the point corresponding to the encoding is within distance $r_q\le d_q$, it remains the closest point and the recovery procedure applies. Smaller instances can be handled directly, and the reduction uses only the fixed rational constants.
\end{remark}

\subsection{A Euclidean obstruction}
\label{pa2:sec-euclidean-obstruction}

To extend this construction to the Euclidean norm, we would need every nonzero lattice vector with $\mu=0$ to be much longer than an encoding. We show that this requirement fails for the lattice constructed here. For every circuit to which the construction applies, including an unsatisfiable one, we instead construct a nonzero vector with $\mu=0$ whose Euclidean norm is much smaller than $\sqrt{R_{\rm sel}W_H}$, the norm an encoding would have. We form an integer array with nonzero entries at only one point, choosing them so that all moments through order $T$ vanish modulo $P$, then apply the Hadamard map $E$.

\begin{lemma}[A short vector with vanishing moments]\label{pa2:lem-collision}
Let $P$ be prime and $T\ge0$, and let $b=\lceil(T+1)\log_2P\rceil$.
If $P\ge2(b+1)$, there is a nonzero $c\in\{-1,0,1\}^P$ with
\[
 \begin{gathered}
 \sum_w c_w=0\quad\text{over }\Z,\\
\sum_w c_w w^j=0\quad\text{over }\F_P\quad(0\le j\le T),\\
\norm c_2^2\le2(b+1).
 \end{gathered}
\]
\end{lemma}
\begin{proof}
Choose $b+1$ disjoint pairs of field elements. Each selection of one element from every pair has exactly $b+1$ elements and determines $T+1$ power sums in $\F_P^{T+1}$. Since $b\ge(T+1)\log_2 P$, there are $2^{b+1}>P^{T+1}$ selections. Thus two distinct selections have the same power sums of orders $0$ through $T$ in $\F_P$. Let $c$ be the difference of their indicator vectors. Then $c\ne0$, its coordinate sum is zero over $\Z$ because the selections have equal size, and its support has size at most $2(b+1)$.
\end{proof}

\begin{theorem}\label{pa2:thm-euclidean-counterexample}
For every circuit to which the construction in Section~\ref{pa2:sec-code} applies, satisfiable or not, the lattice $E(\Lambda)$
contains a nonzero integer vector $v$ with
\[
 \frac{\norm v_2}{\sqrt{R_{\rm sel}W_H}}
 \le\sqrt{\frac{2(\lceil(T+1)\log_2P\rceil+1)}{|\Xi|}}
 =O_a\left(\frac{\sqrt{\log S}}S\right).
\]
It has $\mu(v)=0$, so both constructed lattices contain $(v,0)$ with the same norm. When the original circuit accepts exactly one input, the vector $v$ is linearly independent of the vector $E(x^y)$ representing the selection.
\end{theorem}
\begin{proof}
The parameters imply $P/(T\log P)\longrightarrow\infty$, so
the choice of $S_0$ in Section~\ref{pa2:sec-code} ensures the hypothesis of Lemma~\ref{pa2:lem-collision}, since $a$ is fixed. Let $b=\lceil(T+1)\log_2P\rceil$, and let $c$ be the nonzero vector from Lemma~\ref{pa2:lem-collision}. Choose a block $B_i$, a label $\ell\in B_i$, and a point $\xi_1\in\Xi$. Define the integer array $z$ by $z_{\ell,\xi_1,w}=c_w$ for every $w\in\F_P$, with all other entries zero. For $0\le j\le T$, $\sum_w c_w w^j=0$ in $\F_P$. By the binomial theorem, $\sum_w c_w((w-\zeta_\ell)/(\xi_1-a_i))^j=0$ in $\F_P$ as well. The zero moment polynomials satisfy both families of code constraints, so $z\bmod P\in\mathcal C$ and hence $z\in\Lambda$.
By orthogonality,
\[
 \norm{E(z)}_2^2=W_H\norm c_2^2\le2W_H(b+1).
\]
Divide by $R_{\rm sel}W_H=g|\Xi|W_H$ and use $1\le g\le S$, $|\Xi|=P-g$, $T=4S^{2a+1}$, and $S^{2a+3}\le P<2S^{2a+3}$. The entries in every fiber of $z$ sum to zero over the integers, so
$\mu(E(z))=0$. The vector $E(x^y)$ representing a compatible selection has $\mu(E(x^y))=1$, proving independence.
Both constructed lattices contain $(E(z),0)$.
\end{proof}

For this construction, a nonzero vector in $E(\Lambda)\cap\ker\mu$ is much shorter in Euclidean norm than the encoding of a compatible selection. Thus the hypothesis of Lemma~\ref{lem:normalized-output} fails when the lemma's parameter $R$ is set to the norm of that encoding. For a concrete example, consider the integer vector indexed by $\F_{101}$ with entries $+1$ at $\{1,2,5\}$, entries $-1$ at $\{0,14,95\}$, and zero elsewhere. Its coordinate sum is zero over the integers, its first and second moments are zero modulo $101$, and its squared norm is six. The asymptotic bound follows from Lemma~\ref{pa2:lem-collision}.

\section{Search versions and additional consequences}\label{sec:consequences-main}

\subsection{Search versions}\label{sec:search-convention}

We prove hardness of the search problems by recovering an input that the circuit accepts from a solution to a constructed instance. The reduction checks the recovered input on the original circuit, preventing false positives regardless of what the oracle returns. We use the following oracle convention. On a uSVP input with $\lambda_2>\Gamma\lambda_1$, the oracle must return a shortest nonzero vector. On a lattice BDD input with $\dist_p(a,\Lambda)<\rho\lambda_1$, it must return a closest lattice point. On a code BDD input for which an error $e$ with $He=\sigma$ and weight at most $w$ exists and $w<\rho d(\ker H)$, it must return such an error. On any other query it may return an arbitrary string. We specify how solutions are represented. For the lattice problems, represent a solution by its integer coefficients in the input basis. Each coefficient is encoded by one sign bit followed by the shortest binary representation of its magnitude. Separate the coefficients by commas. For code BDD, encode each entry of the error vector over $\F_q$ by a fixed binary string of length $\lceil\log_2q\rceil$.

To process an arbitrary string from the oracle, we bound the bit length of a valid solution by a polynomial in the bit length of the query. For the lattice problems, clearing denominators and applying Cramer's rule, as in Proposition~\ref{prop:lattice-up}, bounds the bit length of the coefficients in terms of the number of bits describing the instance sent to the oracle and the norm of the lattice vector they represent. For the search version of uSVP, a shortest vector is no longer than a basis vector. For an input to the search version of lattice BDD with $\dist_p(a,\Lambda)<\rho\lambda_1$, the closest lattice point has norm at most $\norm{a}_p+\rho\lambda_1$, and $\lambda_1$ is at most the norm of a basis vector. Each coordinate has magnitude at most the vector's norm, so the argument applies in every fixed $\ell_p$ norm, including $p=\infty$. An error vector for code BDD has the given block length over fixed $\F_q$, with each entry encoded in a constant number of bits. The reductions read at most one bit beyond the polynomial bound and reject oversized or malformed strings before arithmetic, keeping their computation polynomial for any string returned by the oracle.

\begin{corollary}[Search versions]\label{cor:formal-search-usvp}
Under the oracle convention in this subsection, each of the following search problems is NP-hard under randomized polynomial-time oracle reductions that make at most one oracle call, have no false positives, and succeed with inverse-polynomial probability when the input circuit is satisfiable.
\begin{enumerate}[label=(\roman*)]
\item The search versions of $\Gamma$-uSVP and BDD at radius $\rho\lambda_1$ in $\ell_p$, for every fixed polynomial-time computable $1\le p<\infty$, $\Gamma>1$, and ${0<\rho<1/2}$.
\item The search versions of $n^\epsilon$-uSVP and BDD at radius $n^{-\epsilon}\lambda_1$ in $\ell_p$, for the parameters $p$ and $\epsilon$ of Theorem~\ref{pa2:main}, including $p=\infty$, where $n$ is the rank of the output lattice.
\item The search version of code $\BDDc\rho$ over every fixed finite field $\F_q$, for every fixed polynomial-time computable $\rho\in(0,1/2]$.
\end{enumerate}
\end{corollary}
Independent repetition raises the success probability to at least $2/3$ using polynomially many oracle calls.

\begin{proof}
Let $F$ be the input circuit. For (i), use the construction of Theorem~\ref{thm:formal-lattice} directly. For (ii) and (iii), apply the Valiant--Vazirani reduction, producing a circuit $F'$ on the same input variables, then apply the forward reduction from $\UCS$ in Theorem~\ref{pa2:main} or Theorem~\ref{thm:formal-code}, respectively. In cases handled by direct evaluation or enumeration, return the resulting decision without querying an oracle. Otherwise, keep the basis matrices, the left inverse $E_{\rm left}$ where applicable, and the coordinate indices needed to identify labels and recover a circuit input.

Query the corresponding search oracle once and apply the length and format checks in this subsection. For the lattice problems, form $Bz$ from the returned coefficients $z$ and the basis $B$ of the instance sent to the oracle, then discard the last coordinate. In (i), read the remaining coordinates as a tensor and take its diagonal as in Section~\ref{sec:isolation-recovery}. When the construction isolates a selection with indicator vector $y$ and the oracle returns a solution, this tensor is $y^{\otimes m}$ for BDD and $\pm y^{\otimes m}$ for uSVP. Its nonzero diagonal entries therefore identify the selected labels. In (ii), for each label take the coordinate at $\xi_0$ corresponding to the all-ones column of $H_0$, as in Section~\ref{pa2:sec-exponent-decoding}. In (iii), apply $E_{\rm left}^{\otimes t}$ to the returned vector over $\F_q$ and take the diagonal, as in Section~\ref{sec:code-reduction}. In each case, interpret the nonzero coordinates as selected labels. Reject unless the selected labels contain exactly one label per block, assign consistent values to shared wires, and together specify a Boolean input of the expected length. Evaluate $F$ on the resulting candidate input and accept if and only if $F$ accepts.

If $F$ is satisfiable, then with inverse-polynomial probability the reduction accepts by direct evaluation or enumeration, or succeeds in the isolation step and produces a promised YES instance. For a uSVP instance with threshold $r$, $\lambda_1\le r$ and $\lambda_2>\Gamma r$ imply $\lambda_2>\Gamma\lambda_1$. For lattice BDD, $\dist_p(a,\Lambda)\le r<\rho\lambda_1$. For code BDD, an error of weight at most $w$ exists and $w<\rho d(\ker H)$. When the isolation step succeeds and the reduction makes a query, the oracle must return a solution to the corresponding search problem. Lemma~\ref{lem:normalized-output} and the recovery steps above show that every such solution lets us recover an input accepted by the circuit used in the construction: $F$ in (i), and $F'$ in (ii) and (iii). In (ii) and (iii), an input accepted by $F'$ is also accepted by $F$. If $F$ is unsatisfiable, the final evaluation of $F$ rejects every candidate input, regardless of the random choices or what the oracle returns. Constructing the query, checking the length and format of the returned string, and recovering and checking a candidate input all take polynomial time, and each reduction makes at most one oracle call.
\end{proof}

\subsection{Parameterized GapSVP}

We apply the lattice construction to prove hardness of parameterized GapSVP for every constant approximation factor. The problem $\GapSVP^{(p)}_\Gamma$ receives a lattice basis and a threshold $d$ and distinguishes $\lambda_1\le d$ from $\lambda_1>\Gamma d$. For the parameterized problem, the basis is integer and the positive integer threshold $d$ is the parameter. A fixed-parameter tractable (FPT) algorithm would run in time $f(d)N^{O(1)}$ on inputs of length $N$, for some computable function $f$.

An FPT algorithm for this problem would yield a randomized one-sided-error algorithm for $k$-Clique in time $f(k)n^{O(1)}$ on a graph with $n$ vertices. Bennett, Cheraghchi, Guruswami, and Ribeiro proved two-sided-error hardness for every constant factor when $p>1$ and below $2$ when $p=1$ \cite{BCGR23}. Hirahara and Ogitsuka proved one-sided-error hardness below $2^{1/p}$ and asked whether it extends to every constant factor when $p>1$ and to factors below $2$ when $p=1$ \cite{HO26}. Corollary~\ref{cor:main-gapsvp} answers their question and establishes $\mathsf W[1]$-hardness of $\ell_1$-GapSVP beyond factor $2$, answering Bennett's Open Problem~4.7 \cite{Ben23}.

\begin{corollary}[Parameterized GapSVP]\label{cor:formal-gapsvp}\label{cor:main-gapsvp}
For every fixed polynomial-time computable $1\le p<\infty$ and constant $\Gamma>1$, parameterized $\Gamma$-GapSVP with integer bases in $\ell_p$ is $\mathsf W[1]$-hard under randomized FPT many-one reductions with one-sided error, in the sense of \cite[Definitions~9 and~14]{HO26}. The target parameter is an integer threshold $d=k^{O_{p,\Gamma}(1)}$, where $k$ is the clique size in the input instance of $k$-Clique.
\end{corollary}
\begin{proof}
Handle $k=0$ or $n=0$ directly. Create $k$ blocks, numbered $1,\ldots,k$, with one label for each vertex. For each pair $i<j$ of these block numbers, add a block with one label for each ordered pair of distinct adjacent vertices. Represent vertices by fixed-length binary strings. Require the bits encoding the first endpoint in the block for $i,j$ to agree with the corresponding bits in block $i$, and the bits encoding the second endpoint to agree with those in block $j$. Compatibility forces each block for a pair $i,j$ to choose the edge joining the vertices chosen in blocks $i$ and $j$. Thus compatible selections are in bijection with ordered $k$-cliques. The number of blocks is $g=k+\binom k2$, and $L=O(k^2n^2)$. If a block is empty, there is no $k$-clique, so output a fixed NO instance directly.

Apply Proposition~\ref{prop:compatibility-lattice} with $\rho=1/4$ and $G=2\max\{\Gamma,4\}$. In its construction, the tensor order $m$ depends only on $p,\Gamma$. Lemma~\ref{lem:normalized-output} uses a rational upper bound $u$ on the $\ell_p$ norm of the tensor encoding a selection to set the integer threshold $\widehat r=\lceil4u\rceil\le6g^{m/p}$. Thus the threshold depends only on $k$ and these fixed parameters. On a successful trial for a graph with a $k$-clique, the output has $\lambda_1\le\widehat r$. If the graph has no $k$-clique, every seed produces an output with $\lambda_1>\Gamma\widehat r$. These two bounds on $\lambda_1$ are the YES and NO guarantees for GapSVP. Use the same integer threshold $\widehat r$ in every trial.

Let $B_b$ be the set of labels in block $b$. Then $J=\lceil\log_2\prod_b|B_b|\rceil=O(1+k^2\log(n+1))$, and if a $k$-clique exists, each trial succeeds with probability at least $0.20/(J+1)$. To amplify this probability, run the reduction independently
\[
 R=\left\lceil\frac{4(J+1)}{0.20}\right\rceil
\]
times and take the direct sum of the resulting lattices. The common threshold is already an integer $d=\widehat r=k^{O_{p,\Gamma}(1)}$. The shortest nonzero vector length in a direct sum is the minimum of those in its components, so a graph without a $k$-clique maps to a NO instance for every choice of seeds. If a $k$-clique exists, independence implies that at least one trial succeeds with probability at least $1-e^{-4}>2/3$, and on this event the direct sum contains a vector of length at most $d$. Each trial has size $(kn+2)^{O_{p,\Gamma}(1)}$, and $R=O(1+k^2\log(n+1))$. 
\end{proof}

\subsection{Growing factors}

\begin{corollary}[Growing factors]\label{cor:formal-growing}\label{cor:main-growing}
Fix $1\le p<\infty$ and ${0<\beta<1/2}$, both polynomial-time computable. There is a randomized quasipolynomial-time many-one reduction from \CircuitSAT{} to decision uSVP in $\ell_p$ at factor $(\log(n+2))^\beta$, where $n$ is the rank of the output lattice. It has no false positives and succeeds on a satisfiable size-$s$ circuit with probability at least $0.20/(2s+1)$. Its running time is
\begin{equation}\label{eq:growing-time}
 \exp\!\left(O_{p,\beta}\!\left((\log(s+2))^{1/(1-2\beta)}\right)\right).
\end{equation}
Consequently, a polynomial-time algorithm for either the decision or search version of uSVP at this factor yields a randomized \CircuitSAT{} algorithm with running time bounded by \eqref{eq:growing-time}, no false positives, and success probability at least $2/3$.
\end{corollary}
\begin{proof}
The required factor $(\log(n+2))^\beta$ depends on the output rank $n$, while the bound on $n$ depends on the factor $\Gamma_s$ chosen for the construction. We choose $\Gamma_s$ as a function of the circuit size so that it is at least $(\log(n+2))^\beta$. Let $a=\beta/(1-2\beta)$. Choose a sufficiently large integer $A$ depending on $p$ and $\beta$, and compute an integer $\Gamma_s$ satisfying
\[
 A(\log(s+2))^a\le\Gamma_s\le A(\log(s+2))^a+2.
\]
Compute a rational upper bound on $A(\log(s+2))^a$ with error less than one and round it up to an integer. Apply Proposition~\ref{prop:compatibility-lattice} to the instance from the circuit, with uSVP uniqueness factor $\Gamma_s$, $\rho=1/4$, and $G=2\max\{\Gamma_s,4\}$. Let $\widehat r$ be the threshold in the resulting uSVP instance.

Since $g\le s$ and $L\le4g$, Proposition~\ref{prop:compatibility-lattice} bounds the output dimension and running time by $(s+2)^{K_pG^2}$ for a constant $K_p$ independent of $G$ and $s$. Hence
\[
 \log(n+2)\le K'_p A^2(\log(s+2))^{2a+1}.
\]
Since $a=\beta(2a+1)$ and $1-2\beta>0$, choosing $A$ sufficiently large ensures
\[
 \Gamma_s\ge(\log(n+2))^\beta.
\]
The factor $\gamma=(\log(n+2))^\beta$ depends on the output rank $n$. The preceding bound implies $\gamma\le\Gamma_s$. On a successful output, $\lambda_1\le\widehat r<\gamma\widehat r\le\Gamma_s\widehat r<\lambda_2$. On every output from an unsatisfiable input, $\lambda_1>\Gamma_s\widehat r\ge\gamma\widehat r$. Thus the constructed instances satisfy the uSVP YES and NO conditions at factor $\gamma$. The running time is
\[
 \exp\!\left(O_{p,\beta}\!\left((\log(s+2))^{2a+1}\right)\right)
 =
 \exp\!\left(O_{p,\beta}\!\left((\log(s+2))^{1/(1-2\beta)}\right)\right).
\]

As in the proof of Theorem~\ref{thm:formal-lattice}, $J\le2g\le2s$. A single trial succeeds with probability at least $0.20/(2s+1)$ and has no false positives. Repeat independently $O(s)$ times. For the decision version of uSVP, accept if the algorithm accepts any of the constructed instances. For the search version of uSVP, try to recover a candidate input from what the oracle returns on each trial and accept only if it satisfies the original circuit, as in Corollary~\ref{cor:formal-search-usvp}(i). The same length checks and recovery operations take time polynomial in the number of bits describing each lattice instance sent to the oracle, even when the oracle returns arbitrary strings. These repetitions raise the probability of acceptance on a satisfiable circuit above $2/3$ without changing the bound in \eqref{eq:growing-time}.
\end{proof}

\section{Construction and parameter details}\label{sec:implementation}

This appendix completes the reductions by constructing bases and thresholds efficiently and making the output lattices full rank. It also checks the dimensions, bit lengths, and running time.

\subsection{Minimum distance of the Vandermonde code}

The lattice construction uses a Vandermonde code because its minimum distance stays large when each column of the parity-check matrix is multiplied by a nonzero field element. Let $P$ be prime, let $1\le h<N\le P-1$, and choose distinct $\alpha_1,\ldots,\alpha_N\in\F_P^*$. Define the $h\times N$ Vandermonde matrix $H$ by
\[
H_{j,i}=\alpha_i^j\qquad(0\le j<h,\ 1\le i\le N).
\]
For $\beta=(\beta_1,\ldots,\beta_N)\in(\F_P^*)^N$, let $C_\beta=\ker(H\operatorname{diag}(\beta))$ and $\delta=h+1$.

\begin{lemma}[Minimum distance]\label{lem:multiplier-mds}
For every $\beta\in(\F_P^*)^N$,
\[
 d(C_\beta)=h+1=\delta.
\]
\end{lemma}
\begin{proof}
Every $h$ columns of $H$ are independent by the Vandermonde determinant, and multiplying columns by nonzero field elements preserves this property. Thus $d(C_\beta)\ge h+1$. Since $N>h$, any $h+1$ columns are dependent. Their dependence defines a nonzero codeword of weight at most $h+1$. Hence $d(C_\beta)=h+1$.
\end{proof}

\subsection{Computing bases, thresholds, and primes}

\begin{lemma}[A basis with small entries]\label{lem:small-basis}
Given $W$ by a basis or linear equations, one can compute a basis of $\Lambda(W,P)$ in polynomial time, with every entry of its basis matrix having absolute value at most $P$. Given a nonempty affine subspace of $\F_P^D$ by linear equations, one can compute an integer vector with coordinates in $\{0,\ldots,P-1\}$ whose reduction modulo $P$ lies in that subspace.
\end{lemma}
\begin{proof}
Row-reduce a parity-check matrix of $W$ over $\F_P$, discard dependent rows, and permute the columns so that the matrix has the form $[I_a\ F]$. Choose the entries of $F$ in $\{0,\ldots,P-1\}$. In the permuted coordinates, the columns of
\begin{equation}\label{eq:small-basis}
\begin{pmatrix}
PI_a&-F\\
0&I_{D-a}
\end{pmatrix}
\end{equation}
form an integer basis. After undoing the coordinate permutation, these columns form a basis of $\Lambda(W,P)$. Gaussian elimination detects whether the affine equations are consistent and, if so, finds a solution. Replace each coordinate by an integer in $\{0,\ldots,P-1\}$ with that residue modulo $P$.
\end{proof}

\begin{lemma}[Rational thresholds]\label{lem:rational-threshold}
Fix polynomial-time computable real constants $a>0$ and $\eta>0$. Given positive integers $s$ and $m$, one can compute a rational number $r$ such that
\[
\left(1+\frac{\eta}{3}\right)s^{m/a}
<r<
\left(1+\frac{2\eta}{3}\right)s^{m/a}.
\]
The computation takes time polynomial in $m+\log s$, with constants depending on $a$ and $\eta$, and the bit length of $r$ is $O_{a,\eta}(m\log(s+1))$.
\end{lemma}
\begin{proof}
The number $s^{m/a}=\exp((m/a)\log s)$ has rational upper and lower bounds computable in time polynomial in $m+\log s$ to any fixed relative precision. The constants $1+\eta/3$ and $1+2\eta/3$ are polynomial-time computable and have a positive gap. Compute rational intervals enclosing the two products to relative precision depending only on $\eta$, so that the gap between an upper bound for the first and a lower bound for the second is at least $\eta s^{m/a}/6$. Between these bounds, choose a rational with the smallest possible denominator that is a power of two. It satisfies the strict inequalities. The relative precision needed depends only on $a$ and $\eta$, and the resulting number has $O_{a,\eta}(m\log(s+1))$ bits.
\end{proof}

\begin{lemma}[A prime in a numerical interval]\label{lem:prime-interval}
For every integer $Q_0\ge2$, a prime $P\in[Q_0,2Q_0]$ can be found deterministically in time polynomial in the numerical value of $Q_0$. In particular, the procedure is polynomial time whenever $Q_0$ is polynomially bounded in the reduction's input size.
\end{lemma}
\begin{proof}
By Bertrand's postulate, the interval contains a prime. Test the integers from $Q_0$ through $2Q_0$ using a deterministic polynomial-time primality test. There are at most $Q_0+1$ tests, so the running time is polynomial in the numerical value of $Q_0$.
\end{proof}

\subsection{Completing the uSVP and BDD lattices to full rank}\label{sec:output-details}

\begin{proof}[Proof of Lemma~\ref{lem:normalized-output}]
For each integer $h\ge1$, we add basis vectors to the lattice $\{(x,h\mu(x)):x\in\Lambda\}$ to make it full rank. We show that every point of the resulting lattice within distance $\Delta$ of zero or of the BDD target still has this form. We then choose the scale $h$ of the appended coordinate separately for uSVP and BDD.

Write $\Lambda=B\Z^r$ with $B$ of full column rank, and let $\alpha$ be
the row of values of $\mu$ on its columns. Choose $r$ pivot rows $I$ of
$B$ and let $J$ be the remaining rows of $B$. On the real span of the columns of $(B;\alpha)$, the coordinates in $I$ determine the coordinates in $J$ and the last coordinate. These relations define a linear map that vanishes on every column of the stacked matrix $(B;\alpha)$. We choose the integer $Q$ so that lattice points with nonzero coefficients on the added columns lie outside the balls used for uSVP and BDD. We first prove this for $h=1$ and then transfer it to every $h\ge1$ by shrinking the last coordinate. Let
\[
 \begin{gathered}
 F=B_JB_I^{-1},\\
c=\alpha B_I^{-1}.
 \end{gathered}
\]
Define $\pi:\R^{D+1}\to\R^{D-r+1}$ by
$\pi(x,s)=(x_J-Fx_I,s-cx_I)$, and set
\[
 K_\pi=1+\max\{\max_j\norm{F_j}_1,\norm c_1\}.
\]
Here $F_j$ is row $j$ of $F$. The maximum over an empty set is zero. Choose $Q=\lfloor1+K_\pi\Delta\rfloor+1$.
For an integer $h\ge1$, let $L_h$ have basis
\begin{equation}\label{eq:graph-completion}
 G_h=\begin{pmatrix}I_D&0\\0&h\end{pmatrix}
 \begin{pmatrix}B&QE_J&0\\\alpha&0&Q\end{pmatrix},
\end{equation}
where $E_J$ comprises the standard coordinate columns indexed by $J$.
This is a square nonsingular integer matrix.

For $h=1$, the kernel of $\pi$ is the span over the reals of the columns of the
stacked matrix $(B;\alpha)$. The map $\pi$ sends the original columns of $G_1$ to zero and the added columns to $Q$ times the standard basis. Thus $\pi z=Qk$ for each $z\in L_1$, where $k$ is the vector of integer coefficients of $z$ on the added columns. Consider the $\ell_p$ balls of radius $\Delta$ in $\R^{D+1}$. For a point $z\in L_1$ in the ball about zero, $\norm{\pi z}_\infty\le K_\pi\Delta<Q$, so $\pi z=0$. For a point $z\in L_1$ in the ball about $e_{D+1}$, consider its coefficient vector $k$ on the added columns. Since $\pi(e_{D+1})=e_{D-r+1}$, we have $\norm{Qk-e_{D-r+1}}_\infty\le K_\pi\Delta$. If $k\ne0$, a coordinate of $Qk-e_{D-r+1}$ has magnitude at least $Q-1>K_\pi\Delta$, a contradiction. Thus $k=0$.
The coefficients of the lattice point $z$ on the added basis vectors therefore vanish, so its remaining integer coefficients place $z$ in the lattice $\{(x,\mu(x)):x\in\Lambda\}$.
For $h\ge1$, the map $S_h(x,t)=(x,t/h)$ maps the lattice $L_h$ to $L_1$, maps $he_{D+1}$ to $e_{D+1}$, and does not increase any $\ell_p$ norm, including $p=\infty$. It therefore maps lattice points in the balls of radius $\Delta$ about zero and $he_{D+1}$ to lattice points of $L_1$ in the corresponding balls about zero and $e_{D+1}$. Thus every lattice point in either ball in $L_h$ has the form $(x,h\mu(x))$ for some $x\in\Lambda$.

For uSVP take $h_U=\lceil u\rceil$. In case (ii), the vector
$(y,h_U)$ has norm at most
$R+h_U\le13R/4<d$, while $4R\le d\le6R$.
For a lattice vector of norm at most $\Delta$ that is not a real multiple of $(y,h_U)$, write it as $(x,h_Uj)$ with $x\in\Lambda$ and $j=\mu(x)$. Then $x-jy\in\Lambda\cap\ker\mu$ and is nonzero, since otherwise $(x,h_Uj)=j(y,h_U)$. The lower bound in case (ii) therefore gives
\[
 \Delta<\norm{x-jy}_p
 \le\norm x_p+|j|R
 \le2\norm{(x,h_Uj)}_p.
\]
Lattice vectors outside the ball of radius $\Delta$ have norm greater than $\Delta$. Thus every vector
that is not a real multiple of $(y,h_U)$ has norm greater than
$\Delta/2>\Gamma d$, since $\Delta/2>8\tau R\ge8\Gamma R>\Gamma d$.
This proves the YES promise. For search recovery, the last coordinate of every point of $L_{h_U}$ lies in $h_U\Z$. Thus a lattice point $t(y,h_U)$ on the real line spanned by $(y,h_U)$ has $t\in\Z$, and its first $D$ coordinates form the vector $ty\in\Lambda$.
Under (i), every nonzero output vector is
longer than $\Delta>\Gamma d$, proving the NO promise.

For BDD take $h_D=\lceil\Delta\rceil+1$, lattice $L_{h_D}$, and target $t=h_De_{D+1}$. A nonzero vector $(x,h_D\mu(x))$ of norm at most $\Delta$ would have $\mu(x)=0$, since $h_D>\Delta$ and $\mu(x)$ is an integer. Then $x$ would be a nonzero vector in $\Lambda\cap\ker\mu$ of norm at most $\Delta$, contradicting the lower bound in both (i) and (ii). Every lattice point in the ball of radius $\Delta$ about zero has zero coefficients on the added basis vectors, so $\lambda_1(L_{h_D})>\Delta$ in both cases. Since $d\le6R$ and $\rho\Delta>16R$ by $\tau\ge1/\rho$, we have $d<\rho\lambda_1(L_{h_D})$. Under (ii), $(y,h_D)$ is within distance $R<d$ of $t$. Under (i), a lattice point within distance $\Delta$ of $t$ would have the form $(x,h_D\mu(x))$ and have $\mu(x)=1$, forcing $x$ to be a nonzero vector of $\Lambda$ of norm at most $\Delta$. This contradicts case (i), where every nonzero vector of $\Lambda$ has norm greater than $\Delta$. Thus, in case (i), the distance from the target $t$ to $L_{h_D}$ exceeds $\Delta$.

In case (ii), the shortest nonzero vectors are $\pm(y,h_U)$ for uSVP, and the closest lattice point to the BDD target is $(y,h_D)$. Any other lattice point at distance at most $d$ from that target has the form $(x,h_D\mu(x))$ and has $\mu(x)=1$, since $d<h_D$ and $\mu(x)\in\mathbb Z$. Subtracting $y$ from its first $D$ coordinates would produce a nonzero vector in $\Lambda\cap\ker\mu$ of norm at most $d+R\le7R<\Delta$, contradicting the lower bound $\Delta$. The first $D$ coordinates recover $y$, up to sign for uSVP. We can compute $B_I^{-1}$ and choose $Q$ in polynomial time in the input bit length.
\end{proof}

\subsection{Parameters for the lattice reduction with constant factors}\label{sec:lattice-parameters}

\begin{proposition}[From constraints to lattices]\label{prop:compatibility-lattice}
Fix a polynomial-time computable $1\le p<\infty$, polynomial-time computable parameters $\Gamma>1$ and ${0<\rho<1/2}$, and a polynomial-time computable $G\ge2\max\{\Gamma,1/\rho\}$. For an instance of the blocks and compatibility requirements from Section~\ref{sec:common-source}, with $g$ nonempty blocks and $L$ labels, there are two randomized constructions.
\begin{enumerate}[label=(\roman*)]
\item The first applies to an arbitrary instance. If it has a solution, it produces a promised YES instance of $\Up\Gamma$, and likewise of $\Dp\rho$, with probability at least $0.20/(J+1)$, where $\ell_i$ is the number of labels in block $i$ and
\[
 J=\left\lceil\log_2\prod_{i=1}^g\ell_i\right\rceil.
\]
\item The second applies under the promise that the instance has zero or one solution. If it has one, it produces a promised YES instance with probability greater than $0.24$.
\end{enumerate}
An instance with no compatible selection produces a NO instance satisfying the corresponding target problem's conditions for every random seed in either construction. The output dimension is
\[
 (gL+2)^{O_p(G^2)}.
\]
Each coordinate of a basis vector or target and each rational threshold has bit length $O_p(G\log(gL+2))$. For fixed real parameters, the construction runs in polynomial time. When $G$ is supplied as an integer, its running time is also $(gL+2)^{O_p(G^2)}$, with constants depending only on $p$.
\end{proposition}
\begin{proof}[Proof of Proposition~\ref{prop:compatibility-lattice}]
Lemma~\ref{lem:normalized-output} requires lower bounds on the norms of lattice vectors in two cases. Both bounds must be much larger than the norm of an encoding. If no compatible selection has its indicator in the code constructed below, the bound must apply to every nonzero lattice vector. If exactly one compatible selection has its indicator in the code, it must apply to every nonzero lattice vector with $\mu=0$. We also choose a rational upper bound on the norm of an encoding to set the thresholds for the output instances. Apply Proposition~\ref{prop:lp-separation-main} with $32G$ in place of $G$ to choose $m=O_p(G)$, $\varepsilon=1/(10m)$, and $Q_{\rm sep}$.
Set $R=g^{m/p}$, choose
$13R/12<u<7R/6$ by Lemma~\ref{lem:rational-threshold}, and choose a
rational $\widehat G$ with $G\le\widehat G\le9G/8$, taking $\widehat G=G$ when $G$ is an integer. Let
\[
 \begin{gathered}
 \Delta=16\widehat G u,\\
\widehat r=\lceil4u\rceil.
 \end{gathered}
\]
Then $16\max\{\Gamma,1/\rho\}R<\Delta<21GR<32GR$ and
$4R\le\widehat r\le6R$.
Let $g_0$ be the constant from Proposition~\ref{prop:lattice-isolation-main}. For $g<g_0$, enumerate all selections. If a solution exists, output $\Z$ for uSVP and $M_D\Z$ with target zero for BDD. If no solution exists, output $M_U\Z$ for uSVP and $M_D\Z$ with target $M_D/2$ for BDD, where
$M_U=\lfloor\Delta\rfloor+1$ and $M_D=\lfloor\max\{\widehat G\widehat r,2\Delta\}\rfloor+1$. These choices satisfy $M_U>\Delta$ and $M_D>\max\{\widehat r/\rho,2\Delta\}$.
These are promised instances with threshold $\widehat r$.

For $g\ge g_0$, enlarge $Q_{\rm sep}$ so that
$Q_{\rm sep}>64GR$ and set
\[
 Q_0=2+Q_{\rm sep}+(2^{24}gL)^{10m}.
\]
Proposition~\ref{prop:lattice-isolation-main} constructs a prime $P\in[Q_0,2Q_0]$, the copies of the labels, and a code $C$ on them. Use the space $V$ from Proposition~\ref{prop:partition-main}(ii). Its rank-one matrices are $ayy^T$, where $a\ne0$ and $y\in C$ is the indicator of a compatible selection. The space $W=S_m(V)$ contains $y^{\otimes m}$ for every such $y$. Take the lattice
\[
 \begin{gathered}
 \Lambda=\Lambda(W,P),\\
\mu(T)=\sum_{j\in B_1}T_{j,\ldots,j}.
 \end{gathered}
\]
The diagonal sum defining $\mu$ has $|B_1|\le N$ coefficients equal to one and all others zero. For each of these indicators $y$, the tensor $v=y^{\otimes m}$ satisfies
$\mu(v)=1$ and $\norm v_p=R$.
If the original blocks and compatibility requirements have no compatible selection, Proposition~\ref{prop:lp-separation-main} and
$P>32GR$ imply $\lambda_1(\Lambda)>32GR$ for every random choice.
If exactly one compatible selection has its indicator in $C$ for the random choice, Proposition~\ref{prop:lp-separation-main} proves that every nonzero vector of $\Lambda\cap\ker\mu$ has norm greater than $32GR$. Applying Lemma~\ref{lem:normalized-output} with the chosen $u$ constructs instances of both problems with threshold $d=\widehat r$: NO instances in the first case and YES instances in the second. The success probabilities are at least those guaranteed by Proposition~\ref{prop:lattice-isolation-main}. 

In the following bounds, all implicit constants depend only on $p$ and are independent of the integer $G$ used in Corollary~\ref{cor:formal-growing}. $m=O_p(G)$,
$P,N\le(gL+2)^{O_p(G)}$, and the dimension of $\Lambda(W,P)$ is
$D=N^m\le(gL+2)^{O_p(G^2)}$.
First discard dependent linear equations defining $V$, leaving at most $N^2$ equations. The equations requiring every matrix slice to lie in $V$, together with the symmetry equations defining $S_m(V)$, number $O(m^2D)$. Since $m\le\log_2D$, the linear system has size polynomial in $D$. Gaussian elimination takes time polynomial in $D$ and $\log P$, with an exponent independent of $G$. The full-rank construction of Lemma~\ref{lem:normalized-output} adds only one dimension.
Use the basis $B$ from Lemma~\ref{lem:small-basis}. Since $\Lambda(W,P)$ has full rank, all rows of $B$ are pivot rows in Lemma~\ref{lem:normalized-output}. Write $\mu(x)=c_0x$. The row of values of $\mu$ on the basis vectors is $\alpha=c_0B$, so $c=\alpha B^{-1}=c_0$. Consequently, $K_\pi=1+\norm c_1\le1+N$. The values of $\mu$ on the basis vectors have absolute value at most $NP$.
Thus $Q$, the integers $\lceil u\rceil$ and $\lceil\Delta\rceil+1$ that scale $\mu(x)$ in the added coordinate, and every entry of the bases and target we construct have bit length $O_p(G\log(gL+2))$, as does the threshold.
All arithmetic and the deterministic prime search obey the running time bound.
\end{proof}

\needspace{4\baselineskip}
\par\medskip
{\small\noindent\textbf{Statement on AI use.} The main ideas and proof strategies were developed using OpenAI Codex with the GPT-5.6 Sol and GPT-6 Astra models. Codex was also used to search the literature and in the presentation and drafting of this paper. The authors studied, refined, simplified, and verified the resulting material and take responsibility for every claim, proof, and citation.\par}

\end{document}